\documentclass[aos, preprint]{imsart}
\RequirePackage[OT1]{fontenc}
\RequirePackage{amsthm,amsmath}
\allowdisplaybreaks[4]
\usepackage{array}
\usepackage{url}
\usepackage{mathrsfs}
\usepackage{amssymb,bbm}
\usepackage{amsfonts}
\usepackage{amsmath,amssymb}
\usepackage{graphicx}
\usepackage{amscd}
\usepackage{color}
\usepackage{float}
\usepackage{mathrsfs}
\usepackage{latexsym,bm}
\usepackage{array}
\usepackage{threeparttable}
\usepackage{enumerate}
\usepackage{rotating}
\usepackage{booktabs}
\usepackage{rotating}
\usepackage{multirow}
\usepackage{float}
\usepackage{tablefootnote}
\usepackage{subfigure}
\usepackage[ruled,linesnumbered]{algorithm2e}
\usepackage{diagbox}
\usepackage{natbib}
\usepackage{graphicx}
\usepackage{geometry}
\usepackage{bold-extra}

\DeclareFontFamily{U}{mathx}{\hyphenchar\font45}
\DeclareFontShape{U}{mathx}{m}{n}{
      <5> <6> <7> <8> <9> <10>
      <10.95> <12> <14.4> <17.28> <20.74> <24.88>
      mathx10
      }{}
\DeclareSymbolFont{mathx}{U}{mathx}{m}{n}
\DeclareMathAccent{\widecheck}{\mathalpha}{mathx}{"71}

\DeclareMathAccent{\widecheck}{\mathalpha}{mathx}{"71}

\startlocaldefs 
\numberwithin{equation}{section}
\numberwithin{figure}{section}
\numberwithin{table}{section}
\theoremstyle{it}
\newtheorem{thm}{Theorem}[section]
\newtheorem{lemma}{Lemma}[section]
\newtheorem{ass}{Assumption}[section]

\endlocaldefs
\SetKwInput{KwIni}{Initialization}
\SetKwInput{KwPro}{Procedure}
\SetKw{KwendPro}{end procedure}
\def\bftau{\mathbb{\pmb{\tau}}}
\def\bbeta{\mathbb{\pmb{\beta}}}

\begin{document}
\begin{frontmatter}
	\title{ Time series analysis of COVID-19 Infection Curve: \\a change-point perspective}
	%\runtitle{Inference for augmented DAR models}
	
	\begin{aug}
				\author{\fnms{Feiyu} \snm{Jiang}\thanksref{m1}\thanksref{t1}\ead[label=e1]{jfy16@mails.tsinghua.edu.cn}},
		\author{\fnms{Zifeng} \snm{Zhao}\thanksref{m2}\ead[label=e2]{zzhao2@nd.edu}
			\ead[label=u1, url]{http://www.foo.com}}
 \and
		\author{\fnms{Xiaofeng} \snm{Shao}\thanksref{m3}\thanksref{t3}\ead[label=e3]{xshao@illinois.edu}
			\ead[label=u1, url]{http://www.foo.com}}

			\thankstext{t1}{Jiang is supported by the National Key Research and
				Development Program of China (No. 2018YFC1603105), China Scholarship Council (No.201906210093), 	
				National Natural Science Foundation of China (No.11571348 and No.11771239) and acknowledges that  the work was carried out during the visit at Department of Statistics, University of Illinois at Urbana-Champaign. 	}
			\thankstext{t3}{Shao is supported in part by NSF-DMS 1807032.}

		\affiliation{ Tsinghua University\thanksmark{m1}, University of Notre Dame\thanksmark{m2} and  University of Illinois at Urbana Champaign\thanksmark{m3}  }
		
			\address{
		Center for Statistical Science\\ \quad and Department of Industrial Engineering\\
		Tsinghua University\\
		Beijing 100084, China\\
		\printead{e1}\\
	}	
		
		\address
		{Department of Information Technology, Analytics, and Operations\\Mendoza College of Business \\
			University of Notre Dame\\
			Notre Dame, Indiana 46556, U.S.A.\\
			\printead{e2}
		}

		\address
		{Department of Statistics \\
			University of Illinois at Urbana Champaign\\
			Champaign, Illinois 61820, U.S.A. \\
			\printead{e3}
		}
		
	\end{aug}

	\begin{abstract}
	In this paper, we model the trajectory of the cumulative confirmed cases and deaths of COVID-19 (in log scale) via a piecewise linear trend model. The model naturally captures the phase transitions of the epidemic growth rate via change-points and further enjoys great interpretability due to its semiparametric nature.  On the methodological front, we advance the nascent self-normalization~(SN) technique \citep{shao2010self} to testing and estimation of a single change-point in the linear trend of a nonstationary time series. We further combine the SN-based change-point test with the NOT algorithm \citep{baranowski2019narrowest} to achieve multiple change-point estimation. Using the proposed method, we analyze the trajectory of the cumulative COVID-19 cases and deaths for 30 major countries and discover interesting patterns with potentially relevant implications for effectiveness of the pandemic responses by different countries. Furthermore, based on the change-point detection algorithm and a flexible extrapolation function, we design a simple two-stage forecasting scheme for COVID-19 and demonstrate its promising performance in predicting cumulative deaths in the U.S.

    \end{abstract}

\end{frontmatter}

\section{Introduction}
%In the middle of the COVID-19 pandemic, it is crucial to keep track of the number of infections and understand the growth pattern of the virus.  As we all know, the number of confirmed, and death cases has been increasing every day and the growth rate of the infected cases can be very different for different counties, states and countries as well as at different times. The enforcement of social distancing measures and shelter-in-place orders by governments certainly had a positive impact on slowing down the spread of novel coronavirus  but may also have  negative psychological and economic impacts on individuals. By this moment, each county/state/country is on a different trajectory in terms of growth rate, and this could have important policy implications. Specifically, policymakers need decision support systems to decide whether it is a good time to ease the strict social distancing measures, and the information drawn from the daily infected cases would be critical. Such information can also be used to understand the difference of COVID-19 spread for many countries, the degree of flattening of infection curve, the effectiveness of lock-down measure, among others. 

%A motivating application of the piecewise linear trend model \eqref{eq: linear trend} is for modeling the dynamic behavior, such as cumulative confirmed cases, death, recovery, of an infectious disease like the novel coronavirus, COVID-19. 

Since the initial outbreak of the novel coronavirus in Wuhan, China in early January 2020, the COVID-19 pandemic has rapidly spread across the world. Due to the high infectivity of the virus and the lack of immunity in the human population,  the epidemic grows exponentially without intervention, and thus can greatly stress the public health system and bring enormous disruption to economy and society. Thus, a crucial task facing every country is to reduce the transmission  rate and flatten the (infection) curve. Various emergency measures, such as regional lockdown and mass testing, have been taken by different countries and a natural question is whether (and to what degree) these interventions are effective in slowing down the pandemic. Additionally, each country is at a different stage of the epidemic and it is essential for countries to understand its own pattern of virus growth, as such information is critical for important policy decisions such as extending lockdown or reopening. To (at least partially) answer these questions, a natural step is to analyze the trajectory of the infection curve of COVID-19 since the initial outbreak in each country.

In this paper, we propose to model the time series of cumulative confirmed cases and deaths (in log scale) of each country via a piecewise linear trend model (see formal definition later). In other words, we model the mean of the logarithm of cumulative infection as a linear trend with an unknown number of potential changes in the intercept and slope, as it is natural to expect that the spread of COVID-19 may experience several phases, where the initial growth is typically rapid due to absence of immunity and lack of preparation, and the spread may then evolve into phases with slower growth depending on government intervention and public health responses~(i.e. flattening the curve). The estimation of such a model can be formulated as a change-point detection problem. 

In recent years, change-point analysis has become an increasingly active research area in statistics and econometrics thanks to its applications across a wide range of fields, including bioinformatics~\citep{Fan2017}, climate science~\citep{Gromenko2017}, economics~\citep{bai1994least,bai1997estimation,Cho2015}, finance~\citep{fryzlewicz2014wild}, medical science~\citep{chen2011parametric}, and signal processing~\citep{SP2018}; see \cite{perron2006dealing}, \cite{aue2013structural} and \cite{truong2020selective} for some recent reviews. However, most existing change-point literature operates under the piecewise stationarity assumption, where it is assumed that the time series of interest is (potentially) non-stationary but can be partitioned into piecewise stationary segments such that observations within each segment are stationary and share a common parameter of interest such as mean or variance. While the piecewise stationarity assumption is proven to be reasonable and fruitful for many applications, methods developed under this framework cannot handle time series with intrinsic non-stationarity, such as the cumulative infection curve of COVID-19. 

A simple but important class of time series with intrinsic non-stationarity is the piecewise linear trend model, which has the following mathematical formulation. Let the time series $\{Y_t\}_{t=1}^n$ admit 
\begin{align}\label{eq: linear trend}
&Y_{t}=a_t+ b_t (t/n) + u_t,~ t=1,\cdots,n,\\
&(a_t,b_t)=\bbeta^{(i)}=(\beta_{0}^{(i)},\beta_{1}^{(i)})^{\top}, \tau_{i-1}+1\leq t\leq \tau_i,\text{ for }i=1,\cdots,m+1,  \nonumber
\end{align}
where $(a_t, b_t)$ is the linear trend~(intercept and slope) of $\mathbb{E}(Y_t)$ at time $t$, $\{u_t\}$ is a weakly dependent stationary error process, $\bftau=(\tau_1,\cdots,\tau_m)$ denotes the $m\geq 0$ change-points with the convention that $\tau_0=0$ and $\tau_{m+1}=n$, and we require $\bbeta^{(i)}\neq \bbeta^{(i+1)}, i=1,\cdots,m$. In this paper, we set $\{Y_t\}_{t=1}^n$ to be the time series of daily cumulative confirmed cases or deaths (in log scale) of COVID-19. Due to the log transformation, the slope $b_t$ naturally measures the growth rate of the virus at day $t$.

The piecewise linear trend model is intuitive, interpretable and is useful for tracking the dynamics of a pandemic as it naturally segments the spread process into phases with (approximately) the same growth rate. The slope of the last segment can shed light on the current status of the pandemic and provide short-term forecast, while the estimated change-points can be compared with dates when emergency measures such as lockdown were introduced to help assess the effectiveness of different policies. Also, the semiparametric nature of \eqref{eq: linear trend} helps to achieve model flexibility while maintaining simplicity, which is advantageous for modeling the cumulative cases at the early stage of a pandemic as the time series is relatively short, curbing the use of sophisticated fully nonparametric methods.

An important part in estimation of \eqref{eq: linear trend} is  to recover the unknown number $m$ and location $\bftau$ of the change-points. As discussed above, such a problem has mostly been ignored in the change-point literature with only a few exceptions. A CUSUM based detection algorithm is proposed in \cite{baranowski2019narrowest}, and a model selection based procedure is derived in \cite{Fearnhead2019}. However, both methods assume temporal independence of $\{u_t\}$, which can be restrictive as serial dependence is commonly found in time series data. Although \cite{baranowski2019narrowest} briefly discussed possible extensions to temporally dependent series, potentially important issues such as choice of tuning parameters seem not carefully addressed. \cite{bai1998estimating} can detect structural breaks in the linear trend model under serial dependence. However, numerical study (see Section 4) suggests that their method is relatively sensitive to positive temporal dependence, which is indeed exhibited by the COVID-19 data, and may give less favorable estimation performance under small sample size.

%partially due to the more challenging nature of the problem. For example, as noted by \cite{Baranowski2019}, the standard binary segmentation does not work for detecting changes in slope. This is a popular model in economics literature for studying dynamics of macroeconomics time series, such as GDP and unemployment rate, see \cite{Perron1989}, \cite{Stock1999} and \cite{Rho2015}.

%Depending on the application, sometimes it is desirable to additionally impose a ``continuous" path constraint on $\mathbb{E}(Y_{1:n})$, where we require the value at the end of one segment to be equal to the value at the start of the next segment. Mathematically, this requires $a_{\tau_i}+b_{\tau_i}(\tau_i/n)=a_{\tau_{i}+1}+b_{\tau_{i}+1}(\tau_{i}/n)$ at each change-point $\tau_i$ for $i=1,\cdots, m.$

%In the statistics literature, In the econometrics literature,  %Moreover, \cite{Bai1998} cannot recover a continuous piecewise linear path for $\mathbb{E}(Y_{1:n})$, as it contains no mechanism to impose the ``continuous" path constraint on the underlying linear model.

%In this paper, we propose a new change-point detection procedure for the estimation of model \eqref{eq: linear trend} on the basis of self-normalization~(SN, hereafter), a nascent inference technique developed for time series (see  \cite{shao2010self}, \cite{shao2010testing} and \cite{shao2015self}). SN  uses an inconsistent variance estimator to yield an asymptotically pivotal statistic, and does not involve any tuning parameter or involves less number of tuning parameters compared to the traditional procedures.

Based on the self-normalization~(SN) idea in \cite{shao2010self}, we propose a novel SN-based change-point detection procedure for the estimation of \eqref{eq: linear trend} that is robust to temporal dependence both in asymptotic theory and in finite sample. The essential idea of SN is using an inconsistent variance estimator to absorb the unknown serial dependence in the data. See a brief review of SN in Section 2.1 and \cite{shao2015self} for a comprehensive overview of recent developments of SN for low dimensional time series. 

Using the proposed SN method and the piecewise linear trend model, we analyze the time series of cumulative confirmed cases and deaths of COVID-19 (in log scale) in 30 major countries. We find that the spread of coronavirus in each country can typically be segmented into several phases with distinct growth rates and countries with geographical proximity share similar spread patterns, which is particularly evident for continental European countries and developing countries in Latin America. In addition, the transition date from rapid growth phases to moderate growth phases is typically associated with the initiation of emergency measures such as lockdown and mass testing with contact tracing, which partially provides evidence that strict social distancing rules help slow down the virus growth and flatten the curve. Moreover, our analysis further indicates that compared to developed countries, most developing countries are still in the early stages of the pandemic and are generally less efficient in terms of controlling the spread of coronavirus, thus may need more international aids to help contain the epidemic.
%\Revise{However, these public health interventions are found with a delayed effect on  coronavirus-related deaths. The lag between the initiation of external measures and  the decrease in the growth rate of deaths usually lasts for 2-3 weeks,  which emphasizes the importance of the policy continuity.} 

Combining the SN-based change-point detection algorithm with a flexible extrapolation function, we further design a simple two-stage forecasting scheme for COVID-19. The proposed method is used to forecast the cumulative deaths in the U.S. and is found to deliver accurate prediction valuable to data-driven public health decision-making. 
	
	%  We find  our first-segmenting-then-forecasting  method  provides a  comparably accurate prediction, which provides an alternative perspective besides  complex mechanistic models in   epidemiology.}

%Our analysis further shows that in terms of the coronavirus response efficiency, East Asia and Australia are among the best, Western developed countries are the second; while developing countries in general seem to struggle with the spread of COVID-19. We hope this analysis can help better understand the spread pattern of the pandemic and help evaluate the effectiveness of public health policies for fighting the pandemic.

%Talk about what is the piecewise linear assumption can bring to the table in terms of public health policy, identifying phase changes of spread rate in the data and qualitatively cluster countries and effect of lockdown policies in a purely data-driven way.

%We can further talk about 1. Limitation of the data.  2. BaiPerron does not work with continuity. 3. BaiPerron is O(n^2).

\section{Methodology}
In this section, we propose a novel SN-based method for change-point detection in model (\ref{eq: linear trend}) that is robust against a wide range of temporal dependence. Specifically, an SN-based test statistic is first proposed for testing a single change-point alternative and then modified to consistently estimate the change-point. A multiple change-point estimation procedure is further developed by combining the proposed SN test with the NOT algorithm in \cite{baranowski2019narrowest}.

\subsection{Testing for a single change-point}
We start with a change-point testing problem where for model \eqref{eq: linear trend} we want to test the null hypothesis $H_0$ of no change-point against the alternative $H_a$ of one change-point:
\begin{flalign*}
\begin{split}
H_0:{\bbeta}_1=\cdots={\bbeta}_n=\bbeta\quad \text{v.s.}\quad
H_a: \bbeta_t=\left\{\begin{array}{cc}
\bbeta^{(1)},&1\le t\le \tau\\
\bbeta^{(2)},& \tau+1\le t\le n,\\
\end{array}
\right.\quad \text{such that } \bbeta^{(1)}\neq \bbeta^{(2)},
\end{split}
\end{flalign*}
where $\bbeta_t=(a_t,b_t)$, $\tau=\lfloor\kappa n\rfloor$ is an unknown change-point satisfying $\epsilon<\kappa<1-\epsilon$ for some $0<\epsilon<1/2$ and $\epsilon$ is the commonly used trimming parameter in the change-point analysis (see e.g. \citet{andrews1993tests}).
 
Throughout this paper, we operate under the following mild assumption of $\{u_t\}$, which covers a wide range of weakly dependent error process and is weaker than most existing literature where independence of $\{u_t\}$ is assumed.
\begin{ass}\label{ass}
	The error process $\{u_t\}$ is strictly stationary such that $\mathbb{E}(u_t)=0$, $\mathbb{E}(u_t^4)<\infty$ and the long-run variance satisfies 
	$\Gamma^2=\lim\limits_{n\to\infty}\mathrm{Var}(n^{-1/2}\sum_{t=1}^{n}u_t)\in (0,\infty)$. Denote $\{e_t\}$ as a sequence of i.i.d. random variables with zero mean and unit variance, we further assume that $\{u_t\}$ admits one of the following two representations:
	
	(i). $u_t=\sum_{j=0}^{\infty}c_je_{t-j}$ and $\sum_{j=0}^{\infty}|jc_j|<\infty$.
	
	(ii). $u_t = G({\cal F}_t)$ for some measurable function $G$ and ${\cal F}_t = (e_t, e_{t-1},\cdots)$.
	For some $\chi\in (0,1)$, $\|G({\cal F}_k)-G(\{{\cal F}_{-1},e_0',e_1,\cdots,e_k\})\|_4=O(\chi^k)$ if $k\ge 0$ and $0$ otherwise. Here $e_0'$ is an i.i.d. copy of $e_0$ and $\|X\|_4=(\mathbb{E}(X^4))^{1/4}$ for a random variable $X$.
	
	%	 $\chi\in (0,1)$, $\|G({\cal F}_k)−G(\{{\cal F}_{-1},e_0',e_1,\cdots,e_k\})\|^4 = O(\chi^4)$ if $k\ge 0$ and $0$ otherwise. Here, $e_0'$ is an iid copy of $e_0$.  
\end{ass}

%Similar to \cite{andrews1993tests},
Assumption~\ref{ass}(i) is popular in the linear process literature to ensure the central limit theorem and the invariance principle. Assumption~\ref{ass}(ii) is basically equivalent to the geometric moment contracting condition for the nonlinear causal process (\cite{wushao2004}, \cite{wu2005}), which implies invariance principle.

Earlier works on this testing problem include \cite{andrews1993tests} and \cite{bai1998estimating} where Lagrangian multiplier, Wald, likelihood ratio and $F$ statistics are considered. These tests typically require an estimator of the long-run variance~(LRV) $\Gamma$ due to the unknown temporal dependence of the error process $\{u_t\}$. However, as pointed out in \cite{shao2010testing}, the size and power performance of these tests may depend crucially on the selection of various tuning parameters. In particular, if a data-driven bandwidth parameter is used for the estimation of LRV, an undesirable non-monotonic power phenomenon may occur; see \cite{cv2007} and \cite{shao2010testing}. To avoid the bandwidth selection involved in the estimation of LRV, we instead adapt the idea of self-normalization in \cite{shao2010self}, which was originally proposed for inference of stationary time series and was generalized to change-point testing for piecewise stationary time series in \cite{shao2010testing} and \cite{zhang2018unsupervised}. See \cite{shao2015self} for a review of SN. %for the inference of low-dimensional time series.
 
%By SN, we no longer need to estimate the long run variance as they will cancel out asymptotically in the numerator and the denominator of the test statistic.

To proceed, we first introduce some notations. Given $\epsilon,$ denote $h=\lfloor \epsilon n\rfloor$. For a vector $x$, denote the $l_2$ norm as $\|x\|_2$ and denote $x^{\otimes2}=xx^{\top}$. Define  $F(s)=(1,s)^{\top}$, for $1\leq i<j\leq n$, we denote $\widehat{\bbeta}_{i,j}=\Big[\sum_{t=i}^{j}F(t/n)F(t/n)^{\top}\Big]^{-1}\sum_{t=i}^{j}F(t/n)Y_t$ as the OLS estimator of $\bbeta$ based on $\{Y_{t}\}_{t=i}^j$. For any $1\leq t_1 <k <t_2\leq n$, given the subsample $\{Y_t\}_{t=t_1}^{t_2}$ and a potential change-point $k$, we define a contrast statistic $D_{n}$ where
\begin{align}\label{D}
D_{n}(t_1,k,t_2)=\frac{(k-t_1+1)(t_2-k)}{(t_2-t_1+1)^{3/2}}(\widehat{\bbeta}_{t_1,k}-\widehat{\bbeta}_{k+1,t_2}).
\end{align}
Note that $D_{n}(t_1,k,t_2)$ is a normalized difference between the OLS estimates of $\bbeta$ with pre-$k$ samples $\{Y_t\}_{t=t_1}^{k}$ and post-$k$ samples $\{Y_t\}_{t=k+1}^{t_2}$. Intuitively, a large $\max_{h\leq k\leq n-h}\|D_{n}(1,k,n)\|_2$ leads to the rejection of $H_0$. However, the asymptotic distribution of $D_{n}(1,k,n)$ depends on the unknown LRV of $\{u_t\}$, and as discussed before the accurate estimation of LRV is rather challenging and problematic in practice. %for which a consistent estimation can be difficult owing to the difficulty of choosing the tuning parameter in its consistent estimator. 

%The above test statistic is inspired by \cite{zhang2018unsupervised}, who developed an extension of the SN-based test in  \cite{shao2010testing}  by allowing unknown number of change-points and using contrast statistics. 

To bypass the problematic estimation of LRV, we utilize the self-normalization technique. Define $0< \delta<\epsilon/2$ as a local trimming parameter, we define the self-normalizer $V_{n,\delta}(t_1,k,t_2)=L_{n,\delta}(t_1,k,t_2)+	R_{n,\delta}(t_1,k,t_2)$ where
\begin{align}
\label{L}L_{n,\delta}(t_1,k,t_2)=&\sum_{i=t_1+1+\lfloor n\delta\rfloor}^{k-2-\lfloor n\delta\rfloor}\frac{(i-t_1+1)^2(k-i)^2}{(k-t_1+1)^2(t_2-t_1+1)^2}(\widehat{\bbeta}_{t_1,i}-\widehat{\bbeta}_{i+1,k})^{\otimes 2},\\
\label{R}R_{n,\delta}(t_1,k,t_2)=&\sum_{i=k+3+\lfloor n\delta\rfloor}^{t_2-1-\lfloor n\delta\rfloor}\frac{(i-1-k)^2(t_2-i+1)^2}{(t_2-t_1+1)^2(t_2-k)^2}(\widehat{\bbeta}_{i,t_2}-\widehat{\bbeta}_{k+1,i-1})^{\otimes 2}.
\end{align}
The local trimming parameter $\delta$ is introduced to make sure all the subsample estimates of $\bbeta$ in the self-normalizer $V_{n,\delta}(t_1,k,t_2)$ are constructed with a subsample of size being a positive fraction of n, which is a technical condition necessary in our theoretical analysis. We later discuss the implication of the trimming parameters $(\epsilon, \delta)$.

Based on the contrast statistic $D_n(1,k,n)$ and the self-normalizer $V_{n,\delta}(1,k,n)$, we propose an SN-based test statistic $G_n$ for testing the single change-point alternative where
\begin{equation}\label{Gn}
G_n=\max_{k\in\{h,\cdots,n-h\}}T_{n,\delta}(k),\quad T_{n,\delta}(k)=D_{n}(1,k,n)^{\top}V_{n,\delta}(1,k,n)^{-1}D_{n}(1,k,n).
\end{equation}
Intuitively, due to the presence of the self-normalizer, the LRVs in $D_{n}(1,k,n)$ and $V_{n,\delta}(1,k,n)$ cancel out with each other, leading to a test statistic $G_n$ that is invariant to LRV. This phenomenon is made formal in Theorem \ref{thm_main}.

%Note that $G_n$ is invariant to LRV as asymptotically LRV is canceled out in the contrast statistic $D_{n}(1,k,n)$ and the self-normalizer $V_{n,\delta}(1,k,n)$. %, thanks to the use of recursive estimators and their differences in $L_{n,\delta}(t_1,k,t_2)$ and $R_{n,\delta}(t_1,k,t_2)$. 
 
%The grid parameter $\epsilon$ and the trimming parameter $\delta$ are used as we need to have a consistent estimator for $\bbeta$ in both $D_{n}(t_1,k,t_2)$ and $V_{n,\delta}(t_1,k,t_2)$. These two parameters are different with the tuning parameters in the LRV estimator as they will be accounted for in the limiting distribution (see ).

Denote $\overset{\mathcal D}{\longrightarrow}$ as convergence in distribution and $\mathbf{b}=\bbeta^{(2)}-\bbeta^{(1)}$. Define $Q(r)=\int_{0}^{r}F(s)F(s)^{\top}ds$ and $B_F(r)=\int_{0}^{r}F(s)dB(s)$ where $B(\cdot)$ is a standard Brownian motion. Theorem \ref{thm_main} states the asymptotic behavior of the SN test statistic $G_n$ under $H_0$ and $H_a$ respectively.
\begin{thm}\label{thm_main}
Suppose Assumption \ref{ass} holds. Let $G_n$ be defined in (\ref{Gn}), we have 

(i) under $H_0$, we have
\begin{equation}\label{G_asy}
G_n\overset{\mathcal D}{\longrightarrow} G(\epsilon, \delta):= \sup_{\eta\in(\epsilon,1-\epsilon)}D(\eta)^{\top}V_{\delta}(\eta)D(\eta), 
\end{equation}
where 
$D(\eta)=\eta(1-\eta)\Big\{Q(\eta)^{-1}B_F(\eta)-[Q(1)-Q(\eta)]^{-1}[B_F(1)-B_F(\eta)]\Big\}$ and $V_{\delta}(\eta)=L_{\delta}(\eta)+R_{\delta}(\eta)$ with
$L_{\delta}(\eta)=\int_{\delta}^{\eta-\delta}\frac{r^2(\eta-r)^2}{\eta^2}\big\{Q(r)^{-1}B_F(r)-[Q(\eta)-Q(r)]^{-1}[B_F(\eta)-B_F(r)]\big\}^{\otimes2}dr,$
$R_{\delta}(\eta)=\int_{\eta+\delta}^{1-\delta}\frac{(r-\eta)^2(1-r)^2}{(1-\eta)^2}\times\big\{[Q(1)-Q(r)]^{-1}[B_F(1)-B_F(r)]-
[Q(r)-Q(\eta)]^{-1}[B_F(r)-B_F(\eta)]\big\}^{\otimes2}dr.$

%\begin{flalign*}
%D(\eta)=&\eta(1-\eta)\Big\{Q(\eta)^{-1}B_F(\eta)-[Q(1)-Q(\eta)]^{-1}[B_F(1)-B_F(\eta)]\Big\},\\
%L_{\delta}(\eta)=&\int_{\delta}^{\eta-\delta}\frac{r^2(\eta-r)^2}{\eta^2}\Big\{Q(r)^{-1}B_F(r)-[Q(\eta)-Q(r)]^{-1}[B_F(\eta)-B_F(r)]\Big\}^{\otimes2}dr,\\
%R_{\delta}(\eta)=&\int_{\eta+\delta}^{1-\delta}\frac{(r-\eta)^2(1-r)^2}{(1-\eta)^2}\\&\times\Big\{[Q(1)-Q(r)]^{-1}[B_F(1)-B_F(r)]-[Q(r)-Q(\eta)]^{-1}[B_F(r)-B_F(\eta)]\Big\}^{\otimes2}dr,\\
%V_{\delta}(\eta)=&L_{\delta}(\eta)+R_{\delta}(\eta),
%\end{flalign*}

(ii) under $H_a$, given that $n\|\mathbf{b}\|^2_2\to L$, we have
$$\lim\limits_{L\to\infty}\lim\limits_{n\to\infty}{G}_n=\infty,\quad\mbox{ in probability}.$$
\end{thm}

Due to self-normalization, the limiting distribution $G(\epsilon, \delta)$ in (\ref{G_asy}) is pivotal and invariant to the LRV. The corresponding critical values can be easily obtained via simulation. Table \ref{tab_quant} gives the $1-\alpha$ quantiles of $G(\epsilon, \delta)$ for some combinations of $(\epsilon, \delta)$ (based on 10000 replications).
Note that the limiting null distribution $G(\epsilon, \delta)$ explicitly depends on the choice of $(\epsilon,\delta)$, thus the impact of trimming parameters $(\epsilon,\delta)$ is accounted for at the first order, in the same spirit of the fixed-$b$ asymptotics (\cite{kv2005}). See also \cite{zhou2013inference}. Throughout the paper, we set $(\epsilon,\delta)=(0.1,0.02)$. 
 
\begin{table}[H]
	\caption{Simulated quantiles of $G$}
	\label{tab_quant}
	\begin{tabular}{ccccccc}
		\hline
		$\epsilon$           & \diagbox{$\delta$}{$1-\alpha$} & 90\%    & 95\%    & 99\%     & 99.5\%   & 99.9\%   \\ \hline
		0.1& 0.01     & 14.963 & 19.284 & 32.168  & 36.145  & 45.354  \\
		& 0.02     & 24.959 & 32.727 & 53.645  & 64.898  & 92.982  \\
		& 0.03     & 38.277 & 50.872 & 83.713  & 107.062 & 137.433 \\
		& 0.04     & 54.569 & 76.244 & 116.497 & 144.437 & 182.786 \\
		0.2                  & 0.01     & 4.656  & 5.905  & 9.691   & 12.037  & 14.148  \\
		& 0.02     & 7.217  & 9.404  & 15.486  & 18.389  & 24.079  \\
		& 0.03     & 10.526 & 13.767 & 23.060 & 26.758  & 36.388  \\
		& 0.04     & 14.439 & 19.075 & 33.049  & 37.426  & 49.495  \\ \hline
	\end{tabular}
\end{table}

Give that the null hypothesis $H_0$ is rejected, we estimate the change-point ${\tau}$ by 
%\begin{equation}\label{tauhat}
$\widehat{\tau}= \arg\max_{k\in\{h,\cdots,n-h\}}T_{n,\delta}(k).$
%\end{equation}
%\Zifeng{Give the consistency result of $\widehat{\bftau}$ here.}
%\FY{Do we need to give the consistency of $\widehat{\bftau}$? If we give this, we may diminish the contribution in another paper.}
The following theorem gives the consistency result of $\widehat{\kappa}=n^{-1}\widehat{\tau}$.
\begin{thm}\label{thm_consis}
	Under $H_a$, suppose Assumption \ref{ass} holds, and $n\|\mathbf{b}\|^2_2\to\infty$ as $n\to\infty$. Then, we have that for any $\eta>0$,
	$$
	\lim\limits_{n\to\infty}\mathbb{P}(|\widehat{\kappa}-\kappa|<\eta)=1.
	$$
\end{thm}
Theorem \ref{thm_consis} allows a diminishing change size $\|\mathbf{b}\|_2$ with the sample size $n$ as long as $n\|\mathbf{b}\|^2_2\to\infty$. Note that no consistency result is provided in \cite{shao2010testing} for the change-point location estimation, and our result seems to be the first formal attempt based on the SN technique.   However, it is challenging to obtain an explicit rate of convergence for $\widehat{\tau}$ due to the complicated nature of the self-normalizer $V_{n,\delta}$ and we leave it for future investigation. 

\subsection{Multiple change-point estimation}
%A popular multiple change-point estimation method is based on testing and binary segmentation (BS). The latter procedure applies a test that targets one change-point alternative to each segment (starting with the full sample), split the data into segments and stops until no rejection is reported for each segment. 
To extend single change-point testing to multiple change-point estimation, the classical idea is to combine the change-point test with binary segmentation (BS). Although conceptually and computationally simple, it is well known that BS can cause severe power loss for detecting non-monotonic changes~\citep{Olshen2004}, which is common in real data. Several variants of BS have been proposed to address this drawback, such as wild binary segmentation (WBS) (\cite{fryzlewicz2014wild}) and Narrowest-Over-Threshold (NOT) (\cite{baranowski2019narrowest}). Since NOT is shown to be superior to WBS, we combine the SN-based test with the NOT algorithm to estimate multiple change-points and name our algorithm SN-NOT. 

% testing-based To extend a single change-point test to multiple change-point estimation, the classical idea is to combine the test with binary segmentation~(BS) or its variant~(e.g. Wild BS~(WBS) in ). However,  Thus, we instead combine the SN test with the  and propose the SN-NOT algorithm (Algorithm \ref{alg}) for estimating \eqref{eq: linear trend}.

%In this section, we extend the proposed SN test to multiple change-point detection. A classical idea is to combine the change-point test with binary segmentation~(BS): (1) run the test over the full sample, and if the test is rejected, using (\ref{tauhat}) to estimate the change-point location; (2) divide the full sample into two subsamples and apply the test to each of them. The algorithm is repeated until no test is rejected. 

%However, as pointed out by \cite{baranowski2019narrowest}, BS and its variant wild binary segmentation (WBS) proposed by \cite{fryzlewicz2014wild} can be problematic when detecting non-monotonic changes. 

The essential idea of SN-NOT is to compute the SN test on a large collection of random subsamples of $\{Y_t\}_{t=1}^n$ instead of the entire sample $\{Y_t\}_{t=1}^n$. With high probability, some subsamples will only contain a \textit{single} change-point, where the SN test statistics are expected to exhibit large values, leading to the discovery of a change-point.  

Denote $F_n^{M}=\{(s_i,e_i):i=1,\cdots,M\}$ as the set of $M$ random intervals such that each pair of integers $(s_i,e_i)$ are drawn uniformly from $\{1,\cdots, n\}$ and satisfy $1\leq s_i<e_i\leq n$ and $e_i-s_i+1\geq 2h$. For each random interval $(s,e) \in F_n^M$, we calculate the SN test
$$
G_{n,\delta}(s,e)=\max_{k\in\{ s+h-1,\cdots,e-h\}}T_{n,\delta}(s,k,e),\quad T_{n,\delta}(s,k,e)= D_{n}(s,k,e)V_{n,\delta}(s,k,e)^{-1}D_{n}(s,k,e)^{\top}.
$$
SN-NOT finds the narrowest interval $(s,e)\in F_n^M$ where the test statistic $G_{n,\delta}(s,e)$ exceeds a given threshold $\zeta_n$ and estimates the change-point as $\widehat{\tau}=\arg\max_{k\in\{ s+h-1,\cdots,e-h\}}T_{n,\delta}(s,k,e)$. Note that for large $M$, with high probability there is only one change-point in this narrowest interval, which thus remedies the drawback of BS in detecting non-monotonic changes. Once a change-point $\widehat{\tau}$ is identified, SN-NOT then divides the sample into two subsamples accordingly and apply the same procedure on each of them. The process is implemented recursively until no change-point is detected. In addition to the advantage of detecting non-monotonic changes, SN-NOT broadens the applicability of the NOT algorithm itself by allowing for temporal dependence in the error process thanks to the self normalization technique.

The detailed implementation of SN-NOT is given in Algorithm \ref{alg}. We propose to select the threshold $\zeta_n$ as follows. Generate $B$ sequences of i.i.d $\mathcal{N}(0,1)$ random variables $\{\varepsilon_t^b\}_{t=1}^n$, $b=1,\cdots, B$; for the $b$th sample, we calculate 
$$
\zeta_n^{b}=\arg\max_{i=1,\cdots,M}G_{n,\delta}(s_i,e_i),\quad b=1,\cdots,B.
$$
The threshold $\zeta_n$ is set as the 95\% sample quantile of $\{\zeta_n^{b}\}_{b=1}^{B}$. Since the SN test statistic is asymptotically pivotal, this  threshold is expected to well approximate the $95\%$ quantile of the finite sample distribution of the maximum SN test statistic on the $M$ random intervals under null. Throughout this paper, we set $B=1000$, $M=300$. 

%finite sample distribution of maximized SN-based test statistic applied to $M$ randomly drawn sub-samples from the original data. 

\begin{algorithm}[!h]\label{alg}
	\caption{SN-NOT}
	\KwIn{Data $\{Y_t\}_{t=1}^{n}$, threshold $\zeta_n$, trimming size $d=\lfloor \delta n\rfloor$ and $h=\lfloor\epsilon n\rfloor$, random intervals $F_n^{M}$.}
	\KwOut{Estimated number of change-points $\widehat{m}$ and estimated change-points set $\widehat{\bftau}$}
	\KwIni{SN-NOT$(1,n,\zeta_n)$}
	\KwPro{SN-NOT$(s,e,\zeta_n)$}
	\eIf{$e-s+1<2h$}{Stop}{$\mathcal{M}_{(s,e)}:=\{i:[s_i,e_i]\in F_n^M,[s_i,e_i]\subset[s,e], e_i-s_i+1\geq 2h\}$ \;
		\eIf{$\mathcal{M}_{(s,e)}=\varnothing$}{Stop}
		{$\mathcal{O}_{(s,e)}:=\Big\{i\in\mathcal{M}_{(s,e)}: G_{n,\delta}(s_i,e_i)>\zeta_n\Big\}$\;
			\eIf{	$\mathcal{O}_{(s,e)}=\varnothing$}{Stop}{
	$i^*=\arg\min_{i\in \mathcal{O}_{(s,e)}}|e_i-s_i+1|$\;
	$\tau^*=\arg\max_{k\in\{s_i^*+h-1,\cdots,e_i^*-h\}}T_{n,\delta}(s_{i^*},k,e_{i^*})$	\;
	$\widehat{\bftau}=\widehat{\bftau}\cup \tau^*$, $\widehat{m}=\widehat{m}+1$\;
	SN-NOT$(s,\tau^*,\zeta_n)$\;
	SN-NOT$(\tau^*+1,e,\zeta_n)$\;
	}}	}
\end{algorithm}

\section{Simulation}
In this section, we study the finite sample performance of the SN test in testing single change-point and the SN-NOT algorithm in detecting multiple change-points through numerical experiments. All results are reported based on 1000 replications.

\subsection{Testing size and Power}
We generate the data from model (\ref{eq: linear trend}) with sample size $n=100$, $500$  and $1000$ respectively. For the size performance, we let $\bbeta=(3,0.05n)$ while for the power performance, we let $\bbeta^{(1)}=(3,0.06n)$ and $\bbeta^{(2)}=(3+0.015n,0.03n)$ with the change-point $\tau=n/2$. The error process $\{u_t\}$ is generated via an AR(1) model where $u_t=\rho u_{t-1}+e_t$, $e_t\overset{i.i.d.}{\sim}$ $\mathcal{N}(0,(1-\rho^2)\sigma^2)$ with $\rho=0,\pm0.2,\pm 0.5$ and $\sigma=0.15$. 

For comparison, we also implement the supLM test defined in \cite{andrews1993tests} (using function {\tt sctest} of the R package \texttt{strucchange}) with the same trimming parameter $\epsilon=0.1$. The results are summarized in Table \ref{tab_simu} at significance levels $\alpha=5\%$ and 10\%. It can be seen that when $n$ is small, both methods have distorted sizes. In particular, SN is prone to be conservative  when $\rho$ is negative and oversized when $\rho$ is positive while supLM is undersized in all cases. As $n$ increases, we find that both tests tend to have more accurate sizes. For $n=100$, supLM test has slightly higher power than SN test while for $n=500$ and $n=1000$, SN test beats supLM test under positive $\rho$. Note that both tests are more powerful under negative $\rho$.

\begin{table}[]
	\caption{Size and size-adjusted power  for SN test and supLM test.}
	\label{tab_simu}
	\begin{tabular}{ccccccccccccc}
		\hline
		&          & \multicolumn{5}{c}{SN}                   &     & \multicolumn{5}{c}{supLM}               \\ \cline{3-7} \cline{9-13} 
		$\alpha$ & $\rho$   & -0.5  & -0.2  & 0     & 0.2   & 0.5      &     & -0.5    & -0.2  & 0     & 0.2   & 0.5   \\
		&          &       &       &       &       & \multicolumn{3}{c}{Size} &       &       &       &       \\ \cline{3-13} 
		5\%      & $n=100$  & 0.003 & 0.012 & 0.026 & 0.042 & 0.093    &     & 0.043   & 0.023 & 0.016 & 0.012 & 0     \\
		10\%     &          & 0.008 & 0.028 & 0.053 & 0.091 & 0.160    &     & 0.091   & 0.064 & 0.047 & 0.035 & 0.018 \\
		&          &       &       &       &       &          &     &         &       &       &       &       \\
		5\%      & $n=500$  & 0.022 & 0.033 & 0.036 & 0.045 & 0.057    &     & 0.049   & 0.042 & 0.032 & 0.030 & 0.020 \\
		10\%     &          & 0.051 & 0.064 & 0.074 & 0.085 & 0.105    &     & 0.101   & 0.089 & 0.082 & 0.078 & 0.064 \\
		&          &       &       &       &       &          &     &         &       &       &       &       \\
		5\%      & $n=1000$ & 0.040 & 0.045 & 0.045 & 0.045 & 0.049    &     & 0.042   & 0.037 & 0.036 & 0.034 & 0.025 \\
		10\%     &          & 0.086 & 0.086 & 0.089 & 0.092 & 0.096    &     & 0.108   & 0.096 & 0.090 & 0.079 & 0.067 \\
		&          &       &       &       &       &          &     &         &       &       &       &       \\
		&          &       &       &       &       &    &   Power   &         &       &       &       &       \\ \cline{3-13} 
		5\%      & $n=100$  & 1     & 0.990 & 0.909 & 0.654 & 0.269    &     & 1       & 0.999 & 0.965 & 0.846 & 0.438 \\
		10\%     &          & 1     & 1     & 0.983 & 0.879 & 0.531    &     & 1       & 1     & 0.996 & 0.925 & 0.587 \\
		&          &       &       &       &       &          &     &         &       &       &       &       \\
		5\%      & $n=500$  & 1     & 1     & 1     & 1     & 1        &     & 1       & 1     & 1     & 0.989 & 0.277 \\
		10\%     &          & 1     & 1     & 1     & 1     & 1        &     & 1       & 1     & 1     & 0.998 & 0.568 \\
		&          &       &       &       &       &          &     &         &       &       &       &       \\
		5\%      & $n=1000$ & 1     & 1     & 1     & 1     & 1        &     & 1       & 1     & 1     & 1     & 0.906 \\
		10\%  &        & 1     & 1     & 1     & 1     & 1        &     & 1       & 1     & 1     & 1     & 0.989 \\ \hline
	\end{tabular}
\end{table}

\subsection{Multiple change-point estimation}
We examine the numerical performance of SN-NOT by considering the following DGP with $n=100$:
\begin{flalign*}
Y_t=\left\{\begin{aligned}
&3+3.2(t/n)+u_t,  &1\leq t\leq 20,\\
&5.8+1.8(t/n)+u_t,  &21\leq t\leq 40,\\
&9.8+0.8(t/n)+u_t, &41\leq t\leq 70,\\
&15.05+0.05(t/n)+u_t, &71\leq t\leq 100.
\end{aligned}\right.
\end{flalign*}
The error process $\{u_t\}$ is generated via an AR(1) model where $u_t=\rho u_{t-1}+e_t$, $e_t\overset{i.i.d.}{\sim}\mathcal{N}(0,(1-\rho^2)\sigma^2)$ with $\rho=0,\pm0.2,\pm 0.5$ and $\sigma=0.15$. For comparison, we also implement the multiple change-point detection procedure proposed in \cite{bai1998estimating} (denoted as BP hereafter), which is the most widely used detection algorithm allowing for temporal dependence in the error term of model (\ref{eq: linear trend}). BP is implemented using function {\tt breakpoints} of the R package  \texttt{strucchange}. 

To assess the accuracy of change-point estimation, we define the Hausdorff distance between two sets. Denote the set of true change-points as $\bftau_o$ and the set of estimated change-points as $\widehat{\bftau}$, we define
$d_1(\bftau_o, \widehat{\bftau})=\max_{\tau_1 \in \widehat{\bftau}}\min_{\tau_2 \in \bftau_o}|\tau_1-\tau_2|$ and $d_2(\bftau_o, \hat{\bftau})=\max_{\tau_1 \in \bftau_o}\min_{\tau_2 \in \widehat{\bftau}}|\tau_1-\tau_2|$,
where $d_1$ measures the over-segmentation error of $\widehat{\bftau}$ and $d_2$ measures the under-segmentation error of $\widehat{\bftau}$. The Hausdorff distance is then defined as
$d_H(\bftau_o, \widehat{\bftau})=\max(d_1(\bftau_o, \widehat{\bftau}), d_2(\bftau_o, \widehat{\bftau})).$
%With the above definition, it is straightforward to check that $d_H$ is indeed a distance metric for sets of change-points $\bftau.$
In addition, we  report the adjusted Rand index (ARI) which measures the similarity between two partitions of the same observations. Roughly speaking, a higher ARI (with the maximum value of 1) means  more accurate change-point estimation. For the definition and detailed discussions of ARI, we refer to  \cite{hubert1985comparing}.

Table \ref{tab_est} summarizes the numerical result where we report ARI, $d_1$, $d_2$, $d_H$ and the frequency of $|\widehat{m}-m_o|$ for SN-NOT and BP. It can be seen that SN-NOT is overall better than BP in terms of ARI, $d_H$ and the estimated number of change-points when $\rho\geq 0$. This finding suggests using SN-NOT could be more advantageous for analyzing COVID-19 data, which exhibit positive temporal dependence (see the last column of Table \ref{table_app}).  For applications where negatively correlated error is expected, BP could be a better choice.
 
\begin{table}[!h]
	\caption{Estimation results for SN-NOT and BP }
	\label{tab_est}
	\begin{tabular}{cccccccccccc}
		\hline
		& \multicolumn{5}{c}{SN-NOT}            &  & \multicolumn{5}{c}{BP}                 \\ \cline{2-6} \cline{8-12} 
		$\rho$              & -0.5  & -0.2  & 0     & 0.2   & 0.5   &  & -0.5  & -0.2  & 0     & 0.2   & 0.5    \\ \hline
		ARI                 & 0.844 & 0.852 & 0.849 & 0.828 & 0.784 &  & 0.863 & 0.852 & 0.840 & 0.805 & 0.714  \\
		$d_1$               & 4.817 & 3.846 & 3.953 & 4.765 & 6.049 &  & 2.854 & 3.176 & 3.379 & 3.970 & 4.837  \\
		$d_2$               & 2.949 & 3.170 & 3.574 & 3.964 & 6.032 &  & 2.854 & 3.252 & 3.915 & 5.605 & 10.457 \\
		$d_H$               & 4.830 & 3.877 & 4.141 & 4.960 & 7.152 &  & 2.854 & 3.252 & 3.915 & 5.605 & 10.457 \\
		$\widehat{m}=3$     & 0.902 & 0.955 & 0.950 & 0.922 & 0.808 &  & 1     & 0.989 & 0.930 & 0.775 & 0.337  \\
		$|\widehat{m}-3|=1$ & 0.098 & 0.045 & 0.050 & 0.078 & 0.186 &  & 0     & 0.011 & 0.069 & 0.198 & 0.402  \\
		$|\widehat{m}-3|>1$ & 0     & 0     & 0     & 0     & 0.006 &  & 0     & 0     & 0.001 & 0.027 & 0.261  \\ \hline
	\end{tabular}
\end{table}

\section{Analysis for cumulative confirmed cases and deaths of COVID-19}
In this section, based on the proposed SN-NOT algorithm, we provide detailed in-sample analysis of the cumulative confirmed cases~(Section 4.2-4.3) and deaths~(Section 4.4) of COVID-19~(in log scale) in 30 major countries.

\subsection{Data and method}
We focus on G20 (with 19 sovereign countries{\footnote{G20 is an international forum for the  governments and central bank governors from 19 countries and the European Union. We will view members of the European Union as individual countries because the responses to COVID-19 usually come from the national level.}}) and 11 other  countries leading the total infected cases as of May 27, 2020, including Australia (AUS), Argentina (ARG), Belgium (BEL), Brazil (BRA), Canada (CAN), Chile (CHI), China (CHN), France (FRA), Germany (GER), India (IND), Indonesia(INA), Iran (IRI),   Italy (ITA), Japan (JPN), Mexico (MEX), Netherlands (NED), Pakistan (PAK), Peru (PER), Portugal (POR),  Qatar(QAT), Russia (RUS), Saudi Arabia (KSA), Spain (ESP),   South Africa (RSA), South Korea (ROK), Sweden (SWE), Switzerland (SUI),  Turkey (TUR), United Kingdom (GBR), United States (USA). 

We obtain the data from
 \url{https://ourworldindata.org/coronavirus-source-data} 
maintained by ``Our World in Data", where cumulative measures such as confirmed cases and deaths 
 %~\Zifeng{(do we have the testing data? maybe we can take a look at it)} 
are updated daily for each nation.
% We analyze the time series of cumulative confirmed cases . %By taking the logarithm, we can reduce the size of the data and will obtain the growth rate more directly.
For each country, the logarithm of cumulative confirmed cases (or deaths) $\{Y_t\}$ starts on the date when the cumulative cases (or deaths) exceeded 20 and ends on May 27.

We study the cumulative confirmed cases and deaths~(in log scale) of each country via the piecewise linear trend model \eqref{eq: linear trend}, where given $\{Y_t\}$, the change-points $(\tau_1,\cdots, \tau_{\widehat{m}})$ is estimated by the SN-NOT algorithm. An OLS is then used to recover the linear model for the $i$th estimated segment $\{Y_t\}_{t=\widehat{\tau}_{i-1}+1}^{\widehat{\tau}_i}$,  $i=1,2,\cdots,\widehat{m}+1$. With a slight abuse of notation, denote $\widehat{b}_{i}$ as the estimated slope for the $i$th segment. We define the normalized slope $S_i=\widehat{b}_{i}/n$ for each segment. As can be seen from (\ref{eq: linear trend}), the normalized slope $S_i$ measures $\mathbb{E}[Y_{t+1}-Y_t]$ for the $i$th segment, which can be interpreted  as the ``log-return"  and measures the daily growth rate of the cumulative confirmed cases (or deaths) in the original scale.

Methodologically speaking, for cumulative confirmed cases, the piecewise linearity allows us to assess the growth rate of the coronavirus at any given time and further facilitates short-term forecast. In particular, the estimated slope $S_i$ of each segment indicates the pace of the growth rate during the corresponding period. Moreover, by comparing the slope before and after each change-point, we can quantitatively assess the changes in growth rate, which partially measure the effectiveness of policies taken by the government.

%We assume that the logarithm of the cumulative confirmed infected cases $Y_t$ can be modeled by the piecewise linear trend model defined in (\ref{eq: linear trend}). 

\subsection{Detailed analysis of cumulative confirmed cases in 8 representative countries}
We first conduct a detailed case study for eight representative countries that   either lead confirmed cases (the U.S., Brazil, Russia, and India) in the corresponding continent or  receive most media attention (the U.K., Spain, Italy, and South Korea).

Table \ref{table_app} summarizes the detailed estimation result for each country (in descending order of the cumulative confirmed cases), where we report the starting date of the series, length of the series $n$, the estimated number of change-points, dates of the first, second and latest estimated change-point. The first ($S_1$), the second ($S_2$) and the current normalized slope ($S_{\widehat{m}+1}$) are  also presented.  In addition, we report the lag-1 sample autocorrelation $\widehat{\rho}$ of the error process.
From the table, we can see all of these countries have been affected by the coronavirus for more than two months. The average length of segments between two adjacent change-points is around 13-20 days, indicating that the spread rate can be relatively steady for a  window of 2-3 weeks. The latest change-point for most countries appeared in May except for Brazil. We also note that the current normalized slopes (i.e. growth rate) vary considerably across countries with comparably large values in Brazil and India. Meanwhile, the lag-1 sample autocorrelation $\widehat{\rho}$ are all positive, which suggests the use of SN-NOT instead of BP as discussed in Section 3.2. In Figures \ref{ACF} and \ref{PACF} of the supplementary material, we further plot the lag-1 to lag-30 ACF and PACF of the residuals, which rules out the scenario of long memory and supports the validity of Assumption 2.1. 

\begin{figure}[H]
	\centering
	\subfigure{	\hspace{-6mm}	
		\includegraphics[width=0.52\linewidth]{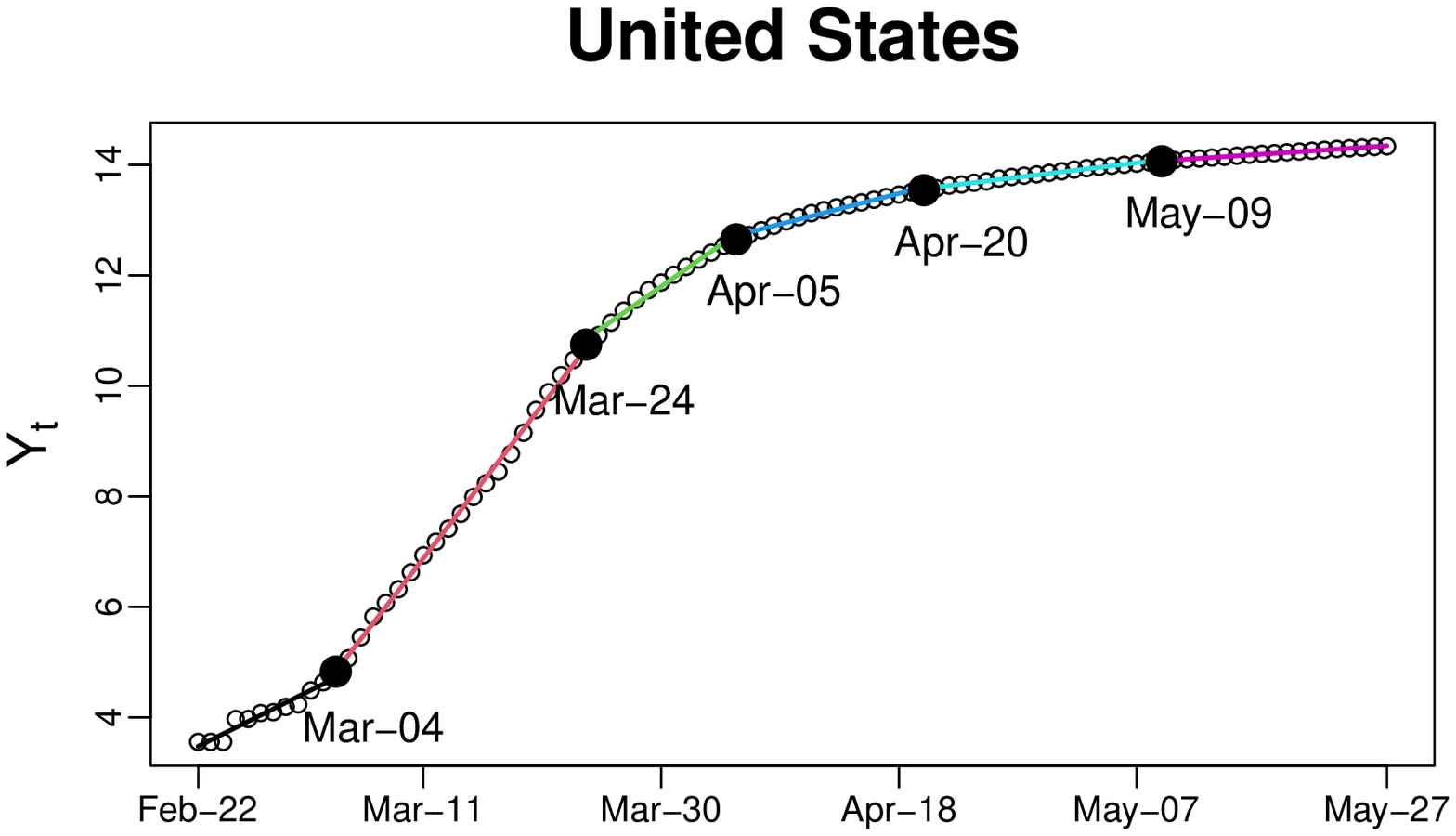}}
		\hfil
	\subfigure{	\hspace{-7mm}
		\includegraphics[width=0.52\linewidth]{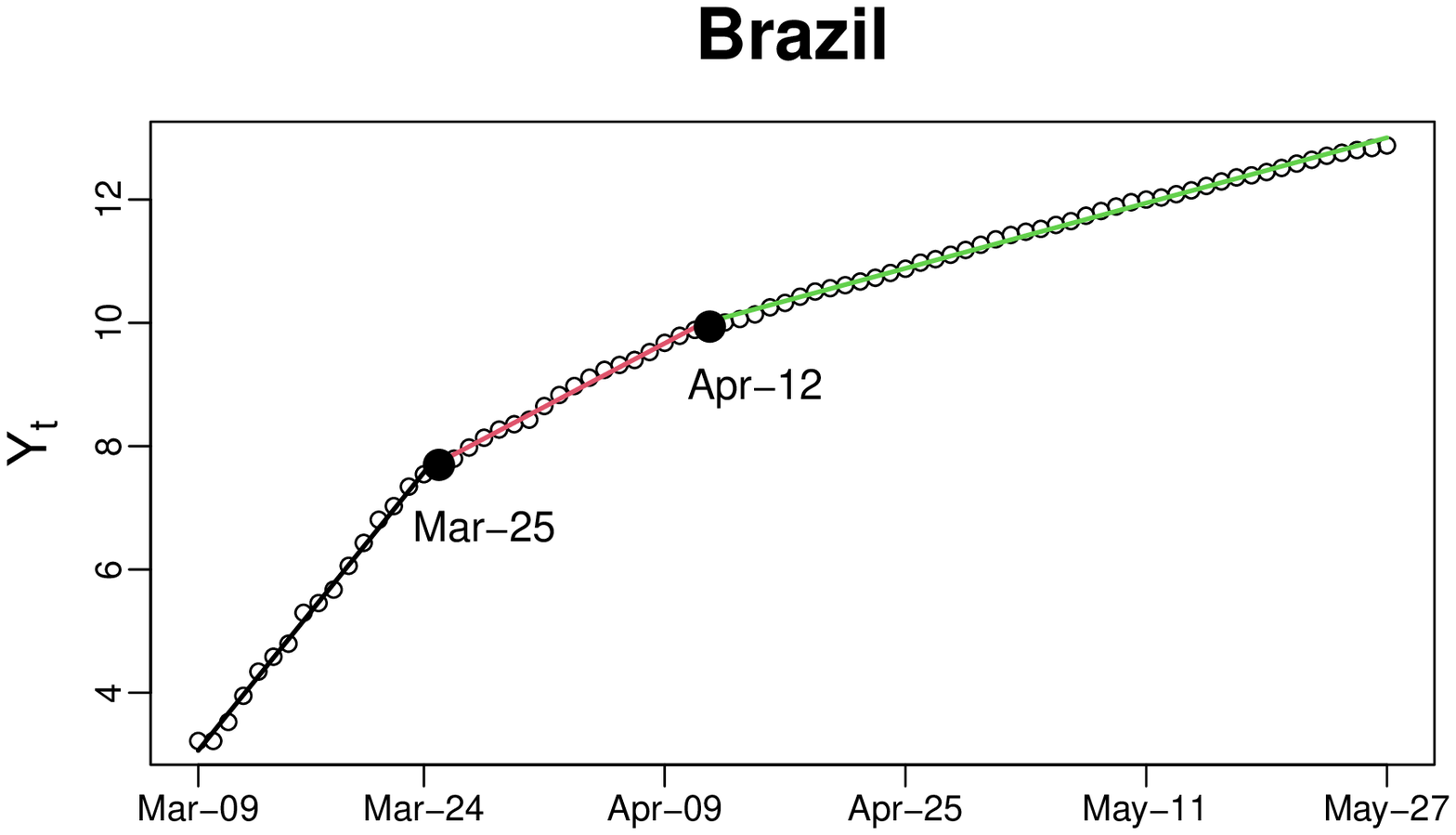}}
	\hfil\vspace{-8mm}
	\subfigure{\hspace{-5mm}
		\includegraphics[width=0.52\linewidth]{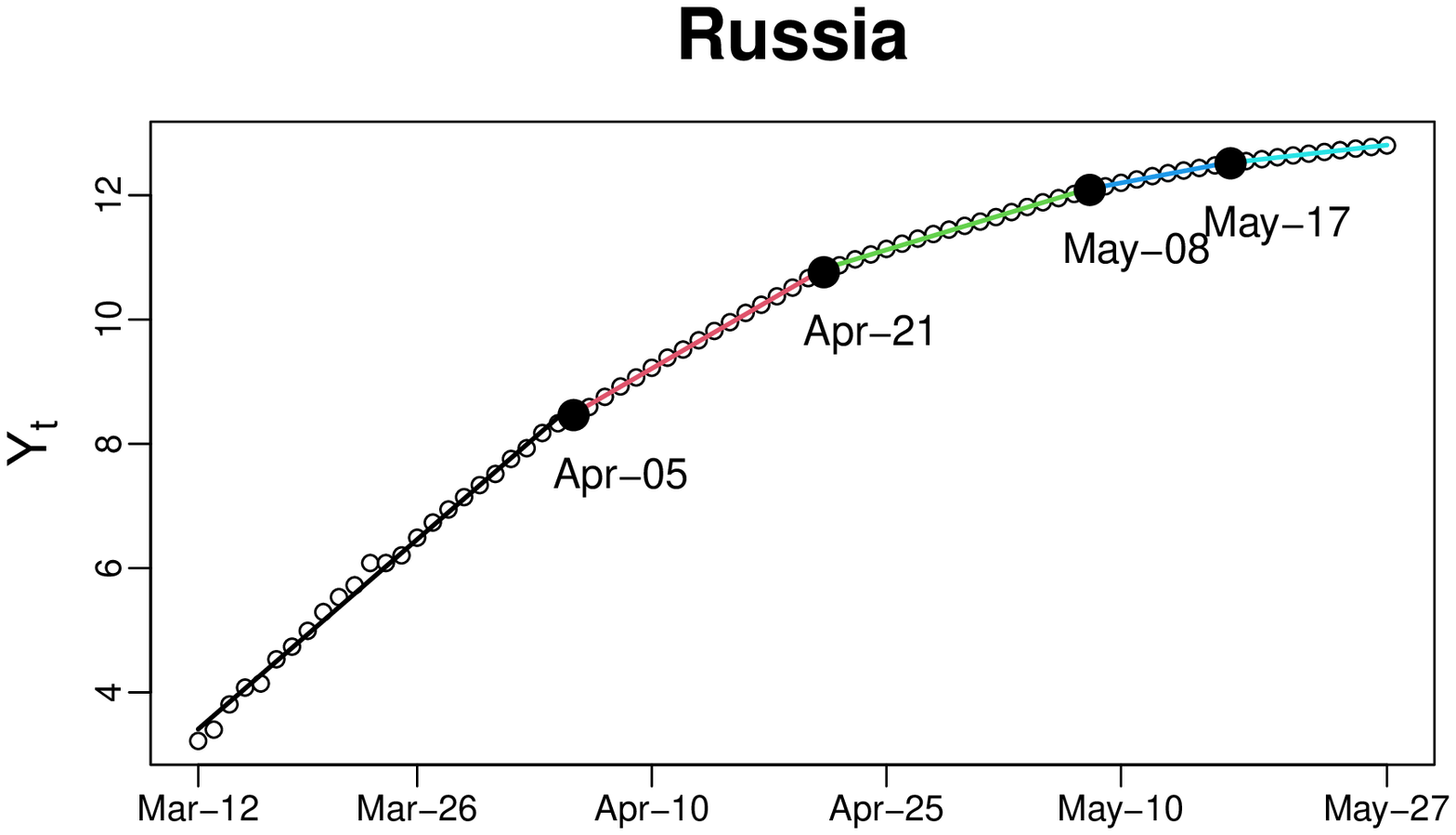}}
	\hfil
	\subfigure{\hspace{-6mm}
		\includegraphics[width=0.52\linewidth]{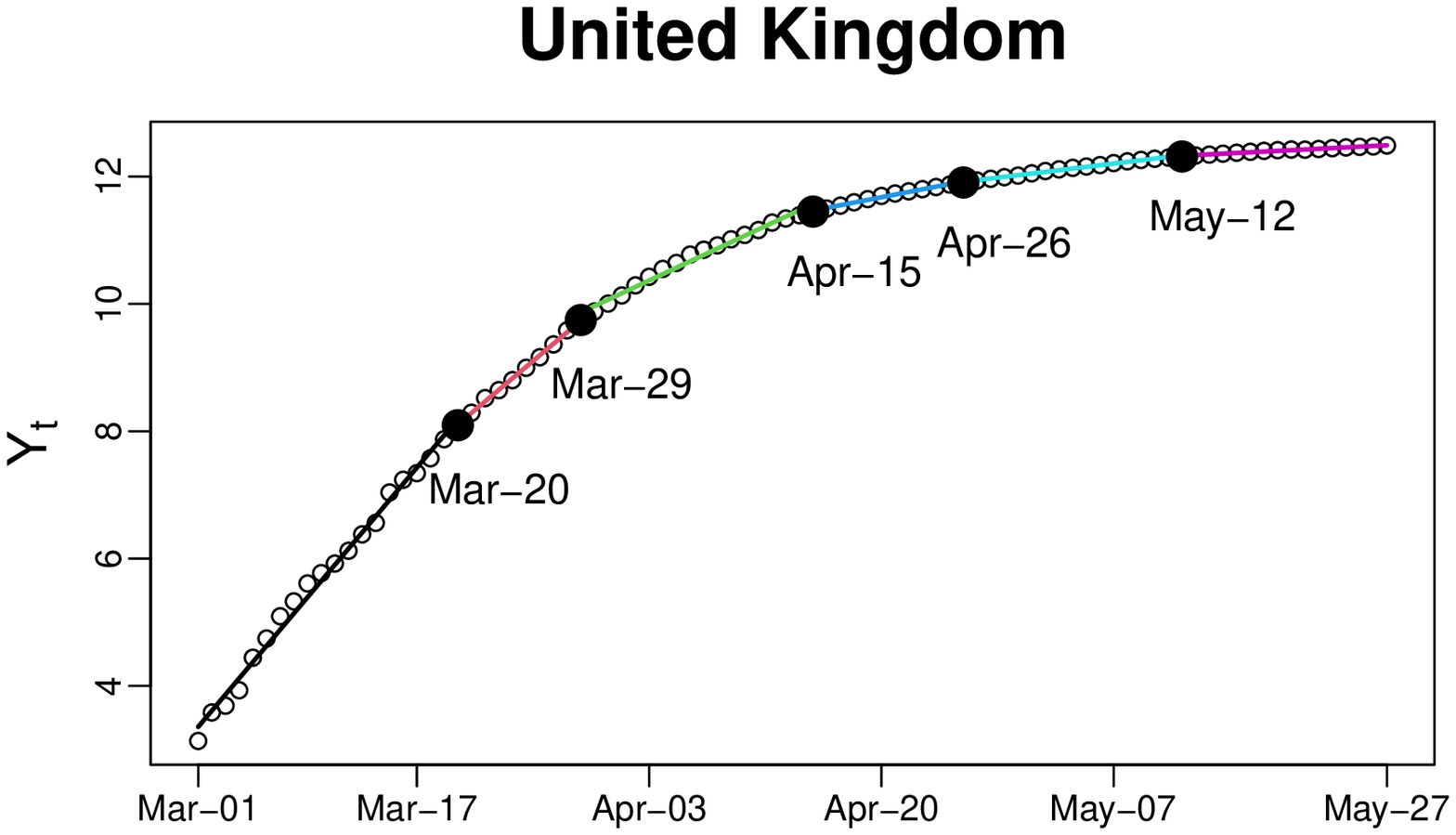}}	
	\hfil\vspace{-8mm}
	\subfigure{\hspace{-5mm}
		\includegraphics[width=0.52\linewidth]{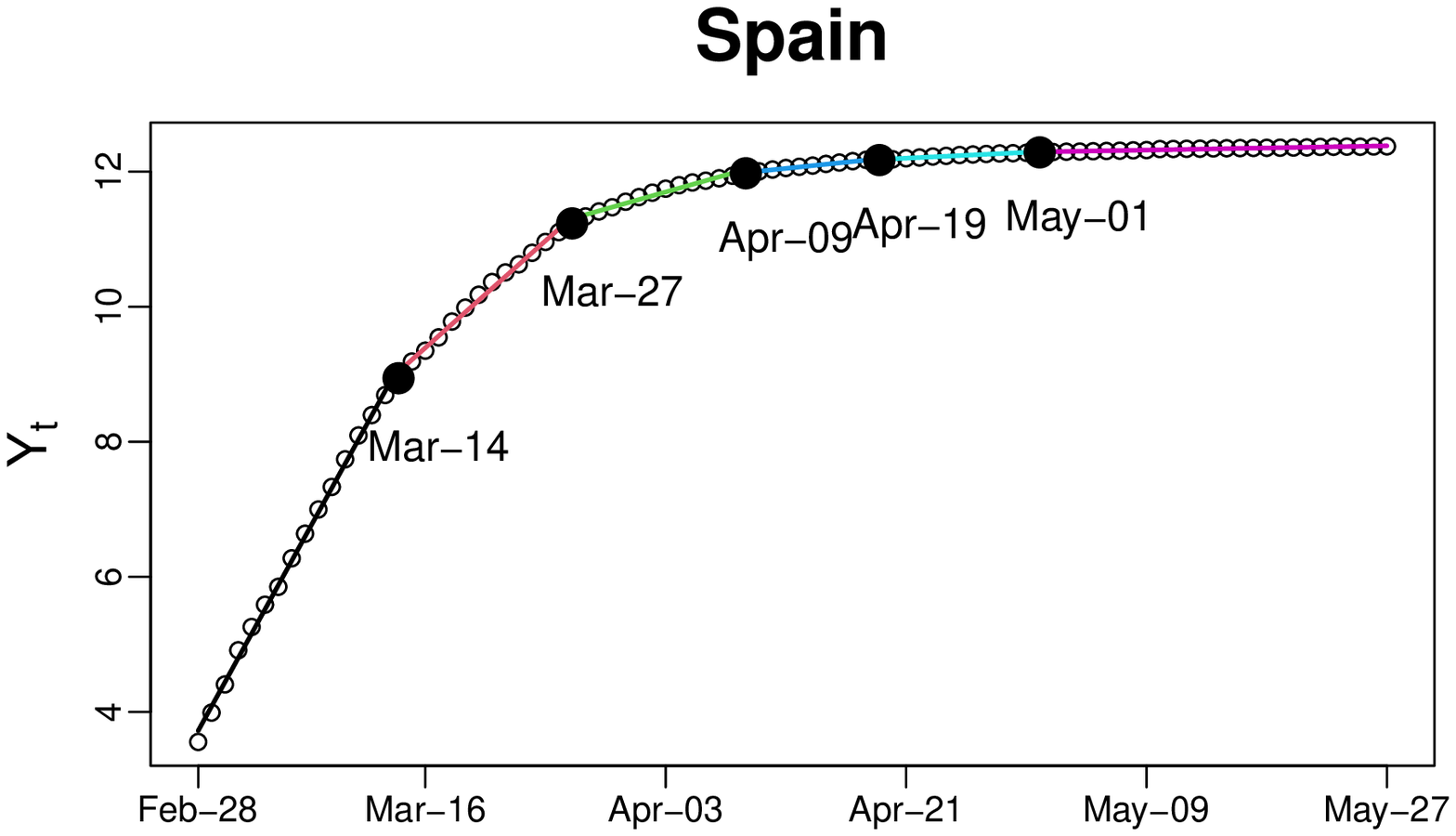}}
	\hfil
	\subfigure{\hspace{-6mm}
		\includegraphics[width=0.52\linewidth]{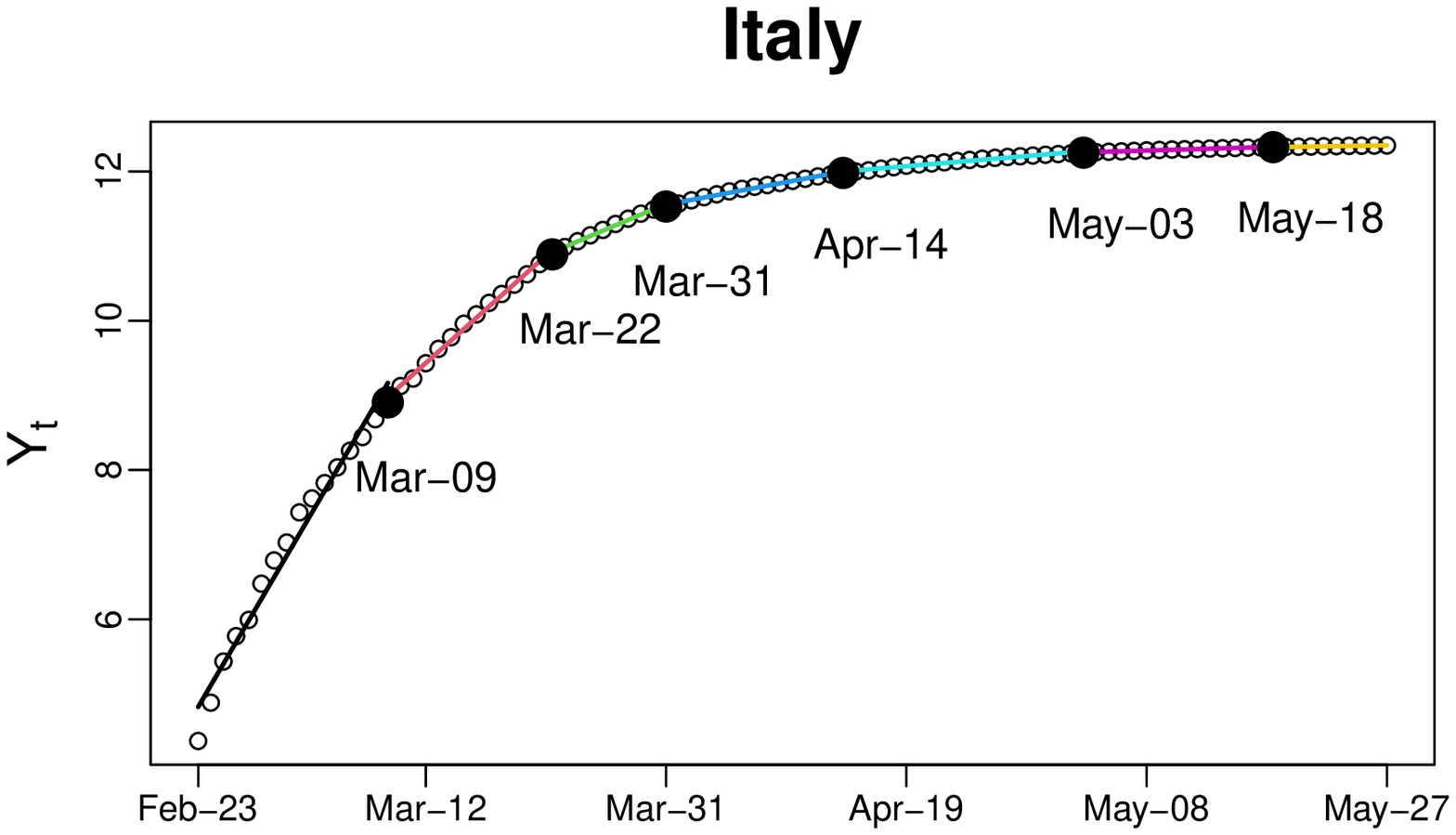}}
	\hfil\vspace{-8mm}
	
	\subfigure{\hspace{-5mm}
		\includegraphics[width=0.52\linewidth]{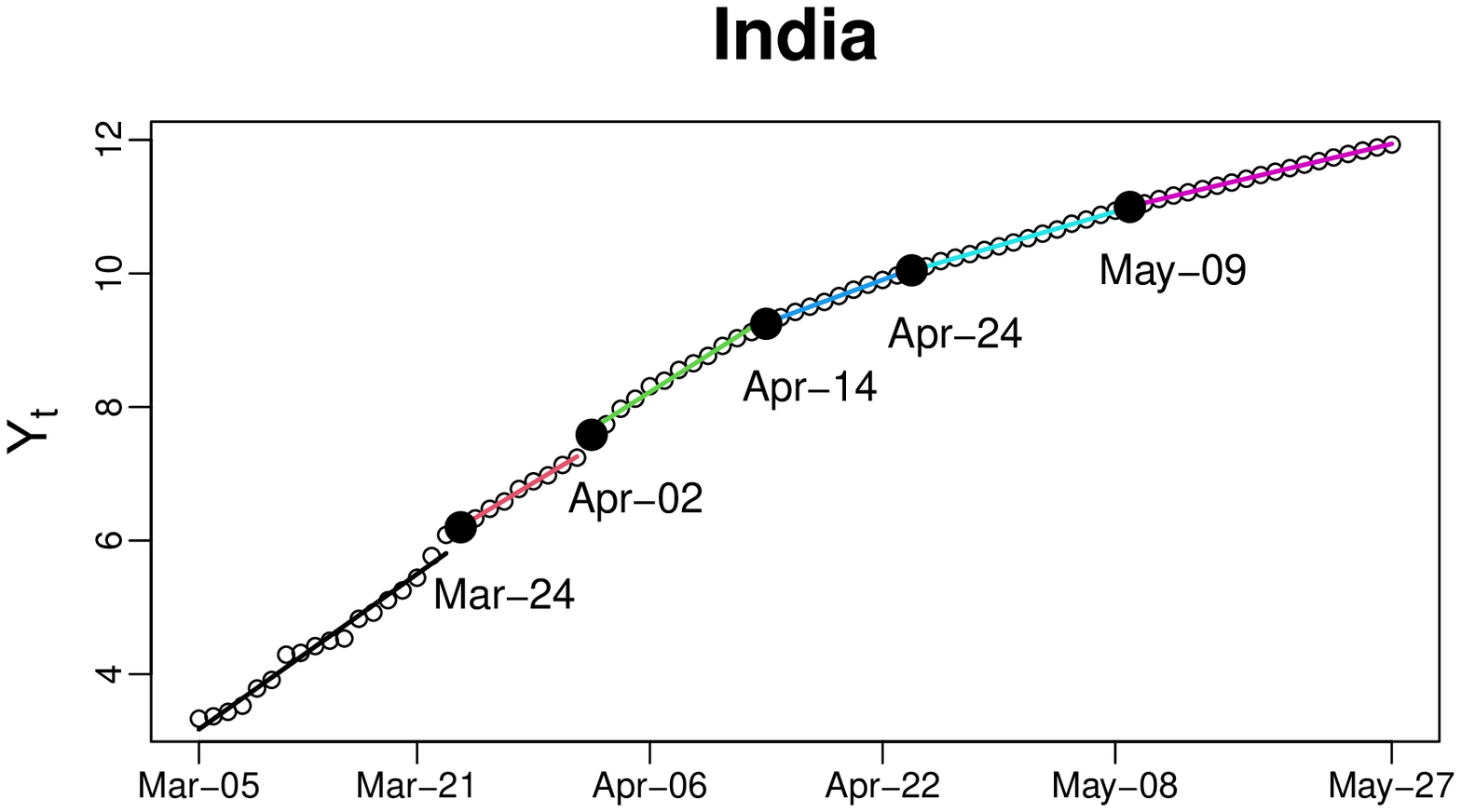}}
	\hfil
	\subfigure{\hspace{-6mm}
		\includegraphics[width=0.52\linewidth]{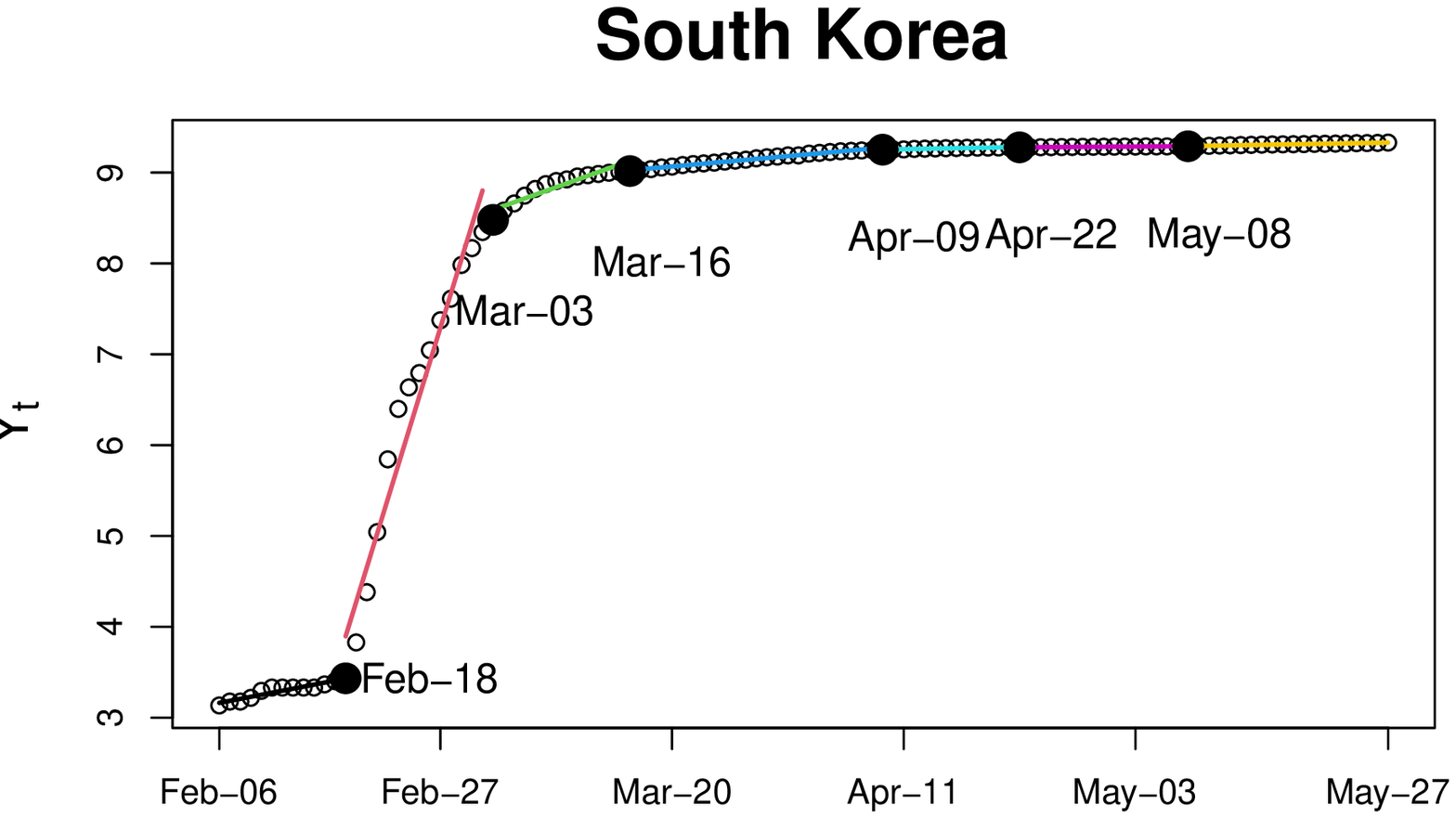}}
	\hfil\vspace{-8mm}
	\caption{Estimated piecewise linear trend for cumulative confirmed cases in 8 representative countries}
    \label{fig_trend}
\end{figure}

Figure \ref{fig_trend} visualizes the estimated piecewise linear models for the eight countries, which gives a more direct perception of how the growth rate  changes over time. Note that the U.S. and  South Korea are the only two countries that witnessed an increase in the slope after the first change-point. For the U.S., the first change-point is March 4, one day after the first confirmed case appeared in New York. Since then, the pandemic underwent an outbreak in the New York state, which has been the leading state in the U.S. in terms of infected cases. The second change-point appeared on March 24, after which the slope began to drop.  This is also noteworthy as on March 20, the U.S. began barring entry of foreign nationals who had traveled to 28 European countries within the past 14 days.  While in South Korea, after February 18, the infected cases increased drastically, and the slope  dropped after March 3. We  find that the first change-point is the day when the first super-spreader in South Korea was diagnosed {\footnote{A member of the Shincheonji religious organization was diagnosed as 31st case in Daegu, see \url{https://foreignpolicy.com/2020/02/27/coronavirus-south-korea-cults-conservatives-china/}}}.  The second change-point, March 3, is when the drive-through testing was made widely available to Korean citizens. 

The growth rate decreased after the first change-point in other countries. For the U.K., the first and second change-points are quite close. In particular, we find the U.K. governments  gradually increased the restrictions on freedom of movement for the general public between these two change-points (March 20 and March 29). This could help explain why both change-points are associated with significant drops in the virus growth rate. In addition, we find that Italy extended the quarantine lockdown from region-focused to nationwide on March 10, one day after the first estimated change-point. For Spain, the first change-point is estimated as March 14, which is one day after Spain declared the nationwide state of emergency. Similar to Italy, the slopes dropped drastically after the first change-point. Generally speaking, the first or second change-point of these countries are closely associated with the date when local or nationwide interventions from the governments were initiated. These countries typically transition from a rapid growth phase to a moderate growth phase after the first or second change-point. This may serve as evidence that government intervention such as lockdown and massive testing could effectively slow down the spread of the coronavirus.  

From Figure \ref{fig_trend}, we also find the situations in Brazil, Russia and India rather somber, as of May 27. Russia is still transitioning from the rapid growth phase to the moderate growth phase, while the fast growing trend in Brazil has not changed since April 12. Even though Brazil managed to bring down the slope by a significant amount at the first change-point on March 25, it seemed the right-wing government took few follow-up effective measures. The situation in India is also grim where the decreases of growth rate at the first and second change-points are quite small and the current growth rate is still high, suggesting  that stricter measures to be taken. In summary, these three countries still have a long way to go in terms of slowing down the spread of COVID-19. 

 %For most countries, the first change-point waited for at least two weeks to come. 
\begin{table}[]
	\caption{Summary of estimated models (\ref{eq: linear trend}) for cumulative confirmed cases in 8 representative countries}
	\label{table_app}
	\begin{tabular}{cccccccc}
		\hline
		Country        & Start  & $n$ & No.CP & 1st CP ($S_1$) & 2nd CP ($S_2$) & Latest CP ($S_{\widehat{m}+1}$) & $\widehat{\rho}$ \\ \hline
		United States  & Feb-22 & 96  & 5    & Mar-04 (0.113)       & Mar-24 (0.292)      & May-09 (0.015)         & 0.492  \\
		Brazil         & Mar-09 & 80  & 2     & Mar-25 (0.301)       & Apr-12 (0.129)       & Apr-12 (0.066)          & 0.438 \\
		Russia         & Mar-12 & 77  & 4     & Apr-05 (0.218)      & Apr-21 (0.146)      & May-17 (0.028)         & 0.573 \\
		United Kingdom & Mar-01 & 88  & 5     & Mar-20 (0.254)      & Mar-29 (0.181)      & May-12 (0.011)         & 0.575  \\
		Spain          & Feb-28 & 90  & 5     & Mar-14 (0.359)      & Mar-27 (0.176)      & May-01 (0.004)         & 0.611 \\
		Italy          & Feb-23 & 95  & 6     & Mar-09 (0.289)      & Mar-22 (0.151)      & May-18 (0.003)         & 0.616  \\
				India           & Mar-05 & 83  & 5     & Mar-24 (0.159)      & Apr-02 (0.142)      & May-09 (0.052)         & 0.375 \\
		South Korea    & Feb-06 & 112 & 6     & Feb-18 (0.022)       & Mar-03 (0.360)      & May-08 (0.002)         & 0.749  \\ \hline
	\end{tabular}
\end{table}

\subsection{Analysis of cumulative confirmed cases in 30 countries}

We further extend the scope of analysis to 30 countries to obtain a relatively complete picture of the pandemic situations around the world. Specifically, we conduct a comparative study based on two important quantities: the maximum normalized slope and the current normalized slope, which are estimated by $S_{max}={\max_{1\leq i\leq \widehat{m}+1}n^{-1}\widehat{b}_i}$  and $S_{cur}=n^{-1}\widehat{b}_{\widehat{m}+1}$ respectively. Combined together, the two measures allow us to obtain an overall picture of the phase when the virus transmitted fastest and the current situation in each country. In particular, $S_{max}$ provides information on the growth rate  at the early stage of the pandemic for a  particular country.  In this phase, often no government regulations are imposed so it depicts the worst scenario if no emergency measure is taken. $S_{cur}$ gives the ongoing epidemic growth rate and  could help make predictions in the short run.

\begin{figure}[!h]
	\centering
\subfigure{\hspace{-5mm}\includegraphics[width=1.06\linewidth]{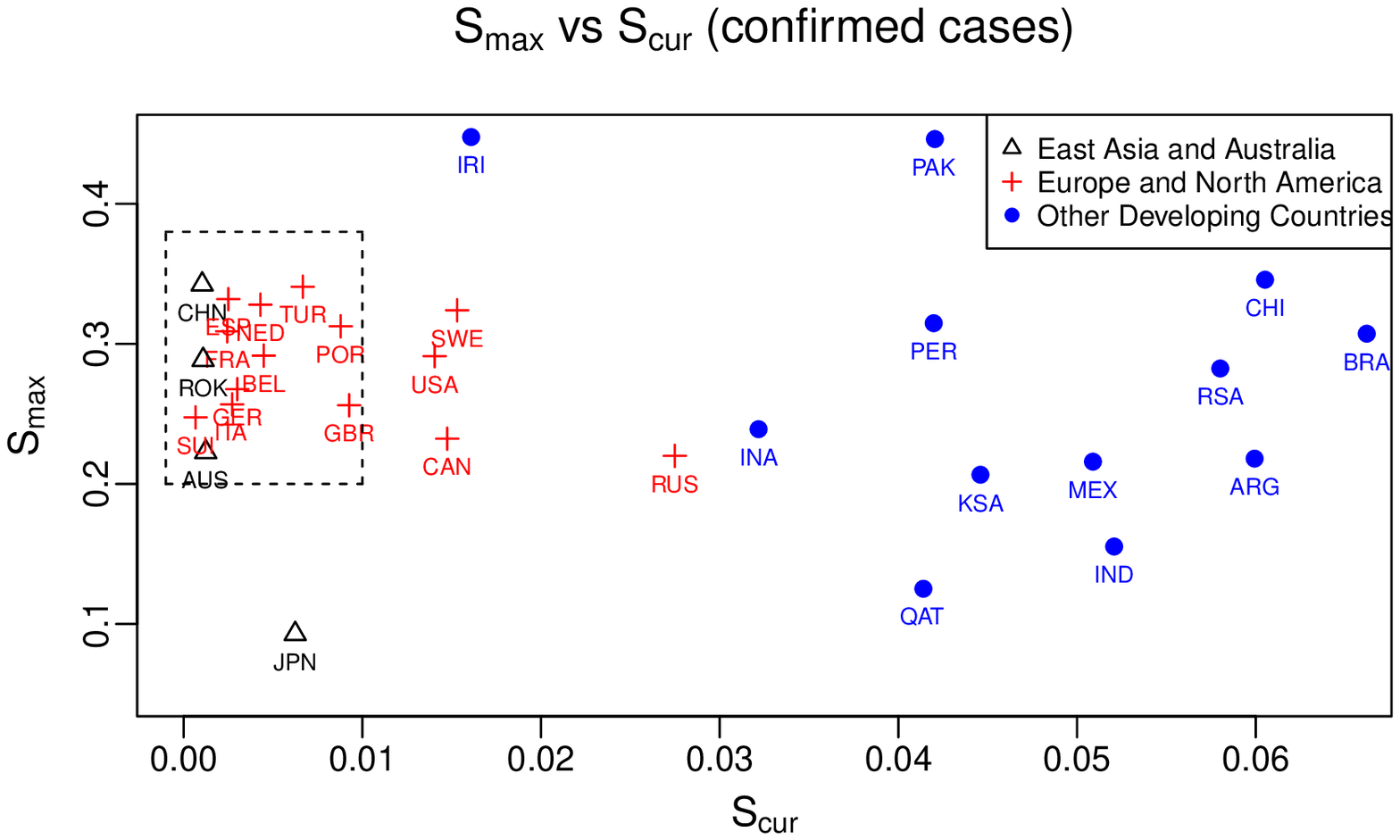}}
		\hfil
	\subfigure{\hspace{-5mm}
		\includegraphics[width=1.06\linewidth]{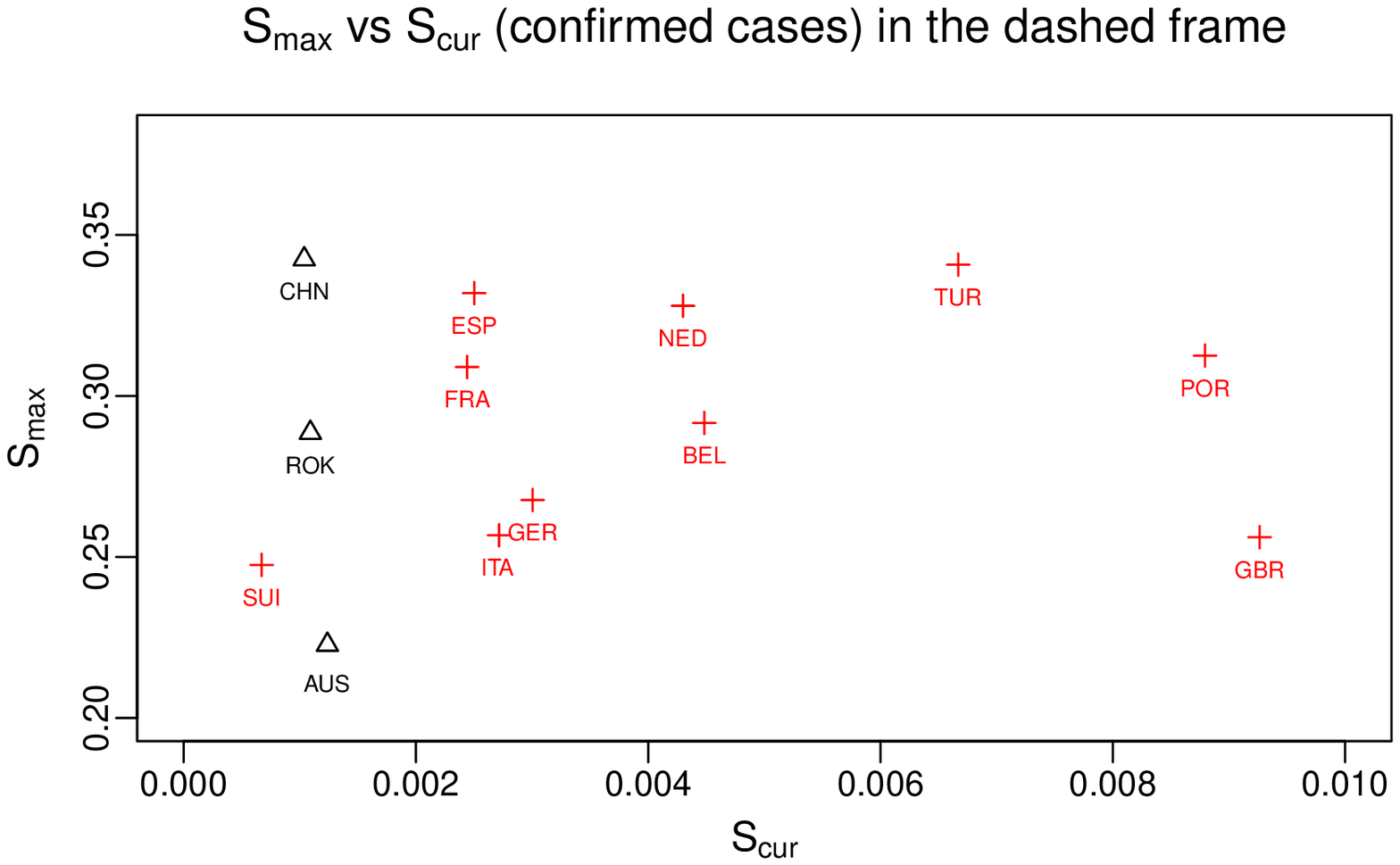}}
	\caption{Plot of maximum normalized slope $S_{max}$ and normalized slope after the latest change-point $S_{cur}$ for cumulative confirmed cases of each country. Black $\Delta$: East Asian Countries and Australia; {\textcolor{red}{red $+$}}: European and  North American Countries; {\textcolor{blue}{blue $\bullet$}}: Other developing countries.}
				\label{fig_slope}
\end{figure}

In Figure \ref{fig_slope}, we plot $S_{max}$ against $S_{cur}$ for each country. Note that by their relative positions in Figure \ref{fig_slope},  the 30 countries can be roughly grouped into three clusters: East Asian countries and Australia, European and North American countries and Other developing countries. We find that countries within the same cluster tend to have similar current growth rate. China, South Korea, and Australia are among the best with $S_{cur}$ close to zero. Most European and North American countries are  in the second tier while countries in continental Europe generally have slower ongoing virus growth than the U.K., the U.S. and Canada. The only exceptions are Sweden and Russia. In fact, Sweden adopted a different strategy than other countries in that no lockdown has been imposed by the government and large parts of its society remain open. Note that Figure \ref{fig_slope} does not take the time effect into account, thus the cluster along the horizontal direction may also be attributed to the cluster of similar eruption time of the virus. This could help explain why Russia is closer to developing countries and why Latin American countries have the largest $S_{cur}$. %Additionally, we find that China, Turkey and Iran have the largest $S_{max}$, while developed countries exhibit comparable magnitude for this index. Japan seems have a pattern different than any other country with the smallest $S_{max}$.
\begin{figure}[!h]
	\centering
	\subfigure{\hspace{-5mm}\includegraphics[width=1.06\linewidth]{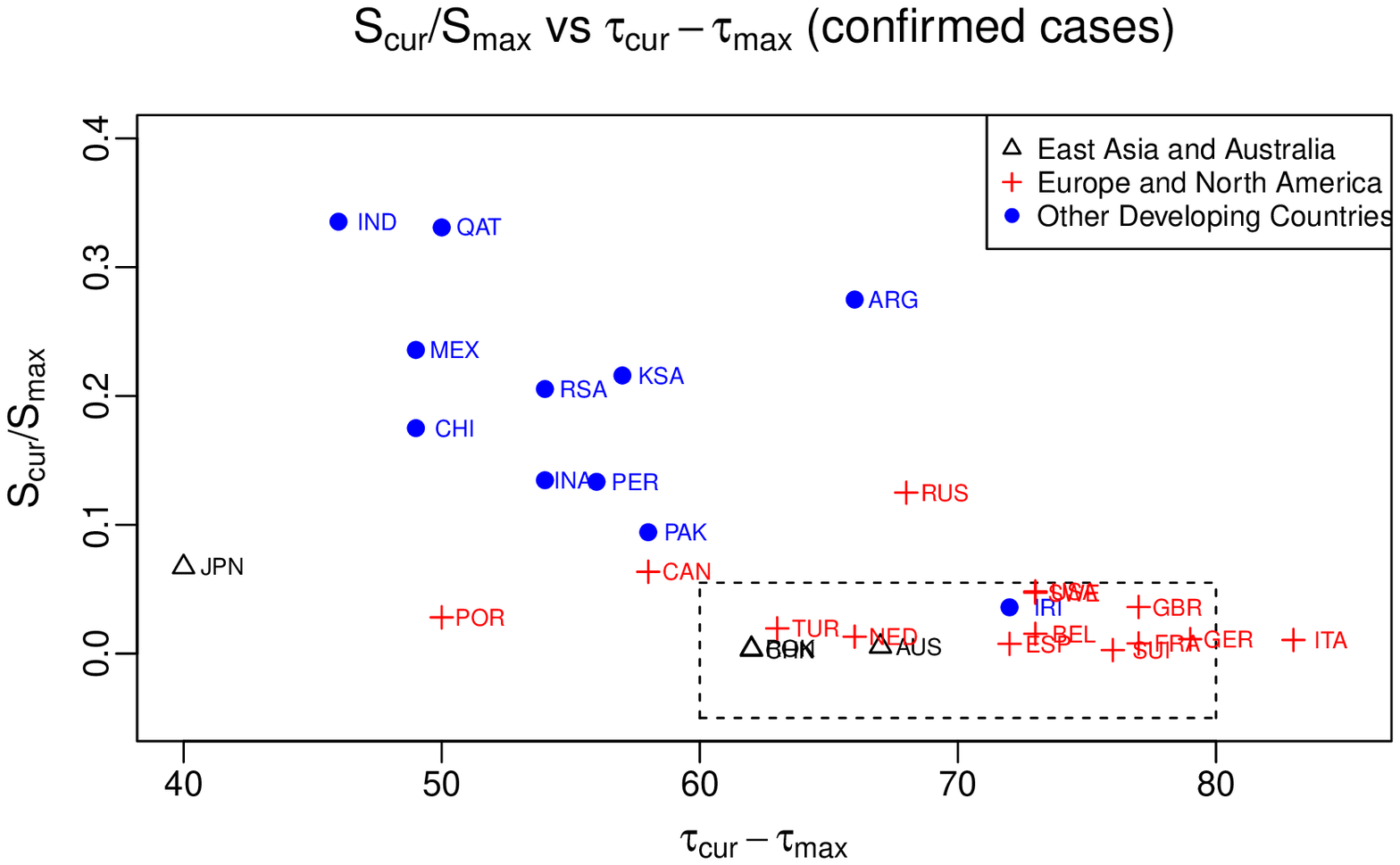}}
	\hfil
	\subfigure{\hspace{-5mm}
		\includegraphics[width=1.06\linewidth]{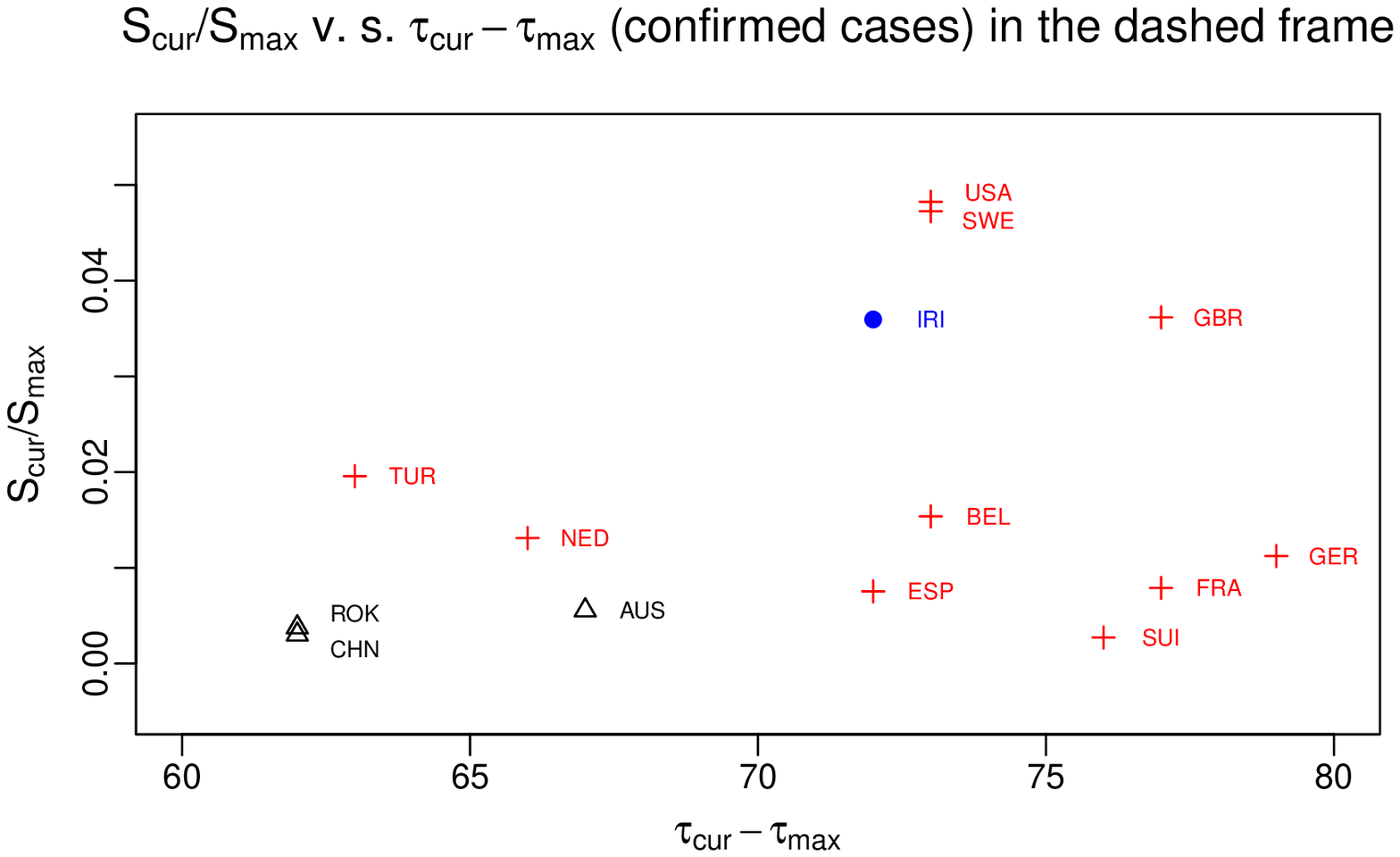}}
	\caption{Plot of ratio between the current normalized slope $S_{cur}$ and maximum normalized slope $S_{max}$ against days from the  start date for the segment with the largest slope to the start date for the latest segment for cumulative confirmed cases of each country.  Black $\Delta$: East Asian Countries and Australia; {\textcolor{red}{red $+$}}: European and  North American Countries; {\textcolor{blue}{blue $\bullet$}}: Other developing countries.}
		\label{fig_ratio}
\end{figure}

%\Zifeng{(Try to rewrite this paragraph for clarity and grammatical mistakes. Try to give a better explanation of the ratio)}
To take the time factor into consideration, in Figure \ref{fig_ratio}, we plot the ratio $S_{cur}/S_{max}$ against the days in between (i.e. $\tau_{cur}-\tau_{max}$ with $\tau_{max}$ as the start date for the segment with the largest slope and $\tau_{cur}=\tau_{\widehat{m}}$ as the latest change-point), which allows us to further understand how the growth rate changes from its peak to the current status with time.
%By examining the ratio, we can understand how flat is the slope in the current phase compared with the worst scenario. \FY{(No idea what is other explanation)}  The time between these two phases then measures how much effort the government has put into to flatten te slopes. 
Horizontally speaking, for the same ratio $S_{cur}/S_{max}$, if country A is to the left of  country B, then A acts faster than B in bringing down the virus growth from its peak value. Vertically speaking, for the same time length $\tau_{cur}-\tau_{max}$, if A is below B, then A is more effective than B in reducing the growth rate.

We again find that most European and North American countries tend to share similar characteristics. The growth rates in the current phases for these countries are less than one-tenth of their peak value, and it took them about two to three months to achieve that.  
From the lower panel in Figure \ref{fig_ratio}, we  find that South Korea, China and Australia outperform other countries as the ratios were brought to near zero in around 65 days. Again, we find that continental European countries (except Russia and Sweden) perform better than U.S, Canada and U.K.  %Despite the ratio in Switzerland is also close to zero, it took a longer time. 

Most developing countries are on the top-left of the plot, suggesting that they are still in the relatively early stage of the pandemic and the situation has not improved much since the beginning of the outbreak. In addition, we find Latin American countries, such as Mexico, Brazil, Chile, and Peru, tend to cluster. Given their geographical proximity, this is not a surprise. We note that developing countries tend to be less efficient in slowing the spread of COVID-19. For example, with roughly the same amount of time, the ratios in India and Argentina are three times larger than developed countries. In summary, more caution and attention should be given to the epidemic in developing countries as they may need more international aids compared to the developed countries.  

\subsection{Analysis of cumulative deaths in 30 countries}
Based on the same methodology, we analyze cumulative deaths in the 30 countries. Note that unlike confirmed cases, public health interventions naturally have a longer lagged effect on coronavirus-related deaths, as severe symptoms may not develop immediately upon infection. Thus, we believe a change-point analysis on cumulative confirmed cases should be preferred in terms of quantifying the effectiveness of emergency policies. Additionally, the criteria for certifying deaths due to COVID-19 vary from nation to nation, thus comparative analysis across countries should be interpreted with caution.

Table \ref{table_app2} summarizes the detailed estimation result for cumulative deaths in the eight representative countries. Notably, for each country, the estimated number of change-points for deaths is smaller than or equal to that for cumulative confirmed cases in Table \ref{table_app}. This is intuitive as the history of cumulative deaths is shorter and number of deaths largely depend on infections (with a lag). Note that the duration between the starting date and the first change-point for cumulative deaths is around 2-3 weeks, which is consistent with that for confirmed cases in Table \ref{table_app}. The same phenomenon also applies to the duration between the first and second change-points. This consistency in part confirms the validity of the change-point estimation results and indicates a 2-3 weeks response lag between changes in growth rate of infections and changes in growth rate of deaths. We note that Italy and Spain have the highest growth rate of cumulative deaths before the first change-point, which highlights the extreme importance of ``flattening the curve", as it is known that the exponential surge of coronavirus cases exhausted the public health system in the two countries at the early stage of the pandemic.

%the devastating effect of coronavirus once the .

%, and their first change-points appear earlier than in other countries.  This could indicate that Spain and Italy may not pay enough attention to the early spread of the virus. 

% This suggests that the timeline of changes in death growth rate has a 2-3 weeks lag behind the confirmed cases. 
%we find that for most countries, the starting dates are at least two weeks later than those of the confirmed cases except for Italy and Spain.  In addition, 
%we find   as this time period usually lasted for roughly two weeks.

\begin{figure}[!h]
	\centering
	\subfigure{	\hspace{-6mm}	
		\includegraphics[width=0.52\linewidth]{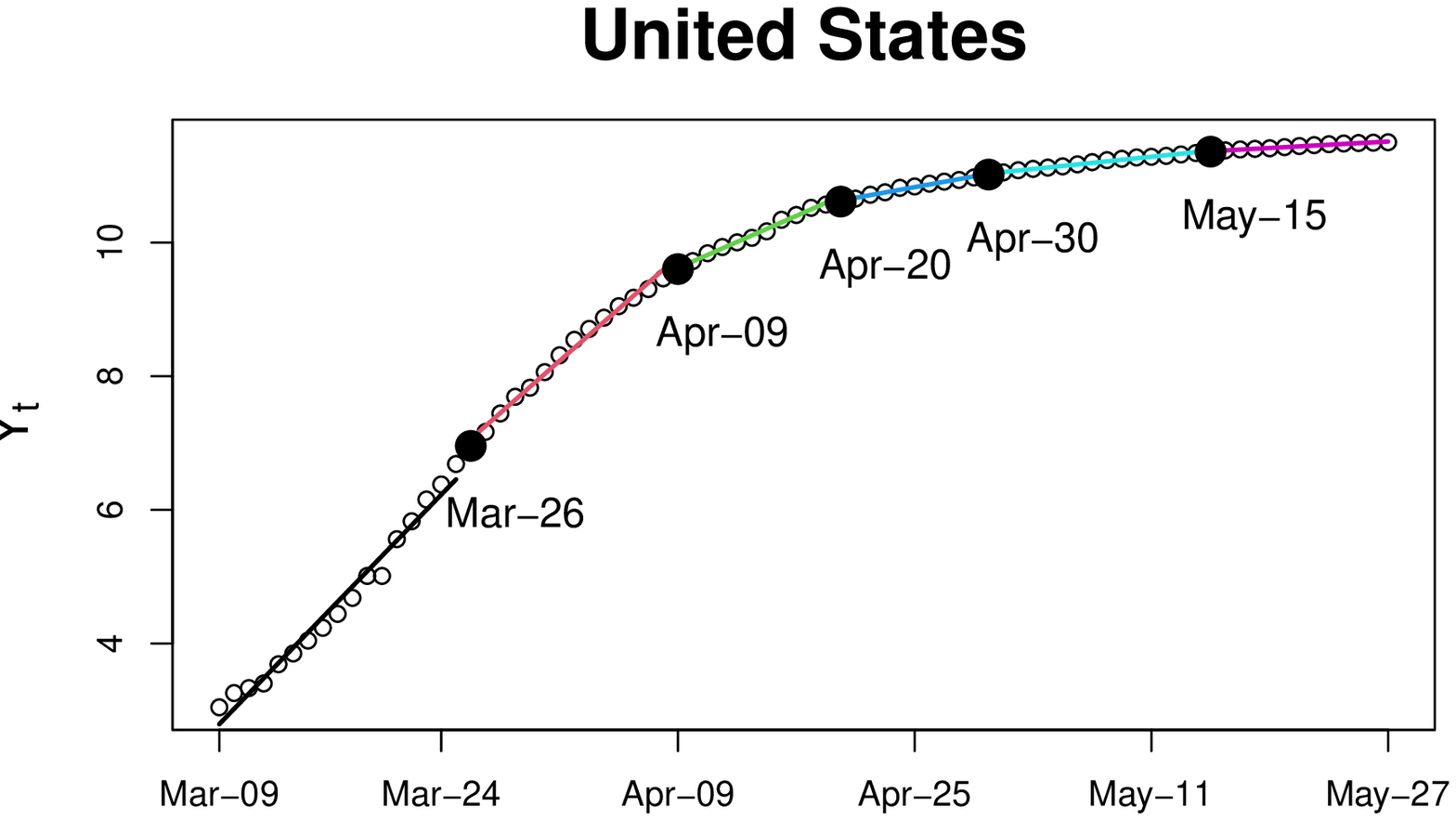}}
	\hfil
	\subfigure{	\hspace{-7mm}
		\includegraphics[width=0.52\linewidth]{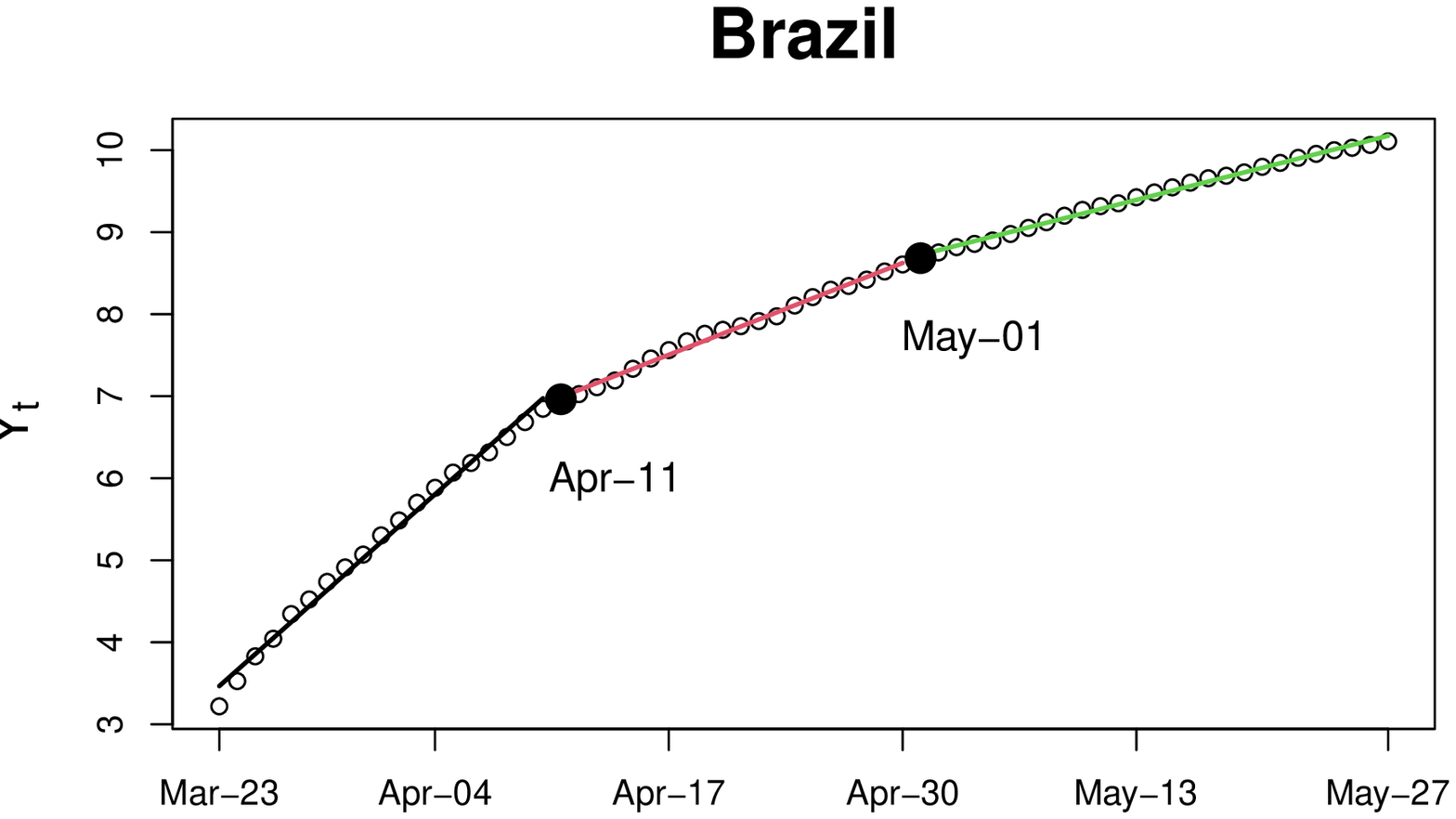}}
	\hfil\vspace{-8mm}
	\subfigure{\hspace{-5mm}
		\includegraphics[width=0.52\linewidth]{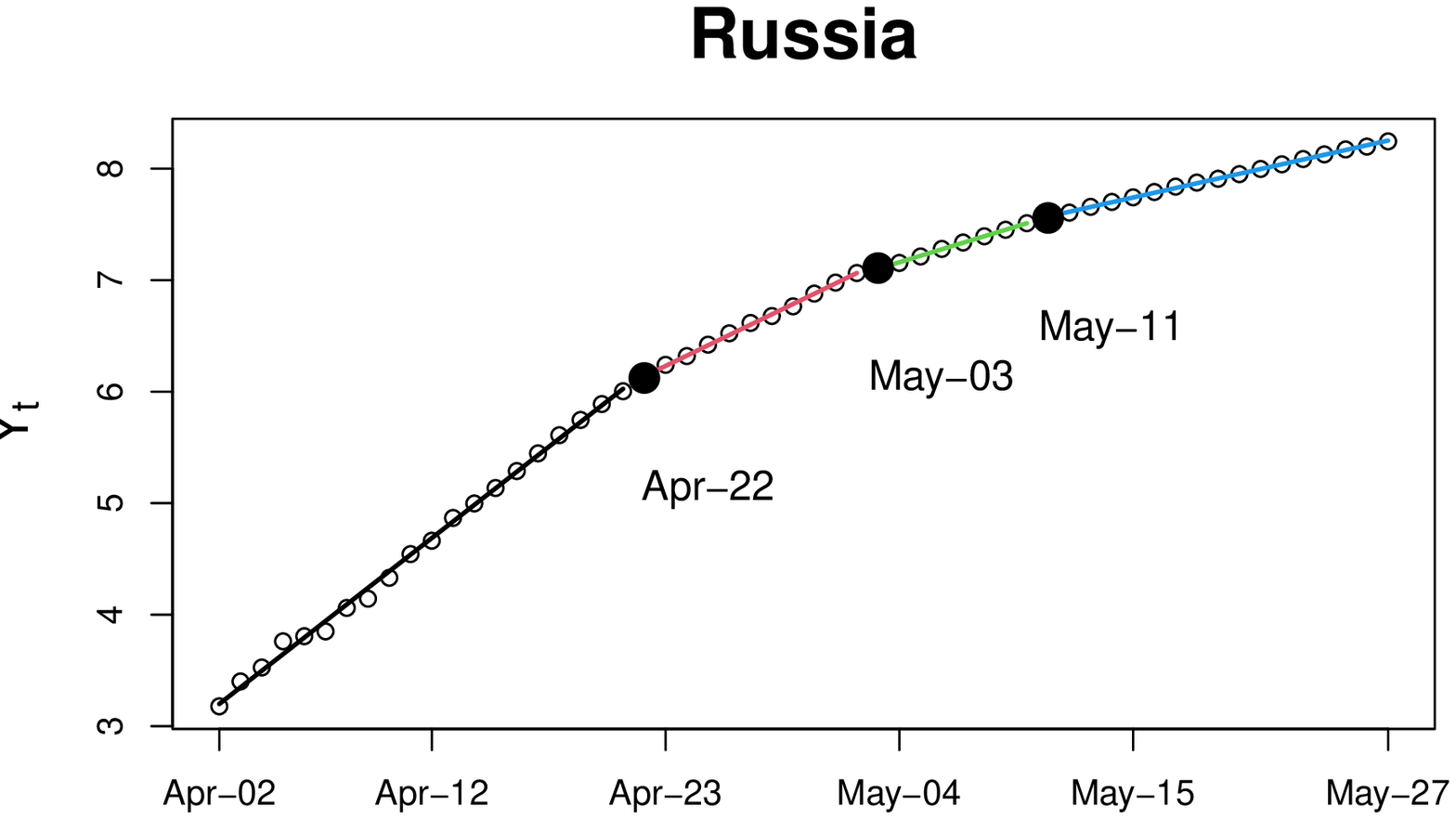}}
	\hfil
	\subfigure{\hspace{-6mm}
		\includegraphics[width=0.52\linewidth]{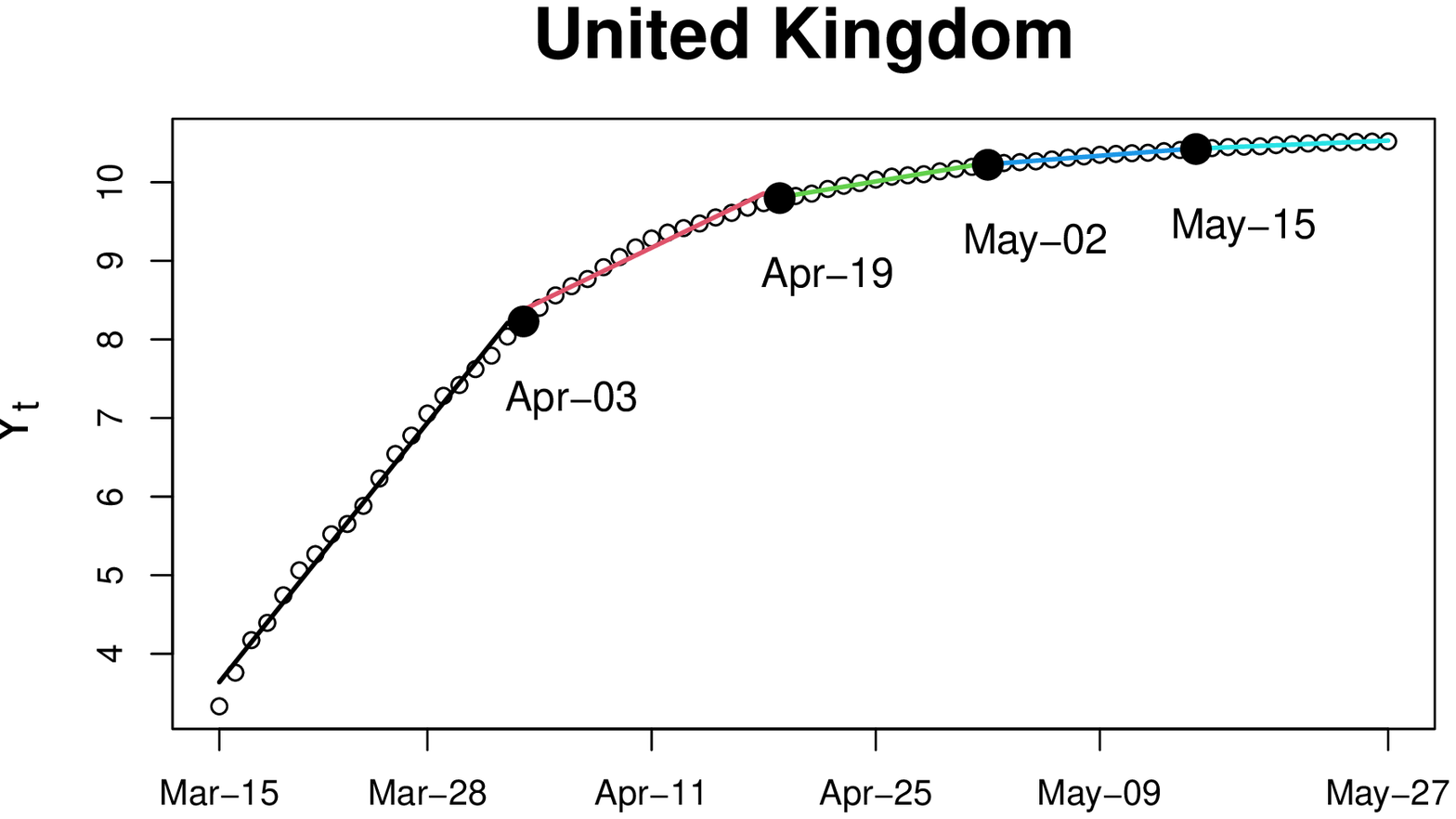}}	
	\hfil\vspace{-8mm}
	\subfigure{\hspace{-5mm}
		\includegraphics[width=0.52\linewidth]{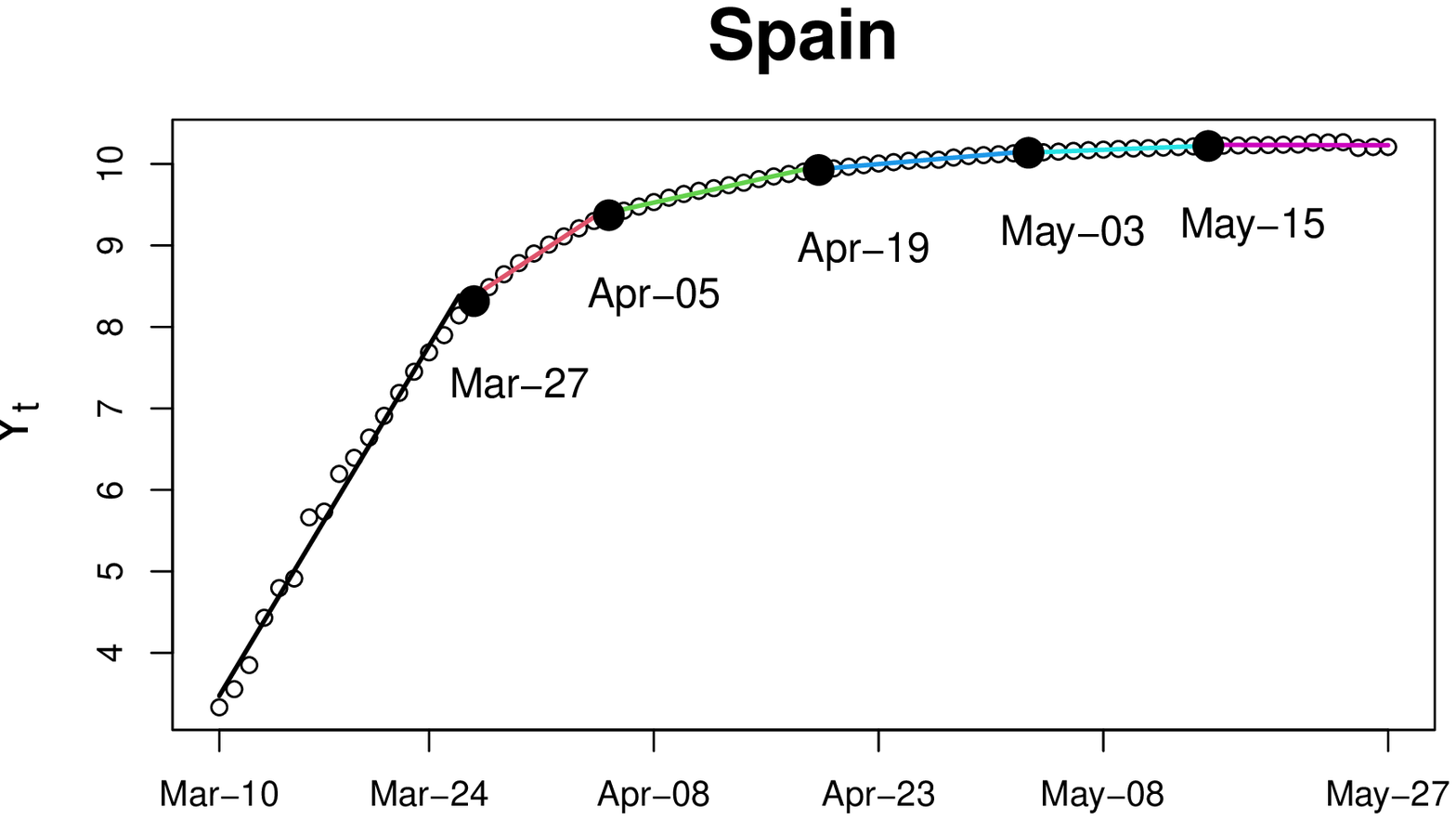}}
	\hfil
	\subfigure{\hspace{-6mm}
		\includegraphics[width=0.52\linewidth]{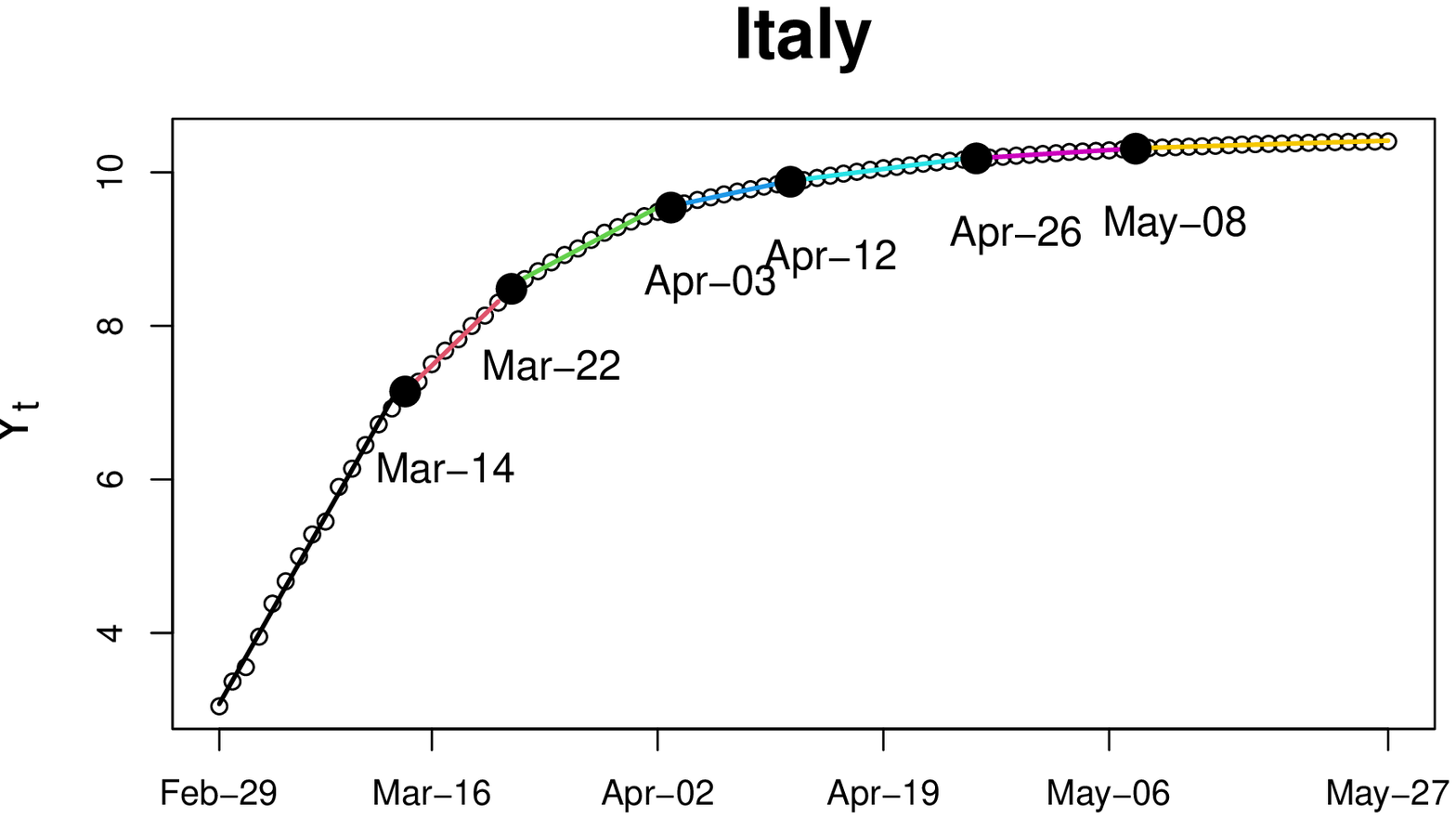}}
	\hfil\vspace{-8mm}
	
	\subfigure{\hspace{-5mm}
		\includegraphics[width=0.52\linewidth]{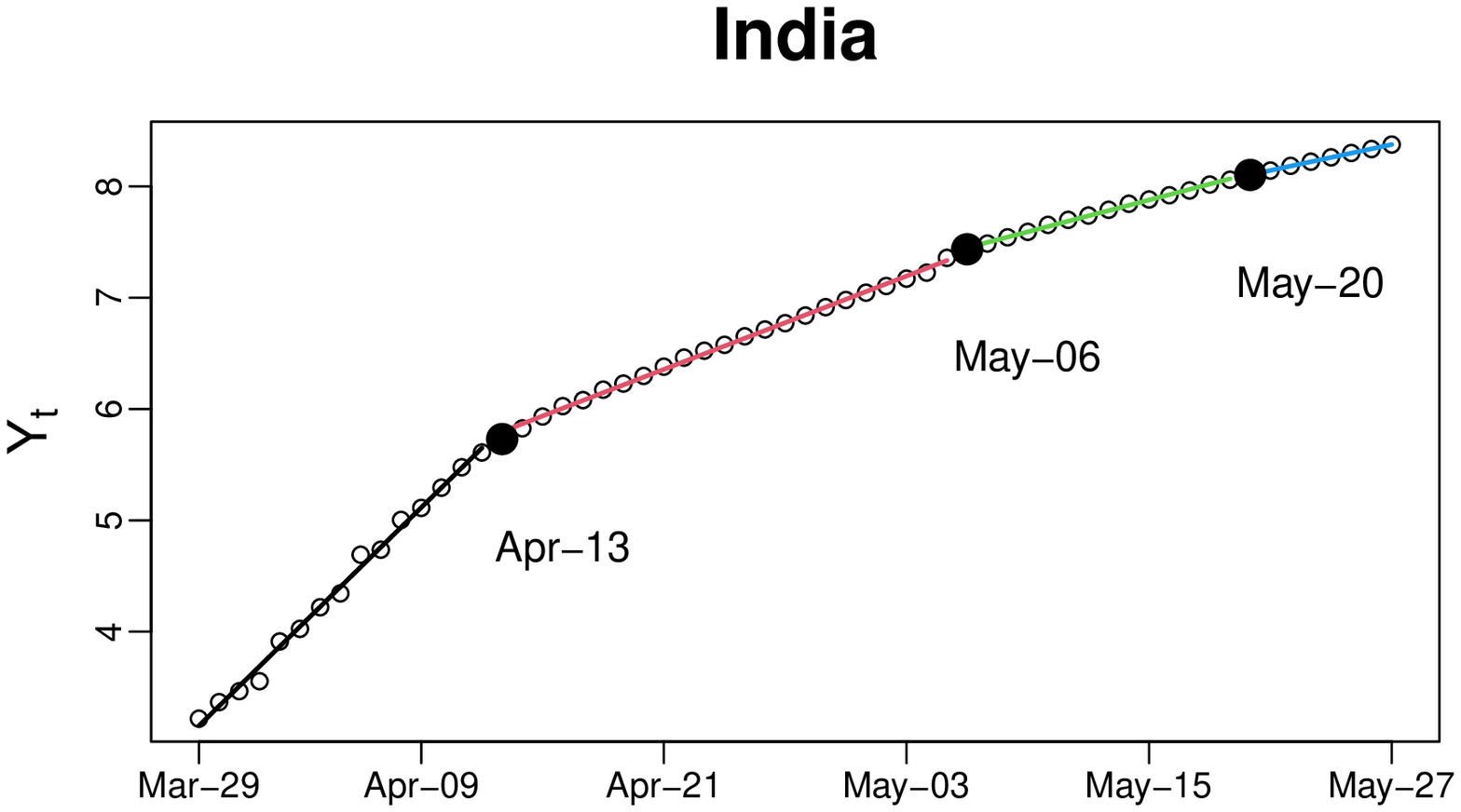}}
	\hfil
	\subfigure{\hspace{-6mm}
		\includegraphics[width=0.52\linewidth]{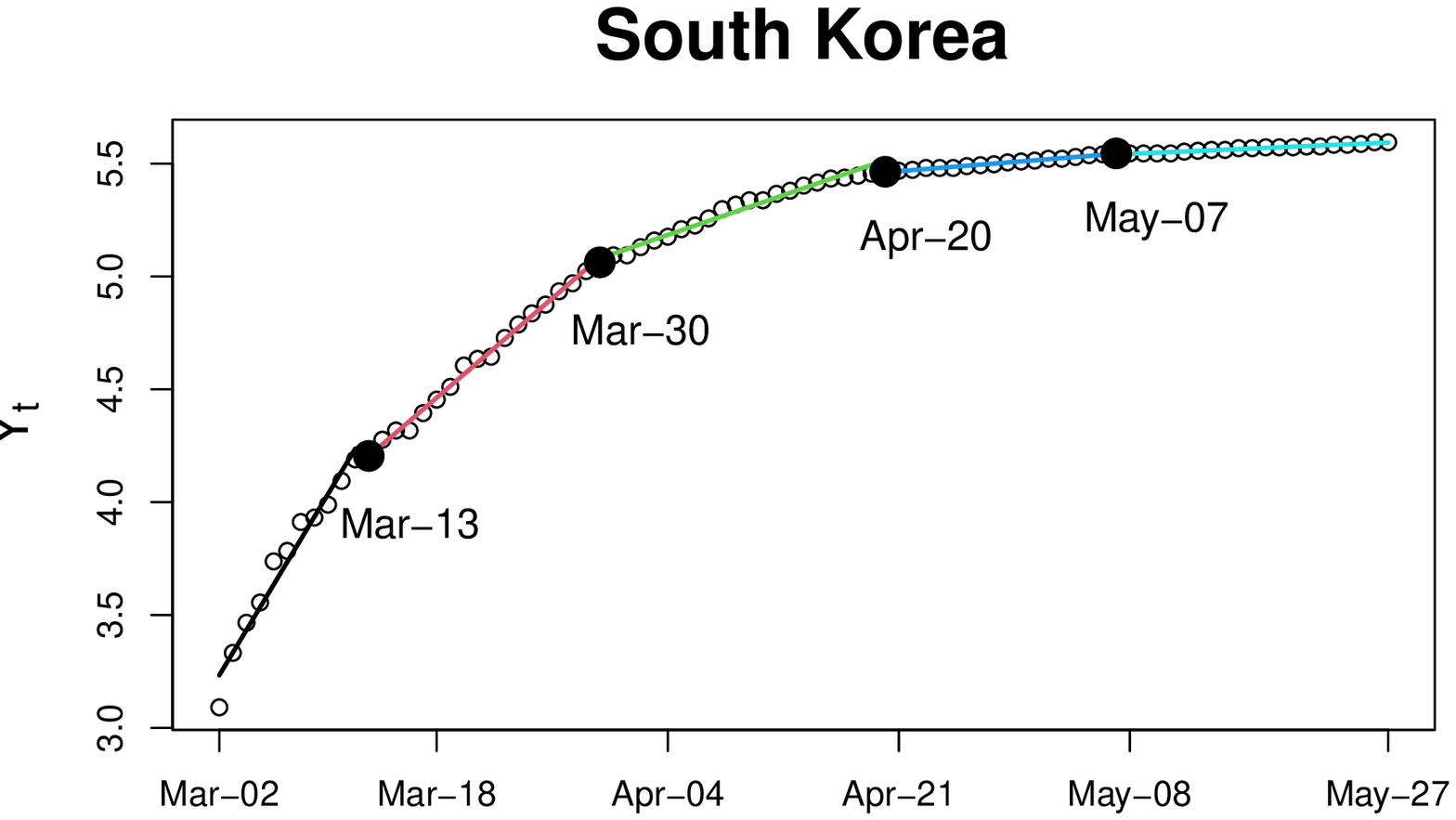}}
	\hfil\vspace{-8mm}
	\caption{Estimated piecewise linear trend for cumulative deaths in 8 representative countries }
	\label{fig_trend2}
\end{figure}

Figure \ref{fig_trend2} further plots the estimated piecewise linear models for cumulative deaths in the eight countries. The pattern exhibited by each country is largely consistent with its pattern in Figure \ref{fig_trend}, except for South Korea. Note that the start date of the cumulative death curve in South Korea is almost 30 days later than the start date of the cumulative confirmed cases, which partially explains the different pattern around its first change-point.

 %We see that  except for the U.S. and South Korea, these patterns are quite similar to the plots for confirmed cases regardless of the delay effect.  For the U.S. and South Korea, the growth rates no longer increase after the first-change point.  
\begin{table}[]
		\caption{Summary of estimated models (\ref{eq: linear trend}) for cumulative deaths in 8 representative countries}
	\label{table_app2}
	\begin{tabular}{cccccccc}
		\hline
		Country        & Start  & $n$ & No.CP & 1st CP ($S_1$)  & 2nd CP ($S_2$) & Latest CP ($S_{\widehat{m}+1}$) & $\widehat\rho$ \\ \hline
		United States  & Mar-09 & 80  & 5     & Mar-26 (0.229) & Apr-09 (0.195)  & May-15 (0.012)                 & 0.556  \\
		Brazil         & Mar-23 & 66  & 2     & Apr-11 (0.195) & May-01 (0.086)  & May-01 (0.056)                 & 0.696  \\
		Russia         & Apr-02 & 56  & 3     & Apr-22 (0.149) & May-03 (0.093)  & May-11 (0.043)                 & 0.366  \\
		United Kingdom & Mar-15 & 74  & 4     & Apr-03 (0.254) & Apr-19 (0.099)  & May-15 (0.008)                 & 0.657  \\
		Spain          & Mar-10 & 79  & 5     & Mar-27 (0.307) & Apr-05 (0.121)  & May-15 (-0.000$^*$)                & 0.507  \\
		Italy          & Feb-29 & 89  & 6     & Mar-14 (0.305) & Mar-22 (0.167)  & May-08 (0.005)                 & 0.287  \\
		India          & Mar-29 & 60  & 3     & Apr-13 (0.179) & May-06 (0.070)  & May-20 (0.039)                 & -0.012  \\
		South Korea    & Mar-02 & 87  & 4     & Mar-13 (0.100) & Mar-30 (0.0518) & May-07 (0.003)                 & 0.363  \\ \hline
	\end{tabular}
\begin{tablenotes}\footnotesize
	\item $*$. Spain revised its death toll downwards on May 25, see {\url{https://english.elpais.com/society/2020-05-26/spanish-health-ministry-lowers-coronavirus-death-toll-by-nearly-2000.html}.}
\end{tablenotes}
\end{table}

\begin{figure}[!h]
	\centering
	\subfigure{\hspace{-5mm}\includegraphics[width=1.06\linewidth]{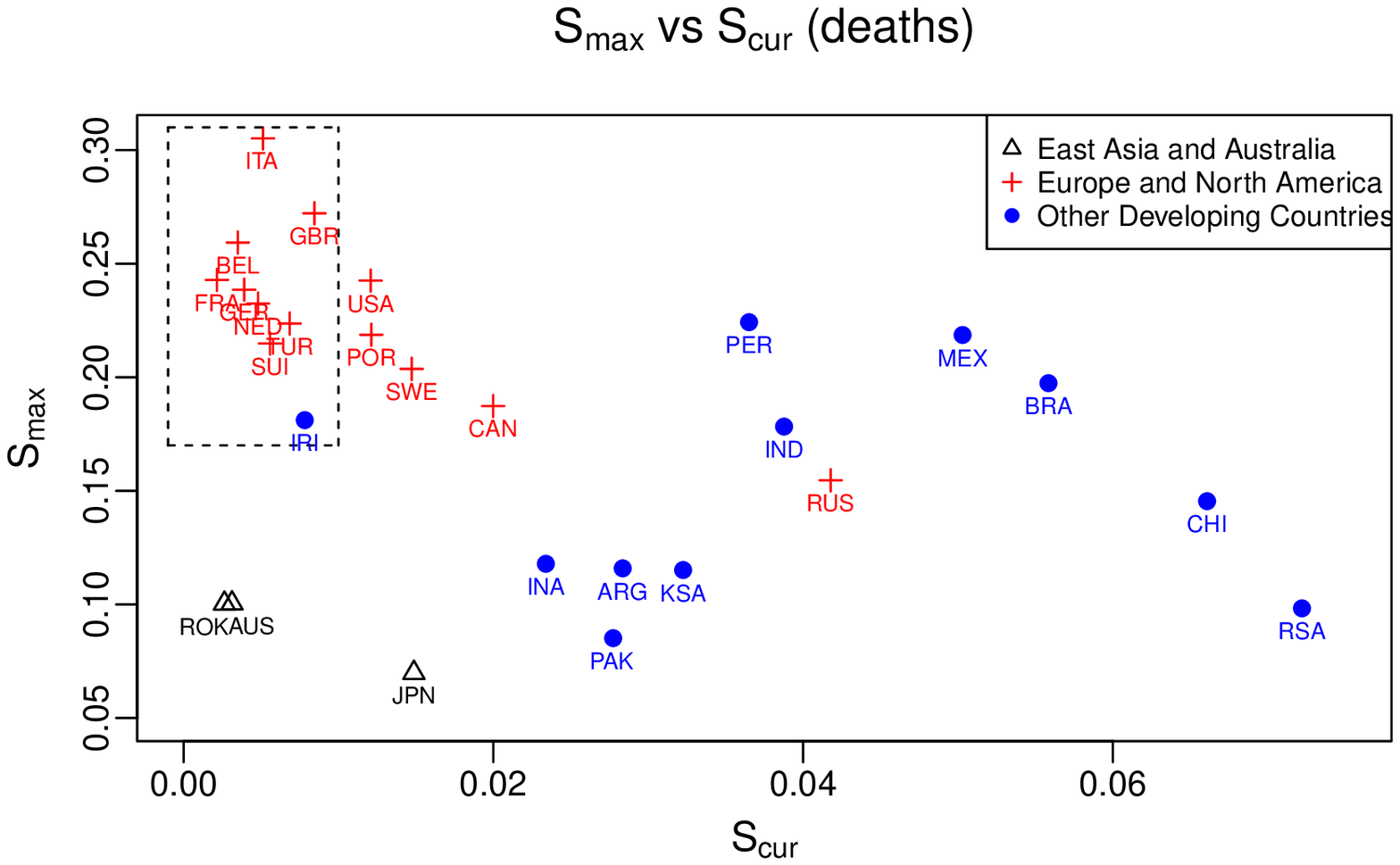}}
	\hfil
	\subfigure{\hspace{-5mm}
		\includegraphics[width=1.06\linewidth]{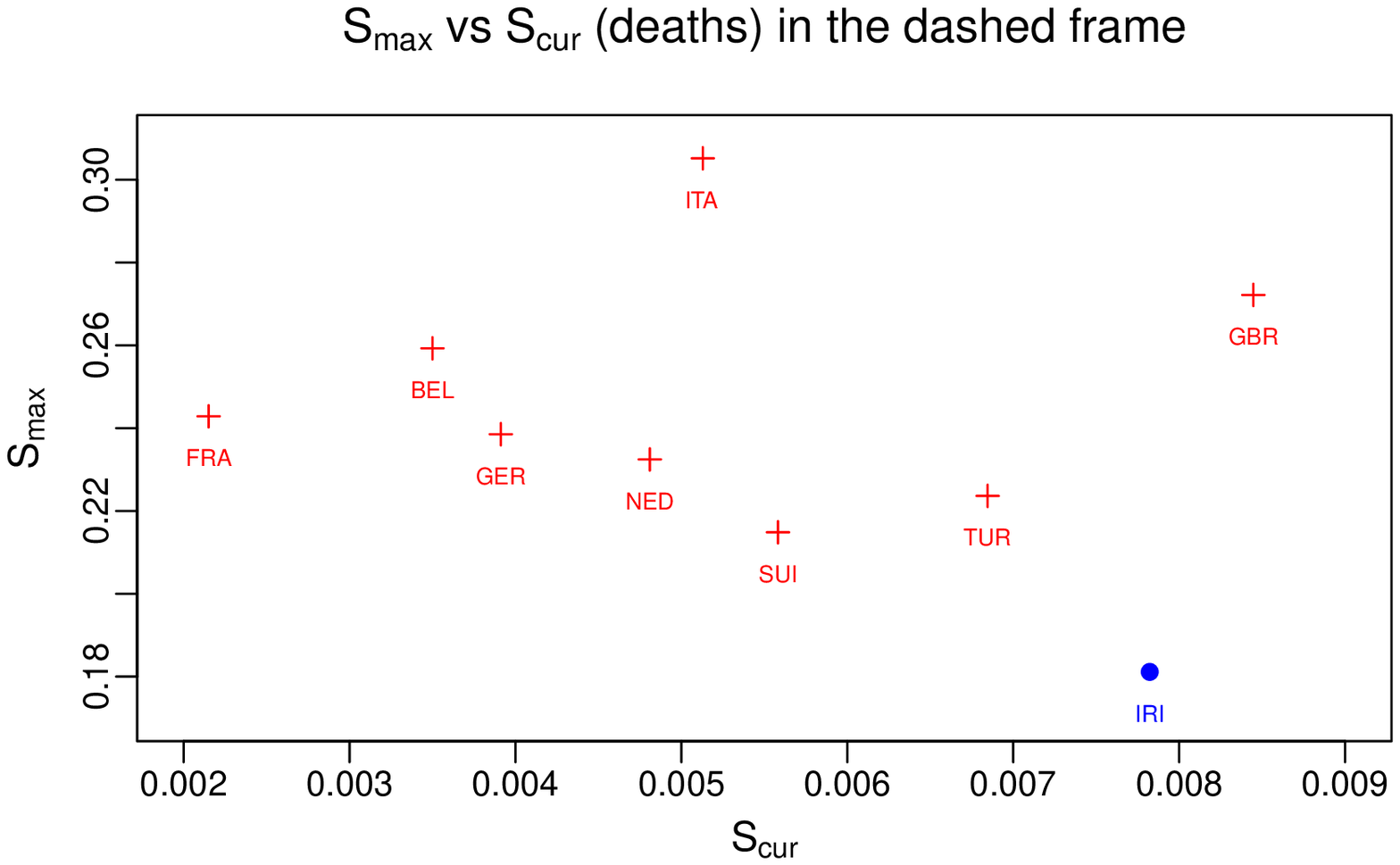}}
	\caption{Plot of maximum normalized slope $S_{max}$ and normalized slope after the latest change-point $S_{cur}$ for cumulative deaths of each country.  Black $\Delta$: East Asian Countries and Australia; {\textcolor{red}{red $+$}}: European and  North American Countries; {\textcolor{blue}{blue $\bullet$}}: Other developing countries.}
	\label{fig_slope2}
\end{figure}

We further conduct a comparative analysis for cumulative deaths in 30 countries. We exclude China, Spain and Qatar in the analysis as the death tolls were either revised or unavailable{\footnote{China revised its death toll upwards on April 17, see {\url{https://www.nytimes.com/2020/04/17/world/asia/china-wuhan-coronavirus-death-toll.html}}. The death toll is not available for Qatar.}}. Figure \ref{fig_slope2} plots $S_{max}$ against $S_{cur}$ for each country. Similar to the results for confirmed cases in Figure \ref{fig_slope}, European and North American countries tend to cluster while developing countries generally have higher ongoing growth rates $S_{cur}$.

Note that South Korea and Australia deliver the best responses with small $S_{max}$ and near-zero $S_{cur}$ for cumulative deaths. However, it is unexpected to see that western developed countries, such as Italy and the U.K., experience the largest maximum growth rate. Since the maximum growth rate always takes place in the first segment of the cumulative death curve, it indicates that the coronavirus may take these countries by surprise and the health systems may not be well prepared for the flood of coronavirus patients in the early stage of the pandemic. Another notable pattern is that Latin American countries tend to have larger values in both maximum and current growth rates than other developing countries, signaling the possibility of Latin America becoming the next epicenter of the COVID-19 pandemic.

%Since these countries have similar medical levels, it could indicate the insufficiency of the medical supply and lack of emergency responses in Latin American is more severe.
 
\begin{figure}[!h]
	\centering
	\subfigure{\hspace{-5mm}\includegraphics[width=1.06\linewidth]{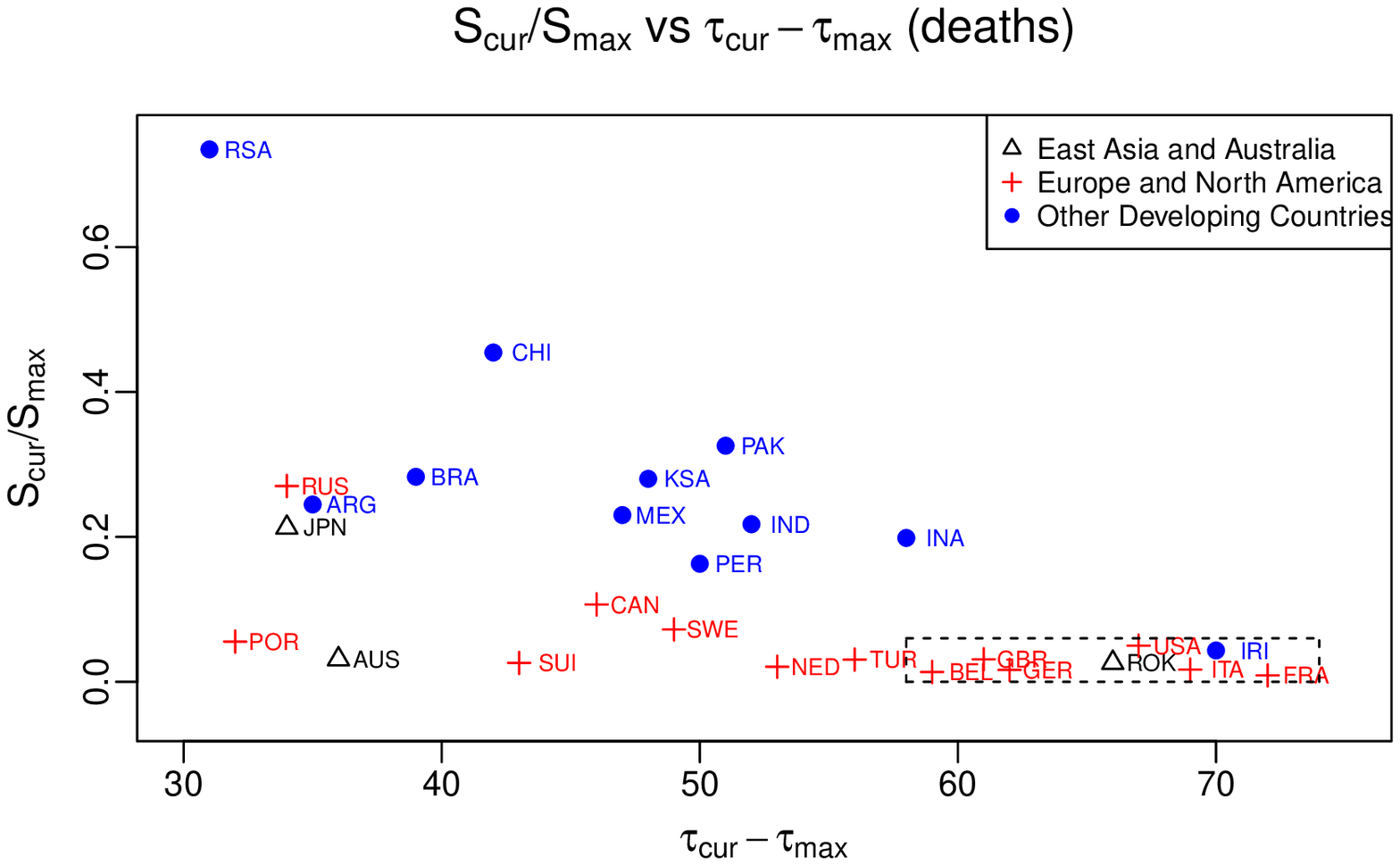}}
	\hfil
\subfigure{\hspace{-5mm}
	\includegraphics[width=1.06\linewidth]{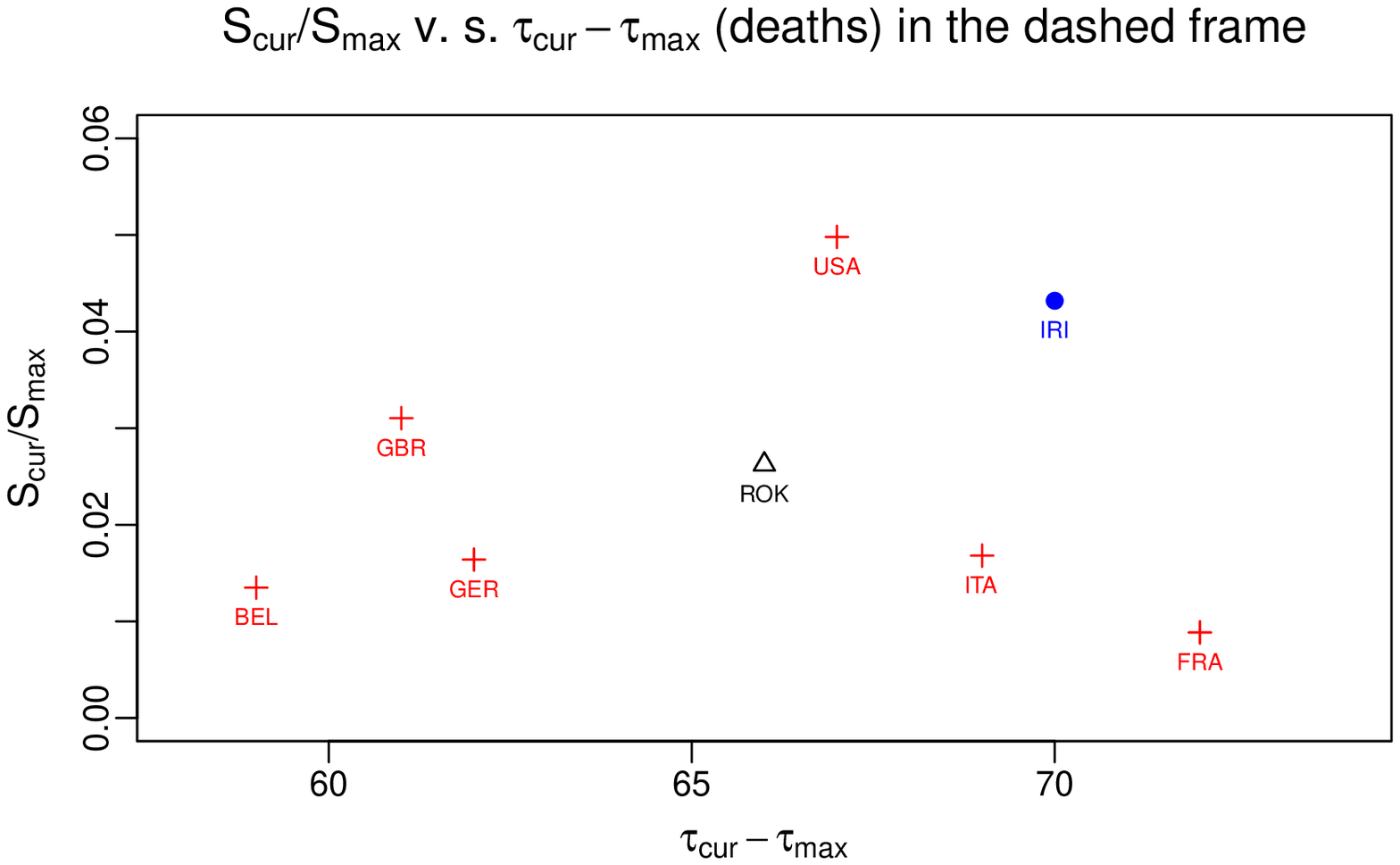}}
	\caption{Plot of ratio between the current normalized slope $S_{cur}$ and maximum normalized slope $S_{max}$ against days from the  start date for the segment with the largest slope to the start date for the latest segment for cumulative deaths of each country.  Black $\Delta$: East Asian Countries and Australia; {\textcolor{red}{red $+$}}: European and  North American Countries; {\textcolor{blue}{blue $\bullet$}}: Other developing countries.}
	\label{fig_ratio2}
\end{figure}

Figure \ref{fig_ratio2} plots $S_{cur}/S_{max}$ against $\tau_{cur}-\tau_{\max}$ for cumulative deaths in each country, where the observed patterns are similar to the ones for cumulative confirmed cases in Figure \ref{fig_ratio2}. Specifically, developing countries again tend to be less efficient in slowing the spread of COVID-19, where with roughly the same amount of time, the ratios $S_{cur}/S_{max}$ in developing countries are noticeably larger than developed countries. 

%, except for Japan, for which the ratio is relatively large. Note that the total number of deaths in Japan is much smaller   than  other countries sharing similar traits like Argentina, Brazil and Russia, more attention should be given to the latter countries.

\section{SN-NOT based forecast for cumulative deaths}
As stated by the Centers for Disease Control and Prevention~(CDC)\footnote{\url{https://www.cdc.gov/coronavirus/2019-ncov/covid-data/forecasting-us.html#why-forecasting-critical}}, accurate forecast of COVID-19 deaths is critical for public health decision-making, as it projects the likely impact of coronavirus to health systems in coming weeks and helps government officials develop data-driven public health policies for controlling the pandemic.

In Section 5.1, we propose a simple and intuitive forecasting scheme for cumulative deaths due to COVID-19 by combining SN-NOT with a flexible extrapolation function. In Section 5.2, we further demonstrate its promising performance in predicting cumulative deaths in the U.S.

%It will  help the government, especially the health department,  to prepare for future impacts from the need for medical assistance such as nurses,  medicine,  ventilators and other medical equipment. 

%Therefore, segmenting the series and making predictions based on the  information in the latest segment should be more appropriate.  Our method can also be applied to other countries as it is purely data-driven and doesn't require additional information such as government policies, local populations,  number of hospitals, etc. 

\subsection{Method}
As suggested by the analysis in Section 4, the spread of coronavirus typically experiences several different stages due to external interventions. While a sophisticated epidemiology model based on differential equations may manage to take into account information about interventions and characterize the entire cumulative death curve, a more natural (and simpler) solution from the change-point aspect is to first segment the time series into periods with relatively stable behavior and then generate forecast based on observations in the last segment, see for example, \cite{pesaran2002market} and \cite{Bauwens2015}.

Following this idea, we propose an SN-NOT based two-stage approach for cumulative deaths prediction. Specifically, in the first stage, given the cumulative deaths~(in log scale) $\{Y_t\}_{t=1}^n$, a piecewise linear trend model is estimated via SN-NOT with change-points $\widehat{\bftau}$. In the second stage, a flexible function $f(t)$ is fitted on the last segment $\{Y_t\}_{t=\widehat{\tau}_{\widehat{m}}+1}^n$ with the assumption that $\mathbb{E}(Y_t)=f(t)$ and the $k$-day ahead forecast for cumulative deaths can be readily made via extrapolation of $\widehat{f}(t)$.

Note that the purpose of the first stage (in-sample) change-point analysis is to identify the most recent segment where $\{Y_t\}_{t=1}^n$ exhibits relatively stable behavior and thus facilitates the second stage (out-of-sample) forecast. As demonstrated in Section 4, the piecewise linear trend model with SN-NOT is sufficient for this task. However, as for prediction in the second stage, any flexible extrapolation function $f(t)$ can be considered, as it is expected that a linear function may only provide a reasonable forecast for short horizons due to its limited flexibility.

In the following, we consider three commonly used extrapolation functions~(in the order of increasing flexibility) in the literature, including the linear function $f(t)=a+b(t/n)$, the quadratic function $f(t)=c+ d(t/n)+e(t/n)^2$ and the logistic function $f(t)=\dfrac{L}{1+\exp\big(-\alpha(t/n-t_0)\big)}$.

Based on $\{Y_t\}_{t=\widehat{\tau}_{\widehat{m}}+1}^n$, a standard OLS can be used to estimate the linear and quadratic functions and a standard nonlinear least square can be used to estimate the logistic function. The $k$-day ahead forecast for $Y_{n+k}$ is formulated respectively as
\begin{alignat*}{2}
&\mbox{SN-NOT + Linear [SNL]:} \quad &&\widehat{Y}_{n+k}=\widehat{a}+ \widehat{b}(1+k/n),\\
&\mbox{SN-NOT + Quadratic [SNQ]:} \quad &&\widehat{Y}_{n+k}=\widehat{c}+ \widehat{d}(1+k/n)+\widehat{e}(1+k/n)^2,\\
&\mbox{SN-NOT + Logistic [SNLG]:} \quad &&\widehat{Y}_{n+k}=\frac{\widehat{L}}{1+\exp\big(-\widehat{\alpha}(1+k/n-\widehat{t}_0)\big)}.
\end{alignat*}
The prediction for cumulative deaths on day $n+k$ is $\widehat{\mathrm{Death}}_{n+k}=\exp(\widehat{Y}_{n+k})$.

%(via the {\tt nls} function in \texttt{R}) 

% We emphasize that $f(t)$ in the second stage does not need to be a linear function. 

%First, we will apply the SN-NOT to estimate the change-points. Equipped with the estimated change-points,  in the second step, we will fit   This two-step forecasting procedure allows us to remove redundant past information (before change-points)  while maintaining enough recent information to  make short-run predictions. % with the observations in the latest segment and make predictions based on the fitted model.

% where $$
%\begin{flalign}
%&\label{SNL}\mbox{[SNL]}\quad\mathbb{E}Y_{t}=a+ b(t/n),\\
%&\label{SNQ}\mbox{[SNQ]}\quad\,\,\,\, \mathbb{E}Y_{t}=c+ d(t/n)+e(t/n)^2,\\
%&\label{SNLG}\mbox{[SNLG]}\quad\mathbb{E}Y_{t}=\frac{L}{1+\exp\big((n \alpha)^{-1}(t-t_0)\big)}, 
%\end{flalign}

%for $t=\widehat{\tau}_{\widehat{m}}+1,\cdots,n$, where $\widehat{\tau}_{\widehat{m}}$ is the estimation for the latest change-point by SN-NOT.

%For  SNL and SNQ, we  obtain the estimates $(\widehat{a},\widehat{b})$ and $(\widehat{c},\widehat{d},\widehat{e})$ by OLS, while a nonlinear least square (via the {\tt nls} function in \texttt{R}) is used to estimate $(\widehat{L},\widehat{\alpha},\widehat{x}_0)$ for SNLG. After fitting the model, the $k$-ahead predictions for the $Y_{n+k}$ are then calculated as 

\subsection{Data and prediction results}
We apply the SN-NOT based prediction method to forecast cumulative deaths in the U.S. and compare its performance with other forecasting models listed on the CDC website\footnote{\url{https://www.cdc.gov/coronavirus/2019-ncov/covid-data/forecasting-us.html}}. Specifically, following the CDC website, the forecast is generated on five dates, April-27, May-04, May-11, May-18 and May-25, and the forecast horizon is 5-day (one-week) ahead and 12-day (two-week) ahead. 
 
%suggested by the Centers for Disease Control and Prevention.  
%The data we use here is different with that used in the previous section in order to make results comparable. Specifically, we use the dataset obtained by ``UT"  by  \cite{UT}, which is publicly available at \url{https://github.com/UT-Covid/USmortality/tree/master/forecasts/archive}. The implementation of SN-NOT depends on the starting date when cumulative deaths exceeded 20 as March 10.

We compare with five forecasting models{\footnote{Other models can be found on the CDC website. The five models are chosen as their predictions are available on all the aforementioned dates while other models only report on some of the recent dates.}} available on the CDC website: ``LANL"  by \cite{LANL}, ``Imperial" by \cite{verity2020estimates},  ``UT"  by  \cite{UT}, ``YYG"  by \cite{YYG} and ``MOBS"  by \cite{MOBS}. These forecasting methods are mainly ensembles of complex mechanistic models~(such as SEIR and SEIS), known as compartmental models in epidemiology, which track the spread of infectious disease via a system of differential equations. To highlight the importance of the first-stage change-point analysis, we additionally report the forecast given by fitting a logistic function on the entire time series without segmentation~(and name it ``Logistic").

Table \ref{table_pre} reports the prediction results and the findings can be summarized as follows.

(1) SNL gives comparable performance to other methods for the 5-day ahead forecast, while it considerably overestimates deaths at the 12-day horizon. In other words, linear extrapolation can only be used for short-term forecasts. This is not surprising as the linear function essentially assumes a constant growth rate for the cumulative deaths. While such an approximation is reasonable for short-term, it may not be able to track the growth rate for a long period to make accurate predictions. SNQ generally performs better than SNL due to its increased flexibility, though it tends to underestimate at the 12-day horizon as the quadratic function may pass its peak for long-horizon extrapolation.

(2) SNLG is consistently a top performer among all models thanks to the flexibility of the logistic function, which ensures the fitted curve is non-decreasing and is capable of tracking both increasing and decreasing growth rate. Note that there is a drastic performance difference between the two-stage SNLG forecast and the pure Logistic forecast, which indicates the value of the first-stage change-point estimation for identifying the most recent segment where cumulative deaths exhibit relatively stable behavior.

In summary, the SN-NOT based two-stage prediction, in particular SNLG, provides decent forecasts for the cumulative deaths in the U.S. Considering that SNLG is solely based on the time series of cumulative deaths, this result is rather promising and further confirms the value and validity of the change-point analysis. Though by no means SNLG can replace the complex mechanistic models built on epidemiology principles, we believe it can serve as a meaningful addition to the existing set of forecasting models for tracking the COVID-19 pandemic.

 %\Revise{especially for developing countries with less data availability.}

%In summary, we find that our proposed three models will deliver reasonable forecasts in the short run. However, SNL and SNQ may work poorly in the long run prediction. SNLG, on the other hand, will make stable  and comparable predictions with other methods. Taken into account that SNLG is purely data-driven and borrows no additional information, this is quite promising. 

%It is also possible to try other models and make nonlinear extrapolations based on the latest segment, but we will not pursue in this direction.  %However, directly using the logistic function for modeling the entire series is disappointing, as is shown in the ``Logistic" model. 

\begin{table}[!h]
	\caption{Prediction performance for cumulative deaths in the U.S. (the top 3 performers on each forecast date are highlighted in bold).}
	\label{table_pre}
	\addtolength{\tabcolsep}{-3.8pt}
	\begin{tabular}{ccccccccccccc}
		\hline
		Date                    & Target             &      & True     & Imperial                   & LANL    & MOBS     & UT       & YYG     & SNLG   & SNL   & SNQ      & Logistic \\ \hline
		\multicolumn{13}{c}{End-of-Week}                                                                                                                                                          \\ \hline
		\multirow{2}{*}{Apr-27} & \multirow{2}{*}{May-02} & Forecast  & 66527    & {66837}                      & 69410   & 63029    & 58720    & 73317   & {65067}   & 70376   & {63775}    & 55480    \\
		&                         & Rel.error & $\slash$ & \bf{0.47\%}                     & 4.33\%  & -5.26\%  & -11.74\% & 10.21\% & \bf{-2.19\%} & 5.79\%  & \bf{-4.14}\%  & -16.61\% \\
		\multirow{2}{*}{May-04} & \multirow{2}{*}{May-09} & Forecast  & 78946    & {79511}                    & {78755}   & 77035    & 70646    & {77522}  & 77178   & 85775   & 75703    & 62930    \\
		&                         & Rel.error & $\slash$ & \bf{0.72\%}                     & \bf{-0.24\%} & -2.42\% & -10.51\% & \bf{-1.80\%} & -2.24\% & 8.65\%  & -4.11\%  & -20.29\% \\
		\multirow{2}{*}{May-11} & \multirow{2}{*}{May-16} & Forecast  & 88893    & 91528                      & 87022   & {88922}    & 87666    &{88767}   & {88128}   & 86331   & 84965    & 69702    \\
		&                         & Rel.error & $\slash$ & 2.96\%                     & -2.10\% & \bf{0.03\%}   & -1.38\%  & \bf{-0.14\%} & \bf{-0.76\%} & 2.88\%  & -4.42\%  & -21.59\% \\
		\multirow{2}{*}{May-18} & \multirow{2}{*}{May-23} & Forecast  & 97220    & 98076                      & 96582   & {97252}    & 96128    & 97625  & {97573}   & 99432   & {97307}    & 75659    \\
		&                         & Rel.error & $\slash$ & 0.88\%                     & -0.66\% & \bf{0.03}\%   & -1.12\%  & 0.42\%  & \bf{0.36\%}  & 2.28\%  &\bf{ 0.09}\%   & -22.18\% \\
		\multirow{2}{*}{May-25} & \multirow{2}{*}{May-30} & Forecast  & 103915   & 104671                     & {104085}  & {104241}   & 104736   & 104436  & {103923}  & 108080  & 103197   & 80887    \\
		&                         & Rel.error & $\slash$ & 0.73\%                     & \bf{0.16\% } &\bf{0.31\%}   & 0.79\%   & 0.50\%  & \bf{0.01\%}  & 4.01\%  & -0.69\%  & -22.16\% \\ \hline
		Date                    & Target              &      & True     & Imperial$^*$                   & LANL    & MOBS     & UT       & YYG     & SNLG   & SNL   & SNQ      & Logistic \\ \hline
		\multicolumn{13}{c}{Two-week} \\ \hline
		\multirow{2}{*}{Apr-27} & \multirow{2}{*}{May-09} & Forecast  & 78946    & \multirow{10}{*}{\slashbox{~\\\\\\\\\strut~}{~\\\\\\\\\strut~}} & {84837}   & 70156    & 65903    & {77336}   & {74244}   & 93730   & 67565    & 58341    \\
		&                         & Rel.error & $\slash$ &                            & \bf{7.46\%}  & -11.13\% & -16.52\% & \bf{-2.04\%} & \bf{-5.96\%} & 18.73\% & -14.42\% & -29.72\% \\
		\multirow{2}{*}{May-04} & \multirow{2}{*}{May-16} & Forecast  & 88893    &                            & {90078}   & 85827    & 78243    & {87608}   & {85896}   & 109953  & 79531    & 64625    \\
		&                         & Rel.error & $\slash$ &                            & \bf{1.33\%}  & -3.45\%  & -11.98\% & \bf{-1.45\%} & \bf{-3.37\%} & 23.69\% & -10.53\% & -29.21\% \\
		\multirow{2}{*}{May-11} & \multirow{2}{*}{May-23} & Forecast  & 97220    &                            & 93997   & {97513}    & {96232}    & 98365   & {96136}   & 124580  & 89205    & 70719    \\
		&                         & Rel.error & $\slash$ &                            & -3.32\% &\bf{ 0.30\%}   & \bf{-1.02\%}  & 1.18\%  & \bf{-1.11\%} & 28.14\% & -8.24\%  & -27.26\% \\
		\multirow{2}{*}{May-18} & \multirow{2}{*}{May-30} & Forecast  & 103915   &                            & {103461}  & {104443}   & 101060   & 106432  & 105985  & 111822  & {104819}   & 76269    \\
		&                         & Rel.error & $\slash$ &                            & \bf{-0.44\%} & \bf{0.51\%}   & -2.75\%  & 2.42\%  & 1.99\%  & 7.61\%  & \bf{0.87\%}   & -26.60\% \\
		\multirow{2}{*}{May-25} & \multirow{2}{*}{June-06} & Forecast  & 109802   &                            &{110640}  & {110285}   & 111759   & 111799  & {109708}  & 119971  & 106995   & 81251    \\
		&                         & Rel.error & $\slash$ &                            & \bf{0.76\%}  & \bf{0.44\%}   & 1.79\%   & 1.82\%  & \bf{-0.09\%} & 9.26\%  & -2.56\%  & -26.00\%    \\ \hline
	\end{tabular}
\begin{tablenotes}\footnotesize
	\item $*$. Imperial only gives one-week ahead forecast.
\end{tablenotes}
\end{table}

\bibliographystyle{chicago}
\bibliography{reference}

\clearpage
\begin{center}
	\textbf{\large Supplement to ``Time series analysis of COVID-19 Infection Curve: a change-point perspective"}
\end{center}
\appendix

This supplement consists of three parts. Appendix A contains technical proofs.  Appendix B extends the piecewise linear structure of model (1.1) to piecewise polynomial and presents an analysis for cumulative confirmed cases in 8 representative countries using a piecewise quadratic model. Appendix C provides the lag-1 to lag-30 (P)ACF plots of the residuals for cumulative confirmed cases in the 8 countries presented in Section 4.2.

\vspace{2mm}

\section{Proofs}\label{proof}

In what follows, we denote $\Rightarrow$ as the weak convergence on $D[\epsilon,1]$, the space of functions on $[\epsilon,1]$ which are right continuous and have left limits, endowed with Skorohod metric.  Let $X_n\in\mathbb{R}^d$ with dimension $d>0$ be a set of random vector  defined in a probability space $(\Omega,\mathbb{P},\mathcal{F})$. For a corresponding set of constants $a_n$, we say $X_n=O_p^s(a_n)$ if for any $\varepsilon>0$, there exists a finite $M>0$ and a finite $N>0$ such that for $n>N$,
$$
\mathbb{P}(\|X_n/a_n\|_d>M)+\mathbb{P}(\|X_n/a_n\|_d<1/M)<\varepsilon,
$$ 
where $\|\|_d$ denotes the $L_d$ norm.

\textsc{Proof of Theorem \ref{thm_main}}
(i) It is a direct application of  Theorem 3.1 in \cite{rho2015inference}
and continuous mapping theorem.  In particular, the result of (i) in Theorem 3.1 in \cite{rho2015inference} corresponds to the case of (i)  in Assumption \ref{ass} for linear processes while the result of   (ii) in Theorem 3.1 in \cite{rho2015inference} corresponds to the case of (ii)  in Assumption \ref{ass} for nonlinear processes.
%\Zifeng{No need  to cite Rho and Shao (2015), give a direct proof. } \FY{I think it's OK. Otherwise, the proof can be quite long and tedious with repeated stuff in Rho and Shao (2015).}

(ii)
On one hand, note that the continuous mapping theorem indicates that 
\begin{flalign*}
L_{n,\delta}(1,\bftau,n)\Rightarrow\Gamma^2L_{\delta}(\kappa),\quad\mbox{and}\quad
R_{n,\delta}(1,\bftau,n)\Rightarrow\Gamma^2R_{\delta}(\kappa).
\end{flalign*}
and it follows that $
V_{n,\delta}(1,\bftau,n)\Rightarrow\Gamma^2V_{\delta}(\kappa).
$

On the other hand, 
\begin{flalign*}
D_{n}(1,\bftau,n)=&\kappa(1-\kappa)
\sqrt{n}\Big(\widehat{\bbeta}_{1,\bftau}-\widehat{\bbeta}_{\bftau+1,n}+\mathbf{b}\Big)-\kappa(1-\kappa)
\sqrt{n}\mathbf{b},
\end{flalign*}
and it is clear that \begin{flalign*}
&\kappa(1-\kappa)\sqrt{n}\Big(\widehat{\bbeta}_{1,\bftau}-\widehat{\bbeta}_{\bftau+1,n}+\mathbf{b}\Big)\\\Rightarrow& \kappa(1-\kappa)\Gamma Q(\kappa)^{-1}B_F(\kappa)-\Gamma[Q(1)-Q(\kappa)]^{-1}[B_F(1)-B_F(\kappa)]=\Gamma D(\kappa).
\end{flalign*}
Then the continuous mapping theorem indicates that 
\begin{align}\label{asy_0}
\begin{split}
&(n\|\mathbf{b}\|^2_2)^{-1}D_{n}(1,\bftau,n)^{\top} 	V_{n,\delta}(1,\bftau,n)^{-1}D_{n}(1,\bftau,n)\\\Rightarrow& \kappa^2(1-\kappa)^2(\|\mathbf{b}\|_2^{-1}\mathbf{b})^{\top} V_{\delta}(\kappa)^{-1}(\|\mathbf{b}\|_2^{-1}\mathbf{b})=O_p^s(1).
\end{split}
\end{align}
Here the last equality uses the fact that RHS of (\ref{asy_0}) is greater than 0 with probability 1, or equivalently, $L_{\delta}(\kappa)$ and $R_{\delta}(\kappa)$ is positive definite with probability 1, which will hold  by similar arguments in Lemma \ref{lem_invert} using Cauchy–Schwarz  inequality.

Observe that $\max_{k}T_{n,\delta}(k)\geq D_{n}(1,\bftau,n)^{\top}V_{n,\delta}(1,\bftau,n)^{-1}D_{n}(1,\bftau,n)=O_p^s(n\|\mathbf{b}\|_2^2).$   The result follows by noting $n\|\mathbf{b}\|_2^2\to L$ and $L\to\infty$,
\qed
\vspace{4mm}

\textsc{Proof of Theorem \ref{thm_consis}} 
Note that by (\ref{asy_0}), we have shown that with probability tending to one, 	$(n\|\mathbf{b}\|_2^2)^{-1}T_{n,\delta}(\bftau)\geq \kappa^2(1-\kappa)^2(\|\mathbf{b}\|_2^{-1}\mathbf{b})^{\top} V_{\delta}(\kappa)^{-1}(\|\mathbf{b}\|_2^{-1}\mathbf{b})=O_p^s(1)$.

Then, let  $M_{n,\eta}=\{ k: |\frac{k}{n}-\kappa|>\eta\}$, it suffices to show that 
$$
(n\|\mathbf{b}\|_2^2)^{-1}\max_{k\in[h,n-h]\cap M_{n,\eta}}D_{n}(1,k,n)^{\top}	V_{n,\delta}(1,k,n)^{-1}	D_{n}(1,k,n)=o_p(1).
$$
By symmetricity, we can consider $M_{n,\eta}^{(1)}=\{ k: \frac{k}{n}<\kappa-\eta\}$, and on  $\{k\in M_{n,\eta}^{(1)}\}$, we have 
\begin{flalign*}
D_{n}(1,k,n)=&\frac{k(n-k)}{n^{3/2}}\Big\{[Q_n(1)-Q_n(\frac{k}{n})]^{-1}[Q_n(1)-Q_n(\frac{\bftau}{n})](\bbeta^{(1)}-\bbeta^{(2)})\\&+n^{-1/2}Q_n(\frac{k}{n})^{-1}B_{n,F}(\frac{k}{n})-n^{-1/2}\big[Q_n(1)-Q_n(\frac{k}{n})\big]^{-1}\big[B_{n,F}(1)-B_{n,F}(\frac{k}{n})\big]\Big\}.
\end{flalign*}
Let $\nu=\lim\limits_{n\to\infty}\frac{k}{n}:=\lim\limits_{n\to\infty}\frac{k(n)}{n}$, 
then by similar arguments in  $(\mathrm{i})$ of Theorem \ref{thm_main}, we have
\begin{equation}\label{D1}
n^{-1/2}\|\mathbf{b}\|_2^{-1}D_{n}(1,k,n)=\nu(1-\nu)[Q(1)-Q(\nu)]^{-1}[Q(1)-Q(\kappa)]\|\mathbf{b}\|_2^{-1}\mathbf{b}+O_p(n^{-1/2}\|\mathbf{b}\|_2^{-1}).
\end{equation}

Next, since $k<\bftau-n\eta$, we decompose $R_{n,\delta}(1,k,n)$ by 
\begin{flalign*}
R_{n,\delta}(1,k,n)=&\Big[\sum_{i=k+3+\lfloor n\delta\rfloor}^{\bftau+\lfloor n\delta\rfloor-1}+\sum_{i=\bftau+\lfloor n\delta\rfloor}^{n-1-\lfloor n\delta\rfloor}\Big]\frac{(i-1-k)^2(n-i+1)^2}{n^2(n-k)^2}(\widehat{\bbeta}_{i,n}-\widehat{\bbeta}_{k+1,i-1})^{\otimes2}
\\:=&R_{n,\delta,1}(1,k,n)+R_{n,\delta,2}(1,k,n).
\end{flalign*}
It follows easily that $V_{n,\delta}(1,k,n)^{-1}\leq R_{n,\delta}(1,k,n)^{-1}\leq R_{n,\delta,2}(1,k,n)^{-1}$ where for semi-positive definite matrices $A$ and $B$, $A\leq B$ indicates $B-A$ is semi-positive definite.

In addition, we have  
\begin{flalign*}
R_{n,\delta,2}(1,k,n)=&\sum_{i=\bftau+\lfloor n\delta\rfloor}^{n-1-\lfloor n\delta\rfloor}\frac{(i-1-k)^2(n-i+1)^2}{n^2(n-k)^2}(\widehat{\bbeta}_{i,n}-\widehat{\bbeta}_{k+1,i-1})^{\otimes2} ,
\end{flalign*}
where for  $r\in(\kappa,1)$ uniformly,  we can show  \begin{flalign*}
&\sqrt{n}\Big(\widehat{\bbeta}_{\lfloor nr\rfloor,n}-\widehat{\bbeta}_{\lfloor n\nu\rfloor+1,\lfloor nr\rfloor-1}-[Q(r)-Q(\nu)]^{-1}[Q(\kappa)-Q(\nu)]\mathbf{b}\Big)\\\Rightarrow&\Gamma[Q(1)-Q(r)]^{-1}[B_{F}(1)-B_{F}(r)]-\Gamma\big[Q(r)-Q(\nu)\big]^{-1}\big[B_{F}(r)-B_{F}(\nu)\big]=O_p^s(1).
\end{flalign*}

Therefore, if follows that
\begin{align}\label{barR}
\begin{split}
&(n\|\mathbf{b}\|_2^2)^{-1}R_{n,\delta,2}(1,k,n)\\\Rightarrow & \int_{\kappa+\delta}^{1-\delta}\frac{(r-\nu)^2(1-r)^2}{(1-\nu)^2}\Big\{\|\mathbf{b}\|_2^{-1}[Q(r)-Q(\nu)]^{-1}[Q(\kappa)-Q(\nu)]\mathbf{b}\Big\}^{\otimes2}dr:=\overline{R}_{\delta,2}(\nu).
\end{split}
\end{align}

By Lemma \ref{lem_invert}, when $\nu<\kappa$, $\overline{R}_{\delta,2}(\nu)$ is invertible, hence 
\begin{flalign}\label{asy_1}
\begin{split}
&(n\|\mathbf{b}\|_2^2)^{-1}D_{n}(1,k,n)^{\top} 	V_{n,\delta}(1,k,n)^{-1}D_{n}(1,k,n)\\=&  (n\|\mathbf{b}\|_2^2)^{-1}\big[n^{-1/2}\|\mathbf{b}\|_2^{-1}D_{n}(1,k,n)\big]^{\top}	\big[(n\|\mathbf{b}\|_2^2)^{-1}R_{n,\delta,2}(1,k,n)\big]^{-1}\big[n^{-1/2}\|\mathbf{b}\|_2^{-1}D_{n}(1,k,n)\big] 
\\\Rightarrow& (n\|\mathbf{b}\|_2^2)^{-1} \Big\{\nu(1-\nu)[Q(1)-Q(\nu)]^{-1}[Q(1)-Q(\kappa)]\|\mathbf{b}\|_2^{-1}\mathbf{b}\Big\}^{\top}
\overline{R}_{\delta,2}(\nu)\\&\times \Big\{\nu(1-\nu)[Q(1)-Q(\nu)]^{-1}[Q(1)-Q(\kappa)]\|\mathbf{b}\|_2^{-1}\mathbf{b}\Big\}\Rightarrow 0 \end{split}
\end{flalign}
by (\ref{D1}) and (\ref{barR}).
\qed
\vspace{4mm}

\begin{lemma}\label{lem_invert}
	$\overline{R}_{\delta,2}(\nu)$, defined in (\ref{barR}), is invertible for $\nu<\kappa$  and $\|\mathbf{b}\|_2\neq 0$.
\end{lemma}
\textsc{Proof of Lemma \ref{lem_invert}}

Note that 
\begin{flalign*}
[Q(\kappa)-Q(\nu)]=&(\kappa-\nu)\left(\begin{matrix}
1&\frac{\kappa+\nu}{2}\\
\frac{\kappa+\nu}{2}&\frac{\nu^2+\kappa^2+\kappa\nu}{3}
\end{matrix}\right),\\ [Q(r)-Q(\nu)]^{-1}=&12(r-\nu)^{-3}\left(\begin{matrix}
\frac{r^2+\nu^2+r\nu}{3}&-\frac{r+\nu}{2}\\
-\frac{r+\nu}{2}&1
\end{matrix}\right).
\end{flalign*}
We first let $\mathbf{b}=(b_1,b_2)^{\top}$, then $$
[Q(\kappa)-Q(\nu)]\|\mathbf{b}\|_2^{-1}\mathbf{b}=(\kappa-\nu)\|\mathbf{b}\|_2^{-1}\left(\begin{matrix}
b_1+\frac{\kappa+\nu}{2}b_2\\
\frac{\kappa+\nu}{2}b_1+\frac{\nu^2+\kappa^2+\kappa\nu}{3}b_2
\end{matrix}\right):=(w_1,w_2)'.
$$

Therefore we obtain
\begin{flalign*}
&\frac{(r-\nu)(1-r)}{(1-\nu)}[Q(r)-Q(\nu)]^{-1}[Q(\kappa)-Q(\nu)]\|\mathbf{b}\|_2^{-1}\mathbf{b}=\frac{12(1-r)}{(r-\nu)^2(1-\nu)}\left(\begin{matrix}
\frac{r^2+\nu^2+r\nu}{3}w_1-\frac{r+\nu}{2}w_2\\
-\frac{r+\nu}{2}w_1+w_2
\end{matrix}\right)\\:=&\Big(g_1(r,\nu,\kappa,b_1,b_2),g_2(r,\nu,\kappa,b_1,b_2)\Big)^{\top}.
\end{flalign*}
Then,  since 
$$
\overline{R}_{\delta,2}(\nu)=\int_{\kappa+\delta}^{1-\delta} \Big((g_1(r,\nu,\kappa,b_1,b_2),g_2(r,\nu,\kappa,b_1,b_2)\Big)^{\top}\Big((g_1(r,\nu,\kappa,b_1,b_2),g_2(r,\nu,\kappa,b_1,b_2)\Big)dr,
$$
the invertibility of $\overline{R}_{\delta,2}(\nu)$ is equivalent to that $\mathrm{det}(\overline{R}_{\delta,2}(\nu))>0$ (as $\overline{R}_{\delta,2}(\nu)$ is clearly semi-positive definite), i.e.
$$
\int_{\kappa+\delta}^{1-\delta}g_1(r,\nu,\kappa,b_1,b_2)^2dr\int_{\kappa+\delta}^{1-\delta}g_2(r,\nu,\kappa,b_1,b_2)^2dr-[\int_{\kappa+\delta}^{1-\delta}g_1(r,\nu,\kappa,b_1,b_2)g_2(r,\nu,\kappa,b_1,b_2)dr]^2>0,
$$
which is implied by Cauchy–Schwarz inequality as long as 
\begin{flalign}\label{cauchy}
\frac{g_1(r,\nu,\kappa,b_1,b_2)}{g_2(r,\nu,\kappa,b_1,b_2)}=\frac{2(r^2+\nu^2+r\nu)w_1-3(r+\nu)w_2}{-6(r+\nu)w_1+12w_2}
\end{flalign}
is not a constant for all $r\geq \kappa$.

To see this, suppose $\overline{R}_{\delta,2}(\nu)$ is not invertible, then (\ref{cauchy}) is a constant for all $r\geq \kappa$. Note that the numerator and the denominator of RHS of (\ref{cauchy}) can be written in a quadratic form of $r$ as
\begin{flalign}
\label{num}&2w_1r^2+(2\nu w_1-3w_2)r+(2\nu^2w_1-3\nu w_2),\\
\label{den}&0r^2-6w_1r+(-6\nu w_1+12w_2),
\end{flalign}
respectively.

Therefore, comparing  coefficients of the quadratic functions (\ref{num}) and (\ref{den}) w.r.t $r$, it must hold that $w_1=0$, and hence $w_2=0$, i.e.
$$
b_1+\frac{\kappa+\nu}{2}b_2=0,\quad\text{and}\quad \frac{\kappa+\nu}{2}b_1+\frac{\nu^2+\kappa^2+\kappa\nu}{3}b_2=0.
$$
Solving these equations for $b_1$ and $b_2$ we obtain that $b_1=b_2=0$, contradiction.

Hence, $\overline{R}_{\delta,2}(\nu)$ is  invertible.
\qed
\vspace{4mm}

%\section{supLM test}
%Let $\widehat{\beta}_{1,n}$ be the OLS estimator for the full sample $\{Y_t\}_{t=1}^n$,  $\widehat{\Gamma}^2$ be a consistent estimator (for example, HAC-type estimator) for the long-run variance $\Gamma^2$ and $\widehat{M}=\frac{1}{n}\sum_{t=1}^{n}F(t/n)F(t/n)^{\top}$.
%In addition, we let $W_n(\eta,\beta)=n^{-1/2}\sum_{t=1}^{\lfloor\eta n\rfloor}[Y_t-\beta F(t/n)]$ be the estimation function using the subsample before time $\lfloor\eta n\rfloor$, then the supLM test is defined by 
%$$
%\sup_{\eta\in(\epsilon,1-\epsilon)}LM(t)=\sup_{\eta\in(\epsilon,1-\epsilon)}[\eta(1-\eta)]^{-1}W_n(\eta,\widehat{\beta}_{1,n})\widehat{\Gamma}^{-2}\widehat{M}\big(\widehat{M}\widehat{\Gamma}^{-2}\widehat{M}\big)^{-1}\widehat{M}\widehat{\Gamma}^{-2}W_n(\eta,\widehat{\beta}_{1,n}).
%$$
%Under the null, $$\sup_{\eta\in(\epsilon,1-\epsilon)}LM(t)\Rightarrow\sup_{\eta\in(\epsilon,1-\epsilon)}[\eta(1-\eta)]^{-1}\|B^0(\eta)\|_2^2$$
%where $B^0(\eta)=B(\eta)-\eta B(1)$ is a two dimensional Brownian bridge.
%
%In the simulation part, we generate the asymptotic critical value under $H_0$ using $n=5000$ with $\beta=(0,0)^{\top}$ by 10000 replications.  The HAC-type estimator for $\Gamma^2$ is based on quadratic spectral kernel defined by 
%$$
%k(x)=\frac{25}{12\pi^2x^2}(\frac{sin(6\pi x/5)}{6\pi x/5}-\cos(6\pi x/5))
%$$
%with the optimal bandwidth defined in \cite{andrew1991}.

\section{Piecewise polynomial trend model}
In this section, we extend the piecewise linear structure in model (1.1) of the main text to a piecewise polynomial structure. We further apply a piecewise quadratic trend model to analyze the cumulative confirmed cases in 8 representative countries as in Section 4.2.
\subsection{Model formulation and inference}~
We extend the piecewise linear trend model (1.1) by allowing higher order polynomial terms. Specifically, let the time series $\{Y_t\}_{t=1}^n$ admit 
\begin{align}\label{modelp}
&Y_{t}=\bbeta_t^\top F^{(p)}_t+u_t=\beta_{0,t}+\beta_{1,t}(t/n)+\cdots+\beta_{p,t}(t/n)^p+u_t,~ t=1,\cdots,n,\\
&(\beta_{0,t},\cdots,\beta_{p,t})^{\top}=\bbeta^{(i)}=(\beta_{0}^{(i)},\cdots,\beta_{p}^{(i)})^{\top}, \tau_{i-1}+1\leq t\leq \tau_i,\text{ for }i=1,\cdots,m+1,  \nonumber
\end{align}
where $F^{(p)}_{t}=(1,t/n,\cdots, (t/n)^p)^\top$ and $\bbeta_t=(\beta_{0,t},\cdots,\beta_{p,t})^{\top}$ are the coefficients at time $t$ with fixed $p\geq1$. Same as in model (1.1), $\{u_t\}$ is a weakly dependent stationary error process, $\bftau=(\tau_1,\cdots,\tau_m)$ denotes the $m\geq 0$ change-points with the convention that $\tau_0=0$ and $\tau_{m+1}=n$, and we require $\bbeta^{(i)}\neq \bbeta^{(i+1)}, i=1,\cdots,m$. Model \eqref{modelp} extends the piecewise linear model by allowing for polynomial trends and provides more flexibility of modeling observations in each segment.

The estimation procedure of model \eqref{modelp} is essentially the same as the one for model (1.1). Given the grid parameter $\epsilon,$ we let $h=\lfloor \epsilon n\rfloor$.  Define  $F^{(p)}(s)=(1,s,\cdots,s^p)^{\top}$. For $1\leq i<j\leq n$, we denote $\widehat{\bbeta}_{i,j}=\Big[\sum_{t=i}^{j}F^{(p)}(t/n)F^{(p)}(t/n)^{\top}\Big]^{-1}\sum_{t=i}^{j}F^{(p)}(t/n)Y_t$ as the OLS estimator of $\bbeta$ based on $\{Y_{t}\}_{t=i}^j$. Let the trimming parameter satisfy $0\leq \delta<\epsilon/2$. For any $1\leq t_1 <k <t_2\leq n$,  given the subsample $\{Y_t\}_{t=t_1}^{t_2}$ and a potential change-point $k$, we define a contrast statistic $D_{n}^{(p)}(t_1,k,t_2)$, and the self-normalizer $V^{(p)}_{n,\delta}(t_1,k,t_2)=L^{(p)}_{n,\delta}(t_1,k,t_2)+R^{(p)}_{n,\delta}(t_1,k,t_2)$ in the same spirit as (\ref{D}), (\ref{L}) and (\ref{R}) by:
\begin{align*}
D^{(p)}_{n}(t_1,k,t_2)=&\frac{(k-t_1+1)(t_2-k)}{(t_2-t_1+1)^{3/2}}(\widehat{\bbeta}_{t_1,k}-\widehat{\bbeta}_{k+1,t_2}),\\
L^{(p)}_{n,\delta}(t_1,k,t_2)=&\sum_{i=t_1+p+\lfloor n\delta\rfloor}^{k-p-1-\lfloor n\delta\rfloor}\frac{(i-t_1+1)^2(k-i)^2}{(k-t_1+1)^2(t_2-t_1+1)^2}(\widehat{\bbeta}_{t_1,i}-\widehat{\bbeta}_{i+1,k})^{\otimes 2},\\
R^{(p)}_{n,\delta}(t_1,k,t_2)=&\sum_{i=k+2+p+\lfloor n\delta\rfloor}^{t_2-p-\lfloor n\delta\rfloor}\frac{(i-1-k)^2(t_2-i+1)^2}{(t_2-t_1+1)^2(t_2-k)^2}(\widehat{\bbeta}_{i,t_2}-\widehat{\bbeta}_{k+1,i-1})^{\otimes 2}.
\end{align*}

Then, the test statistic targeting against the one change-point alternative is defined as:
\begin{equation*}
G_n^{(p)}=\max_{k\in\{h,\cdots,n-h\}}T^{(p)}_{n,\delta}(k),\quad T^{(p)}_{n,\delta}(k)=D^{(p)}_{n}(1,k,n)^{\top}V^{(p)}_{n,\delta}(1,k,n)^{-1}D^{(p)}_{n}(1,k,n).
\end{equation*}

Define $Q^{(p)}(r)=\int_{0}^{r}F^{(p)}(s)F^{(p)}(s)^{\top}ds$ and $B^{(p)}_{F}(r)=\int_{0}^{r}F^{(p)}(s)dB(s)$ where $B(\cdot)$ is a standard Brownian motion. The following theorem extends  Theorem \ref{thm_main} in the main text.
\begin{thm}
	Suppose Assumption \ref{ass} holds. Then, 
	
	(i) under $H_0$, we have
	\begin{equation*}
	G_n^{(p)}\overset{\mathcal D}{\longrightarrow} G^{(p)}(\epsilon, \delta):= \sup_{\eta\in(\epsilon,1-\epsilon)}D^{(p)}(\eta)^{\top}V^{(p)}_{\delta}(\eta)D^{(p)}(\eta), 
	\end{equation*}
	where 
	$D^{(p)}(\eta)$ and $V^{(p)}_{\delta}(\eta)=L^{(p)}_{\delta}(\eta)+R^{(p)}_{\delta}(\eta)$ have the similar expression as given in Theorem \ref{thm_main}, except $F(\cdot)$ and $Q(\cdot)$ are replaced by $F^{(p)}(\cdot)$ and $Q^{(p)}(\cdot)$ respectively.
	
	%\begin{flalign*}
	%D(\eta)=&\eta(1-\eta)\Big\{Q(\eta)^{-1}B_F(\eta)-[Q(1)-Q(\eta)]^{-1}[B_F(1)-B_F(\eta)]\Big\},\\
	%L_{\delta}(\eta)=&\int_{\delta}^{\eta-\delta}\frac{r^2(\eta-r)^2}{\eta^2}\Big\{Q(r)^{-1}B_F(r)-[Q(\eta)-Q(r)]^{-1}[B_F(\eta)-B_F(r)]\Big\}^{\otimes2}dr,\\
	%R_{\delta}(\eta)=&\int_{\eta+\delta}^{1-\delta}\frac{(r-\eta)^2(1-r)^2}{(1-\eta)^2}\\&\times\Big\{[Q(1)-Q(r)]^{-1}[B_F(1)-B_F(r)]-[Q(r)-Q(\eta)]^{-1}[B_F(r)-B_F(\eta)]\Big\}^{\otimes2}dr,\\
	%V_{\delta}(\eta)=&L_{\delta}(\eta)+R_{\delta}(\eta),
	%\end{flalign*}
	
	(ii) under $H_a$, given that $n\|\mathbf{b}\|^2_2\to L$, we have
	$$\lim\limits_{L\to\infty}\lim\limits_{n\to\infty}{G}^{(p)}_n=\infty,\quad\mbox{ in probability}.$$
\end{thm}

The proof is a simple extension of Appendix \ref{proof}, hence omitted.

\subsection{Analysis of cumulative confirmed cases in 8 representative countries}
\begin{figure}[H]
	\centering
	\subfigure{	\hspace{-6mm}	
		\includegraphics[width=0.52\linewidth]{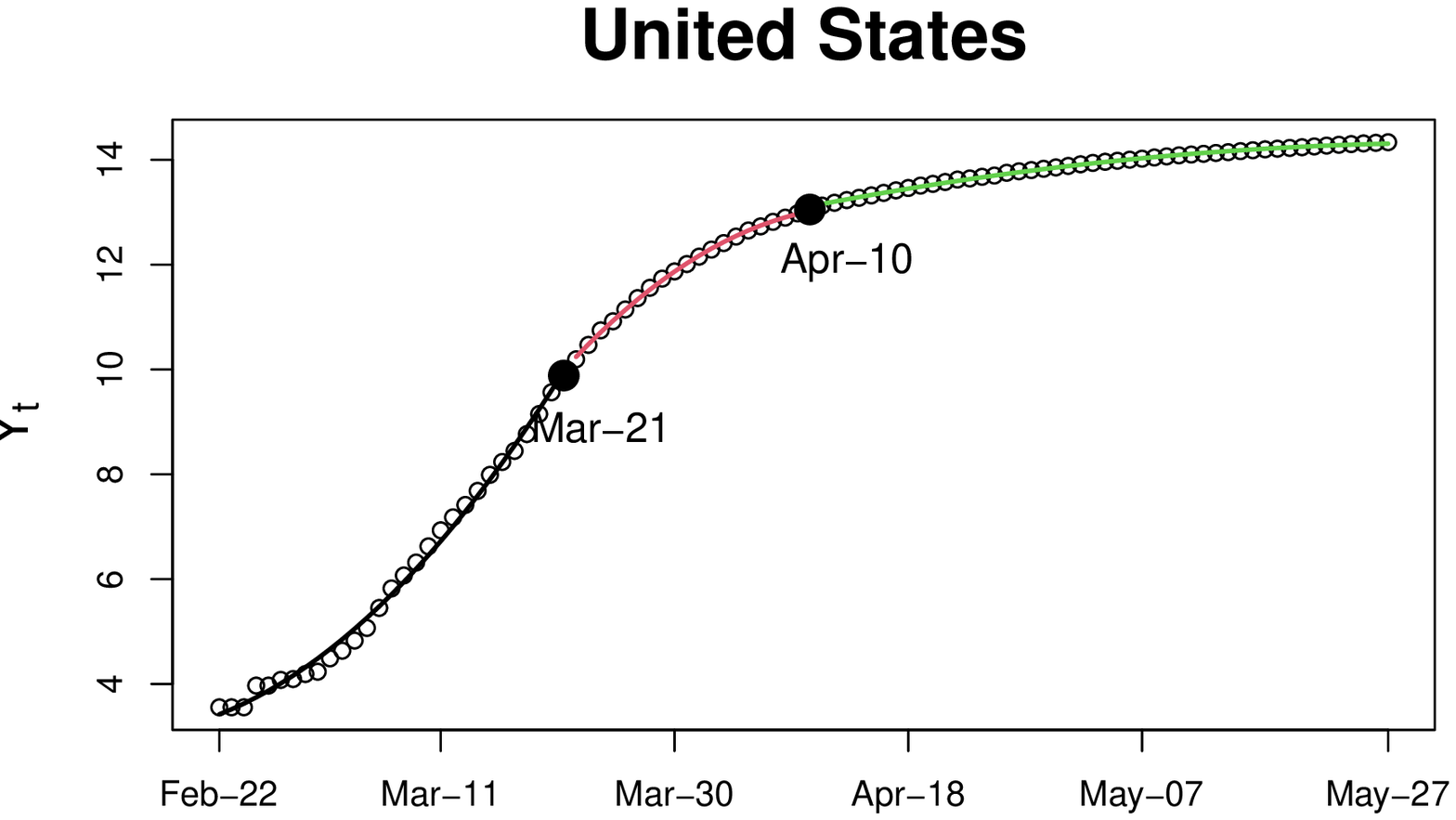}}
	\hfil
	\subfigure{	\hspace{-7mm}
		\includegraphics[width=0.52\linewidth]{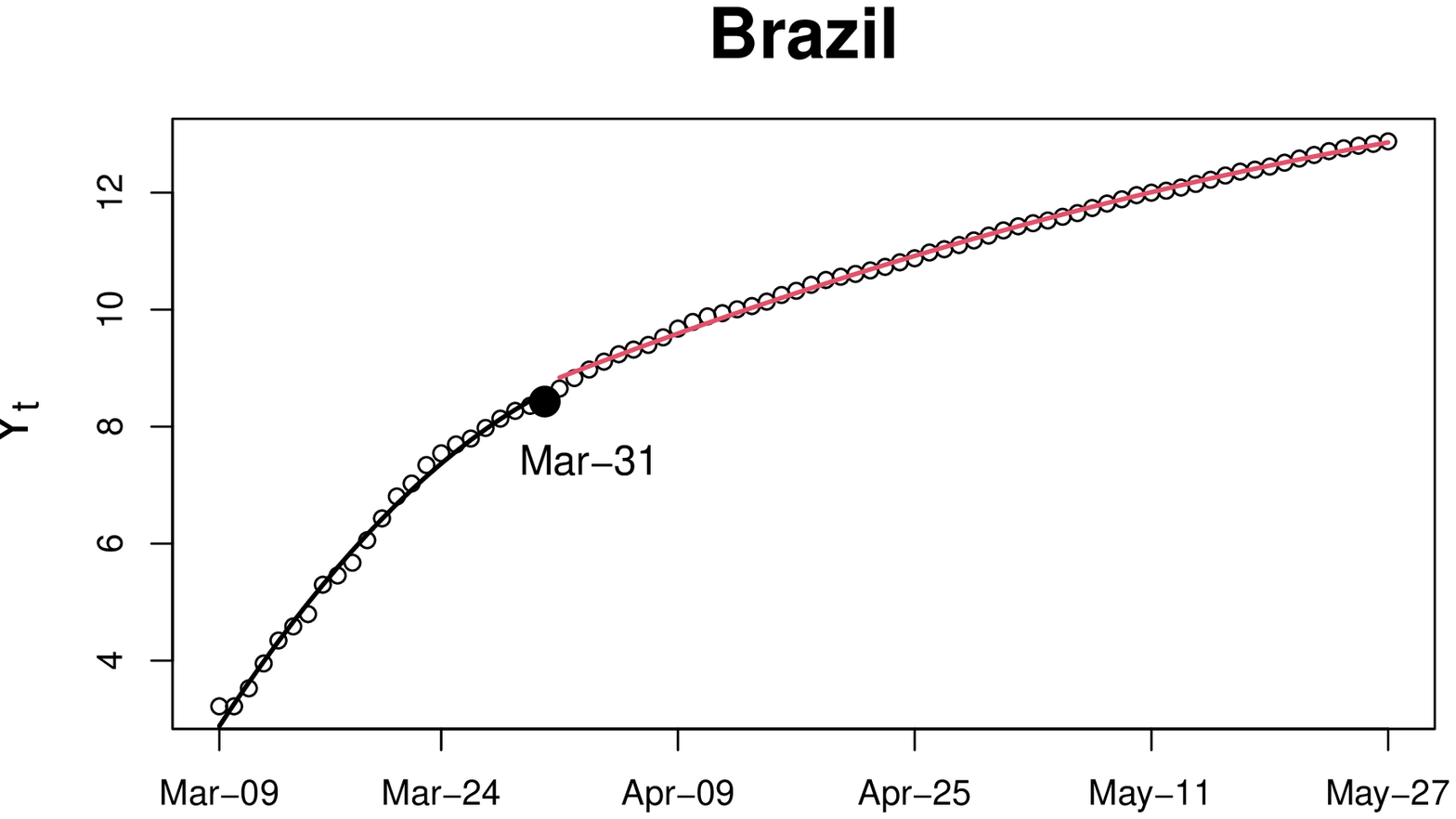}}
	\hfil\vspace{-8mm}
	\subfigure{\hspace{-5mm}
		\includegraphics[width=0.52\linewidth]{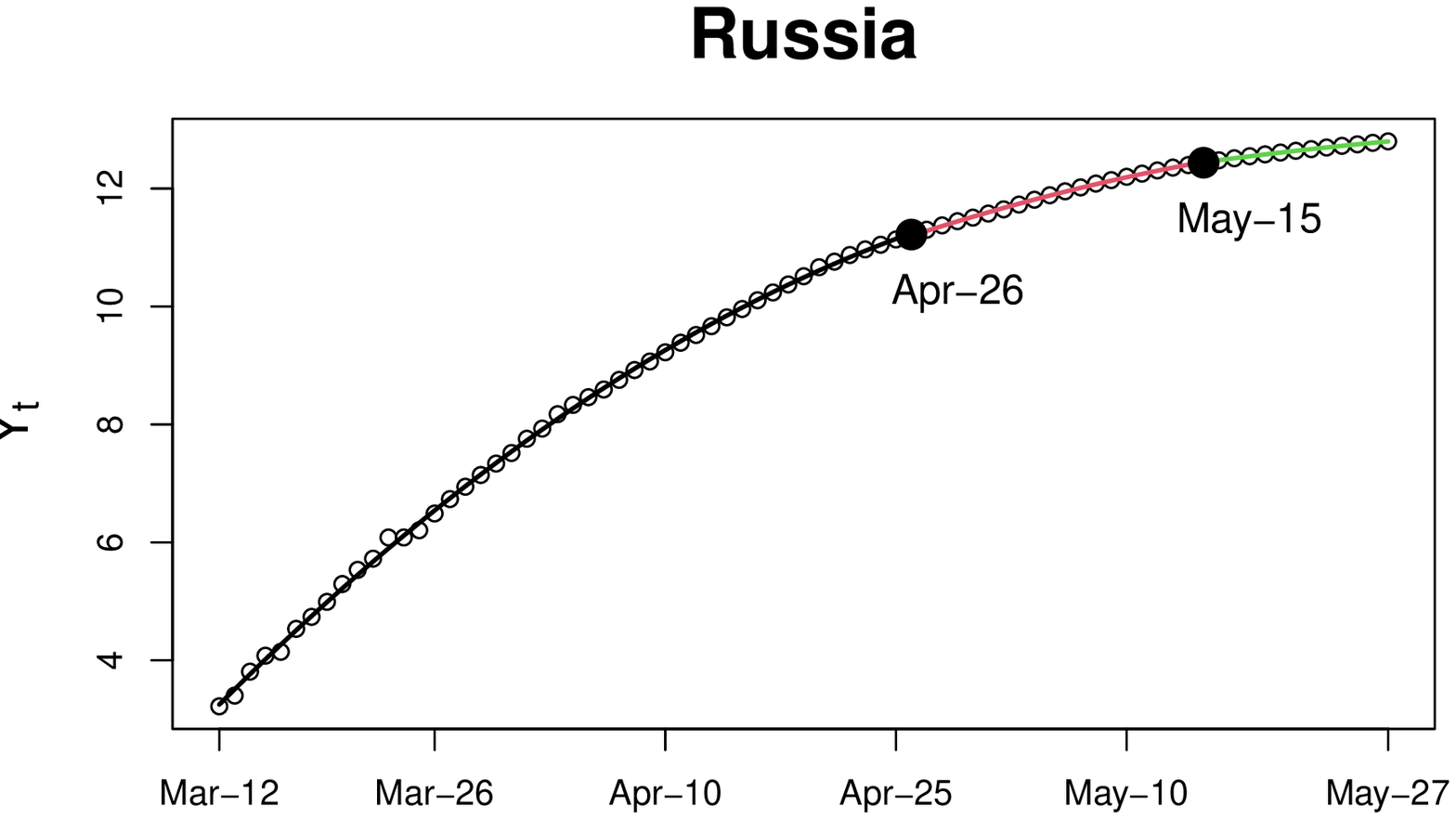}}
	\hfil
	\subfigure{\hspace{-6mm}
		\includegraphics[width=0.52\linewidth]{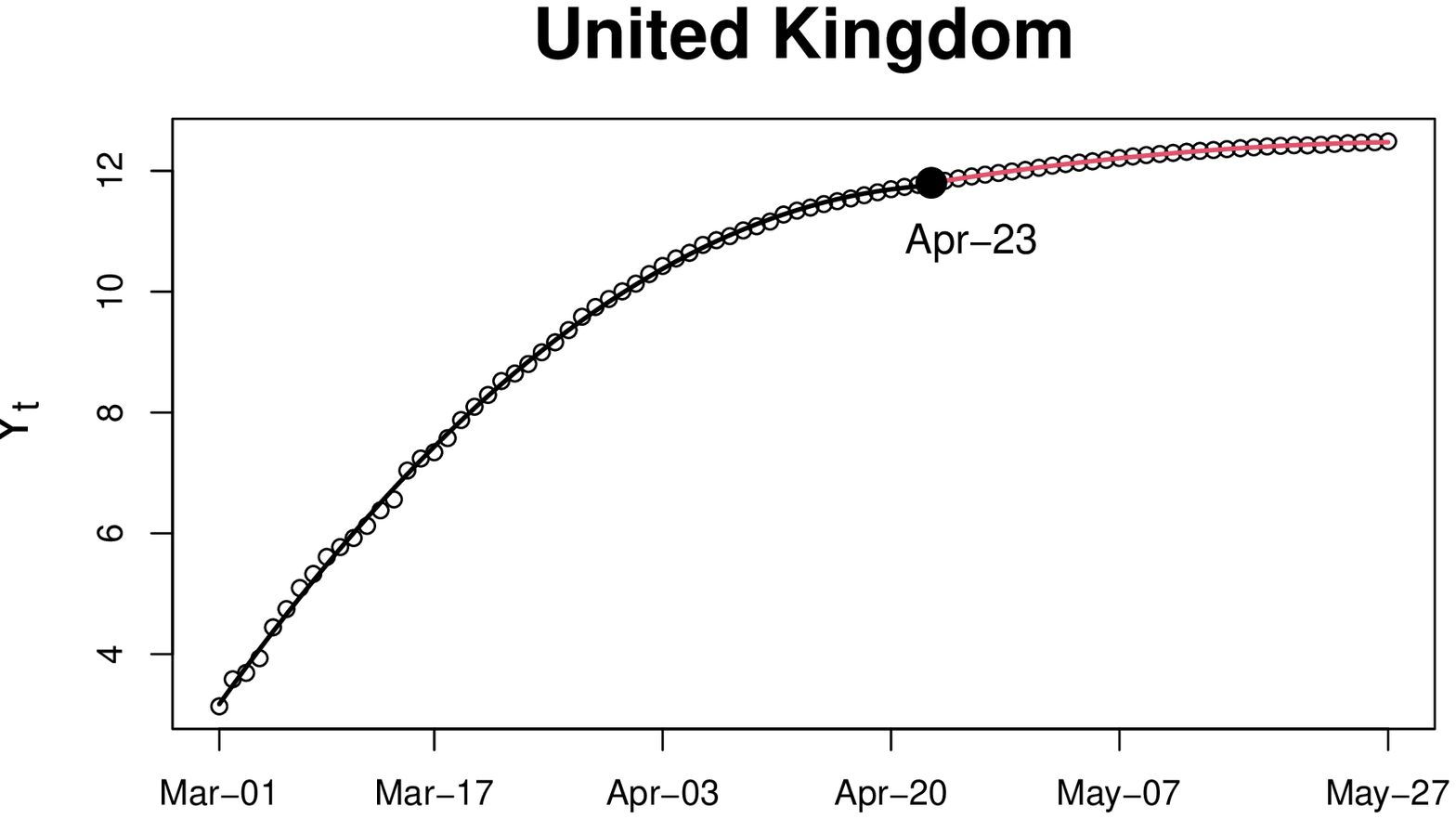}}	
	\hfil\vspace{-8mm}
	\subfigure{\hspace{-5mm}
		\includegraphics[width=0.52\linewidth]{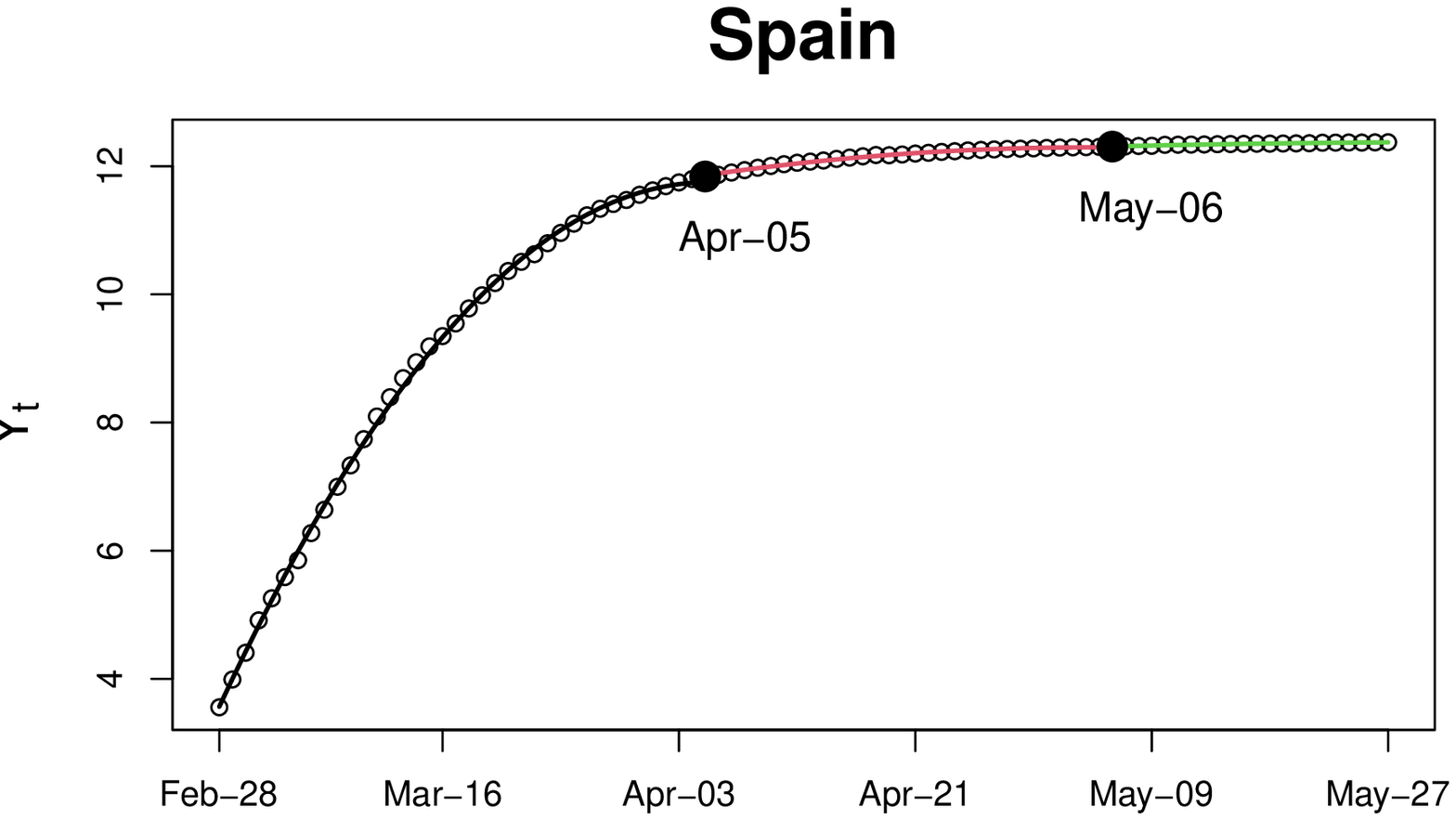}}
	\hfil
	\subfigure{\hspace{-6mm}
		\includegraphics[width=0.52\linewidth]{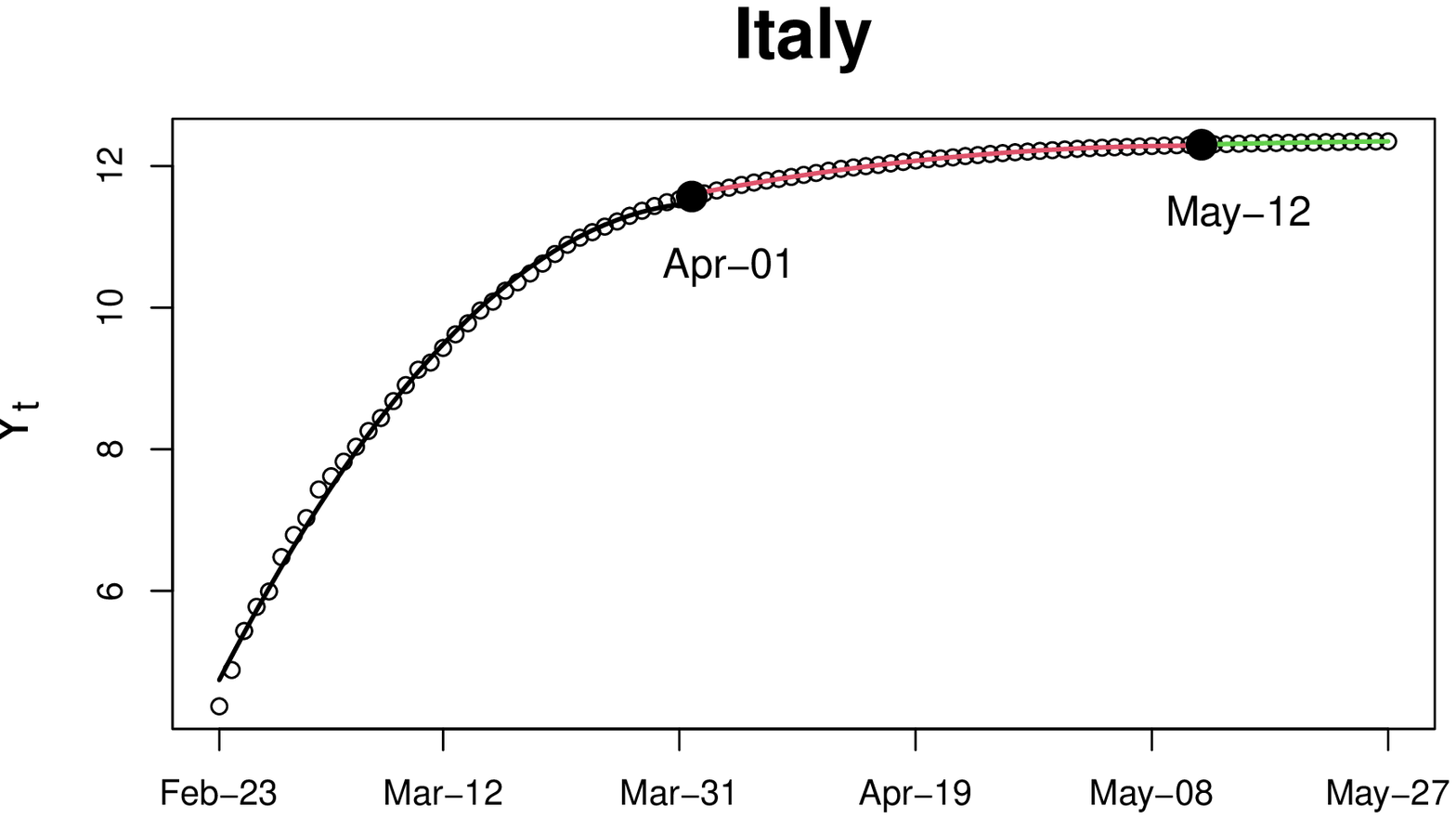}}
	\hfil\vspace{-8mm}
	
	\subfigure{\hspace{-5mm}
		\includegraphics[width=0.52\linewidth]{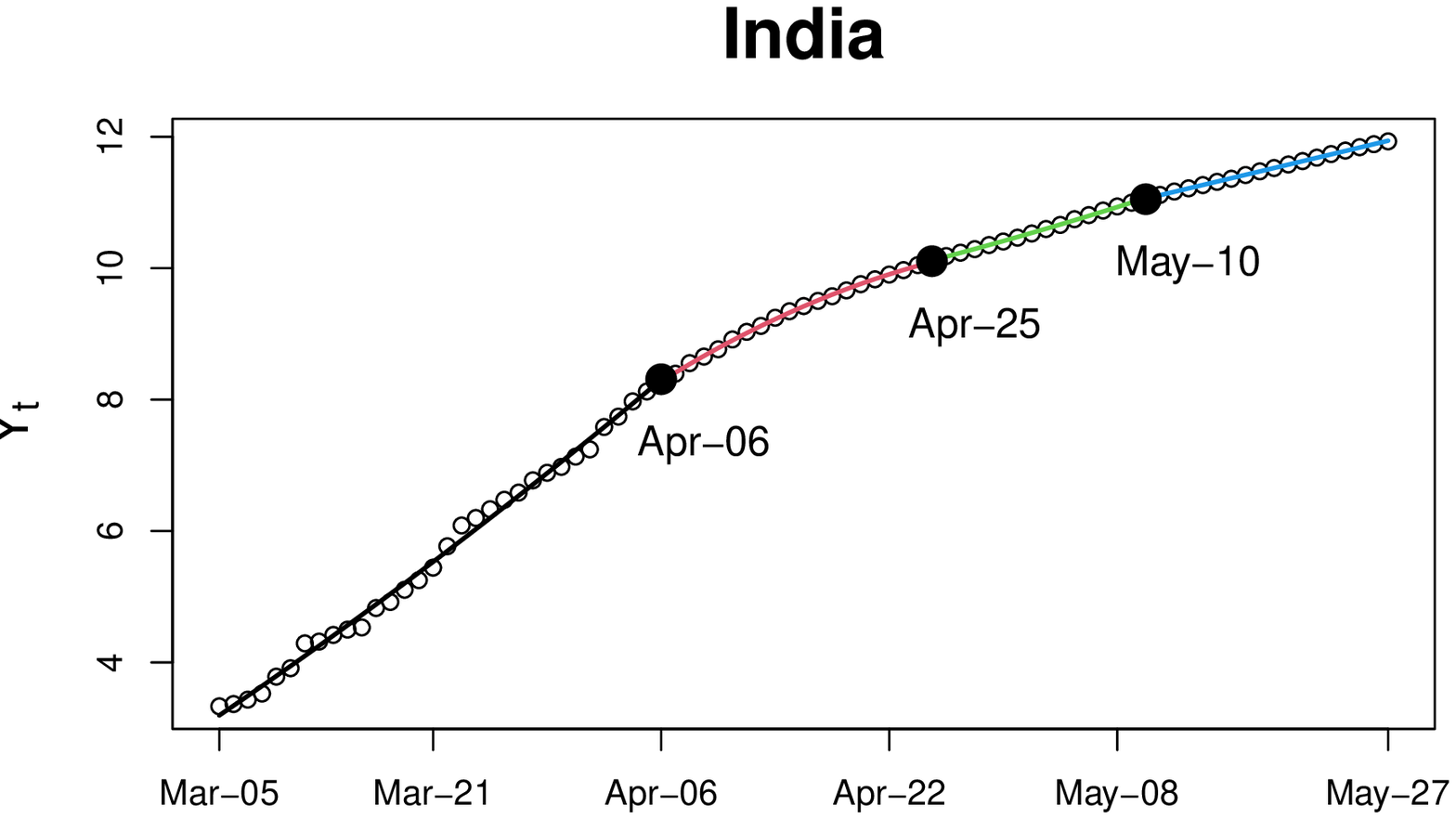}}
	\hfil
	\subfigure{\hspace{-6mm}
		\includegraphics[width=0.52\linewidth]{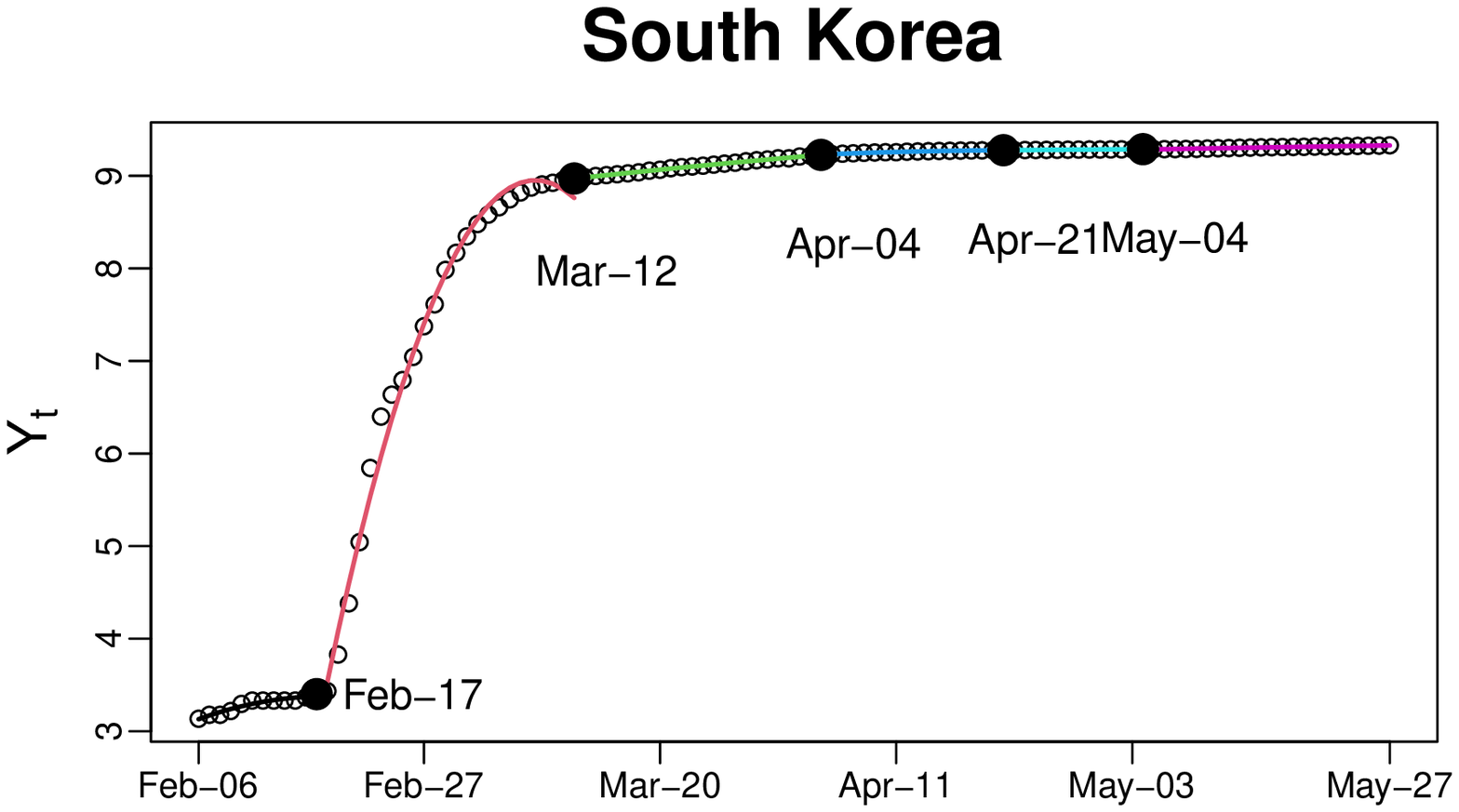}}
	\hfil\vspace{-8mm}
	\caption{Estimated piecewise quadratic trend for 8 representative countries }
	\label{fig_qtrend}
\end{figure}

We use the piecewise quadratic trend model, i.e. model (\ref{modelp}) with $p=2$ to re-analyze the cumulative confirmed cases in the 8 countries as in Section 4.2. Figure \ref{fig_qtrend} gives the estimated models for each country.  As can be seen, compared to Figure 4.1 in the main text, the estimated number of change-points decreases for every country, which is intuitive as more flexibility is brought into the model. For most countries, a piecewise quadratic model with one or two change-points fits the data reasonably well.

However, compared to the piecewise linear trend model, the quadratic model losses its interpretability as the parameters of each segment cannot be naturally linked to growth rate. Thus the meaning of ``change-point" needs a more delicate definition. Moreover, within each segment, the growth rate of the virus still changes from day to day, making it difficult to interpret the behavior of the estimated segments. For example, we find that most estimated change-points can hardly be associated with the initiations of emergency public health measures, as the intervention effect may have been absorbed into the quadratic function. Therefore, we prefer the piecewise linear trend model for the analysis.

%Therefore, we view the piecewise quadratic model more as a possible mathematical extension of the piecewise 

%than giving practical implications.

\section{ACF and PACF plots of residuals (after fitting piecewise linear trend model) for cumulative confirmed cases in 8 countries}
\begin{figure}[H]
	\centering
	\subfigure{	\hspace{-6mm}	
		\includegraphics[width=0.52\linewidth]{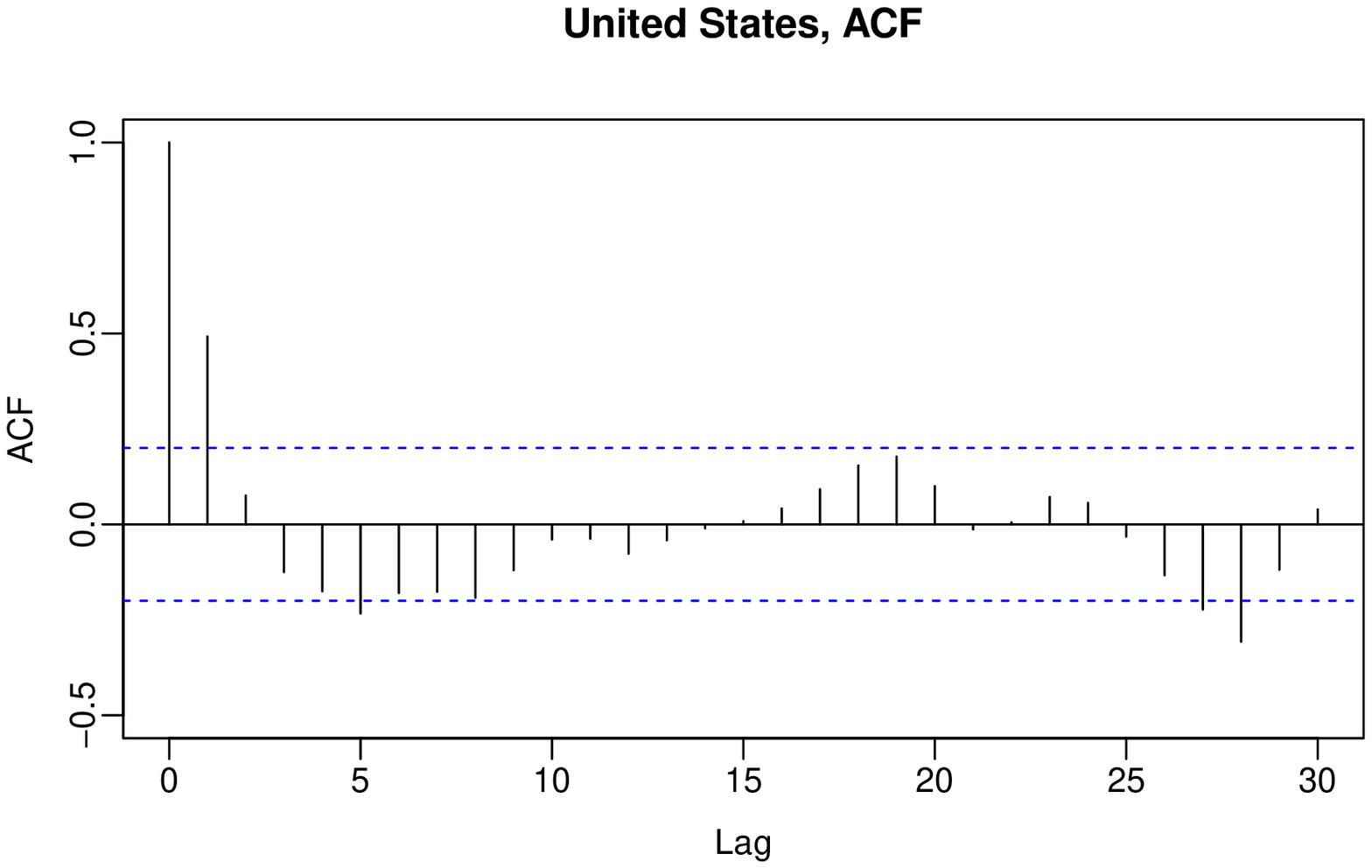}}
	\hfil
	\subfigure{	\hspace{-7mm}
		\includegraphics[width=0.52\linewidth]{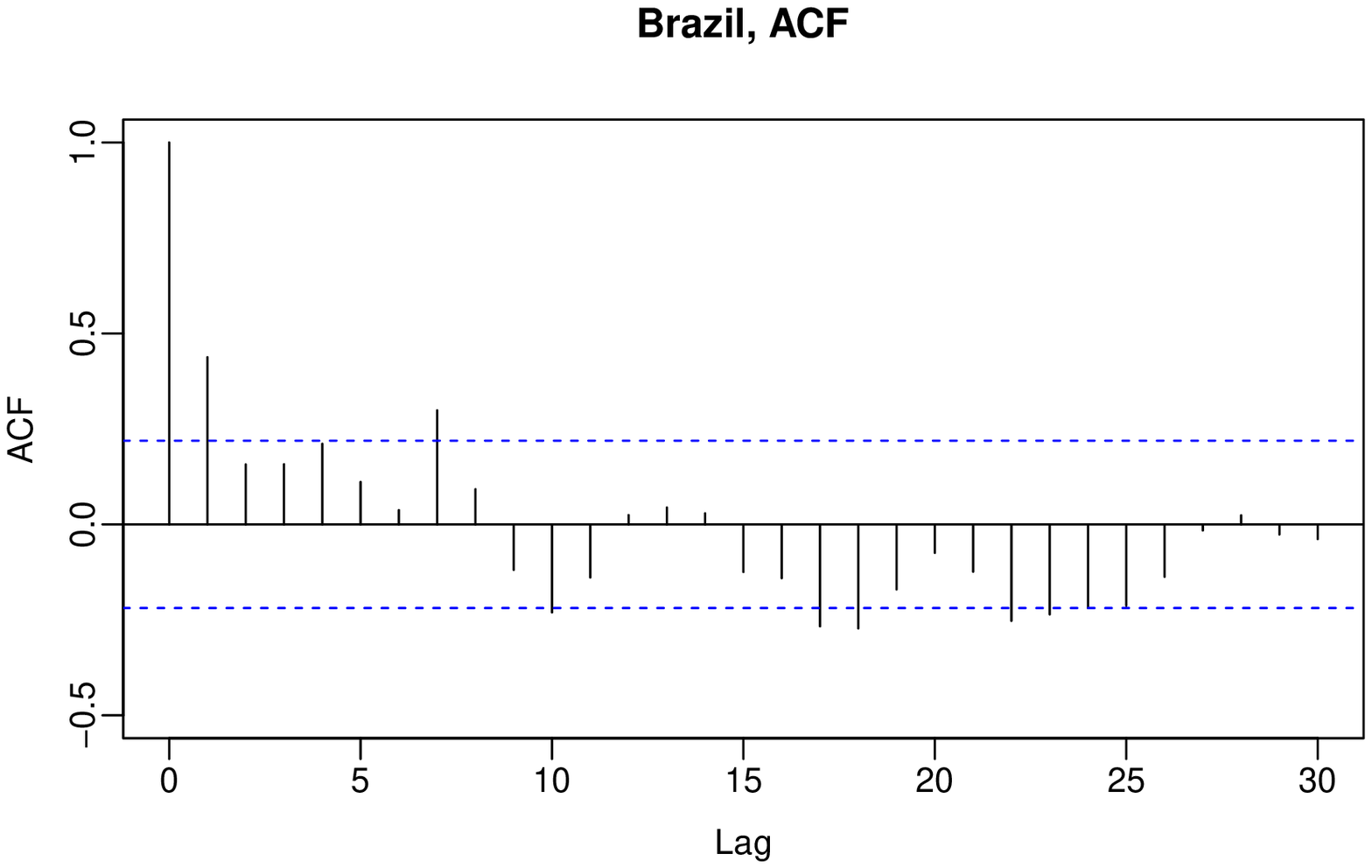}}
	\hfil\vspace{-8mm}
	\subfigure{\hspace{-5mm}
		\includegraphics[width=0.52\linewidth]{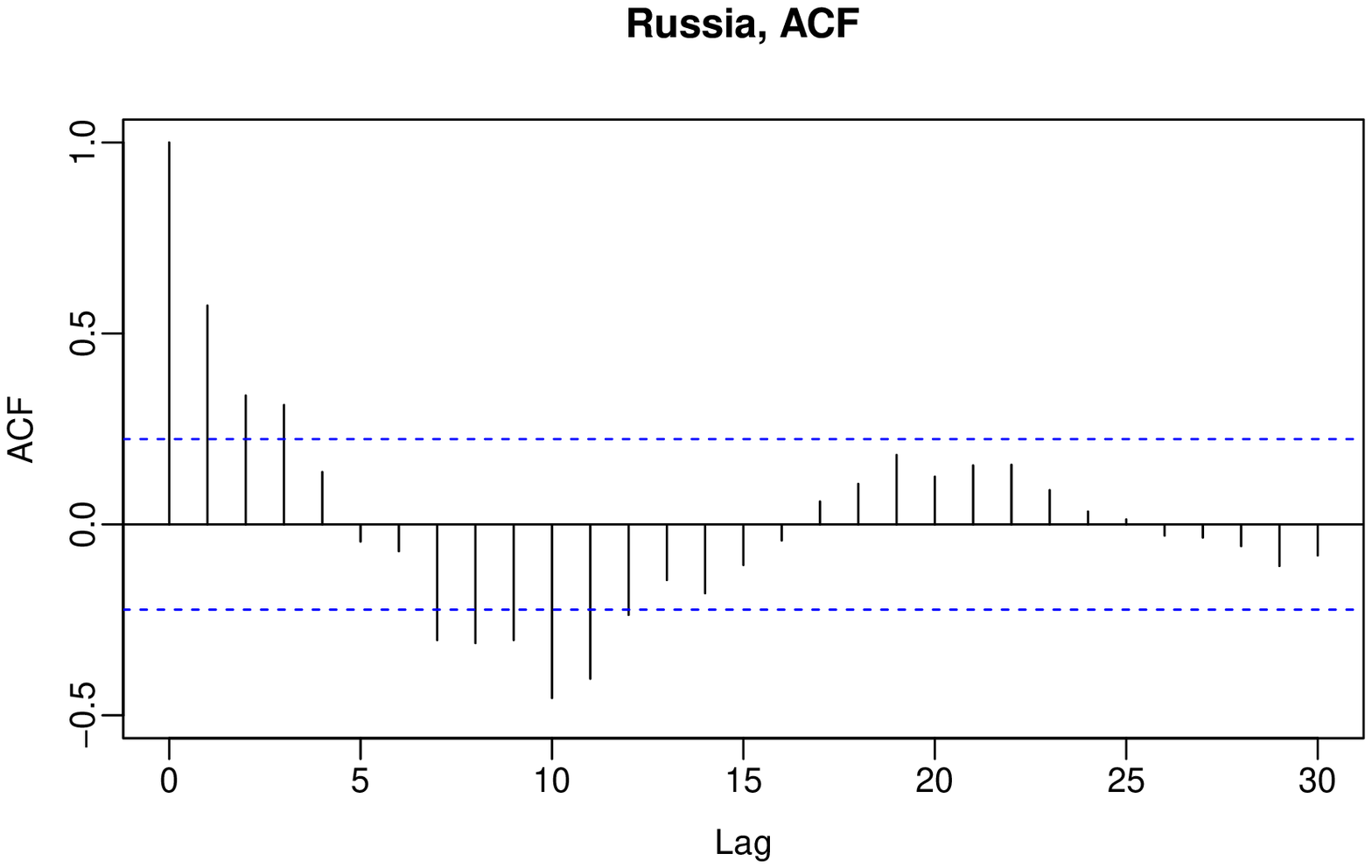}}
	\hfil
	\subfigure{\hspace{-6mm}
		\includegraphics[width=0.52\linewidth]{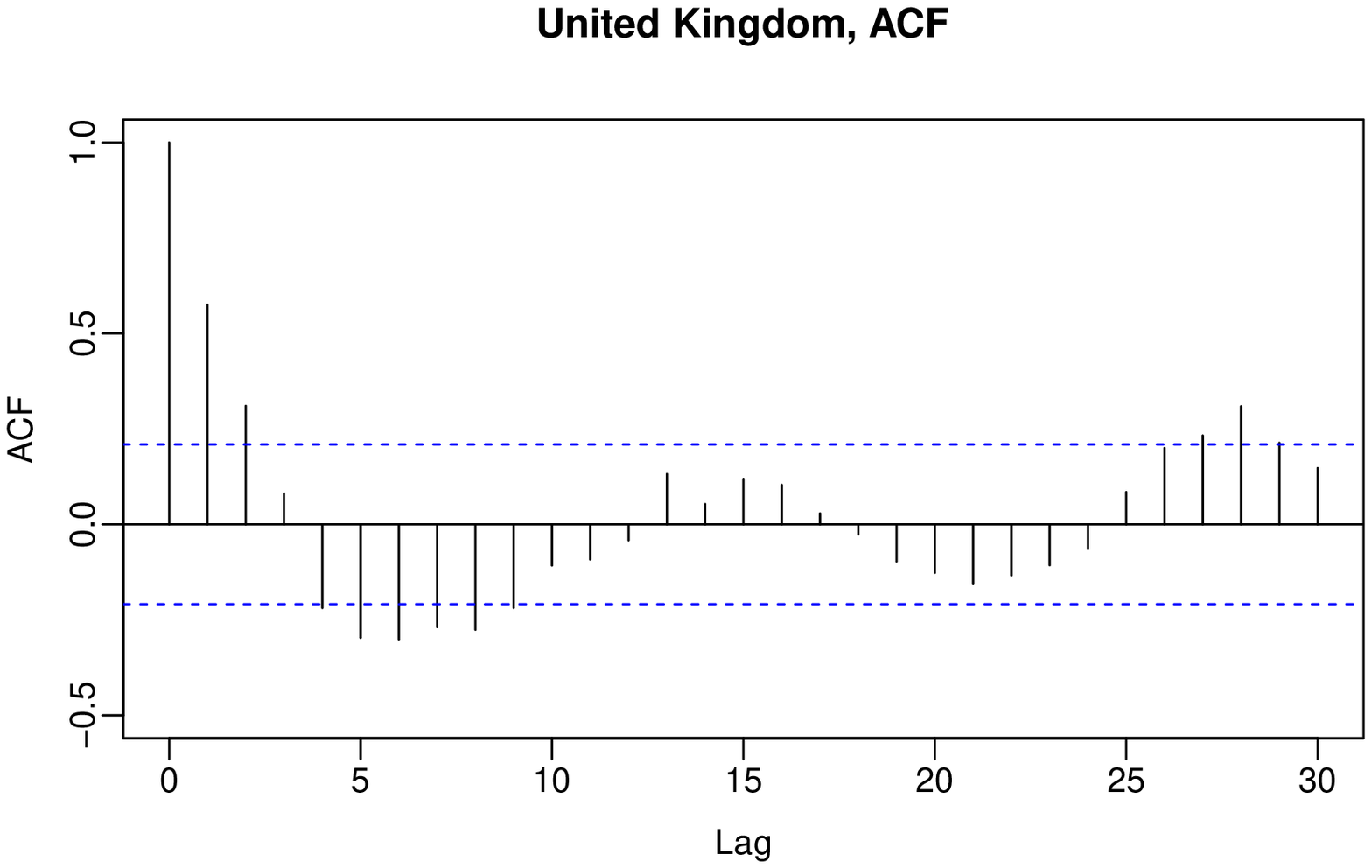}}	
	\hfil\vspace{-8mm}
	\subfigure{\hspace{-5mm}
		\includegraphics[width=0.52\linewidth]{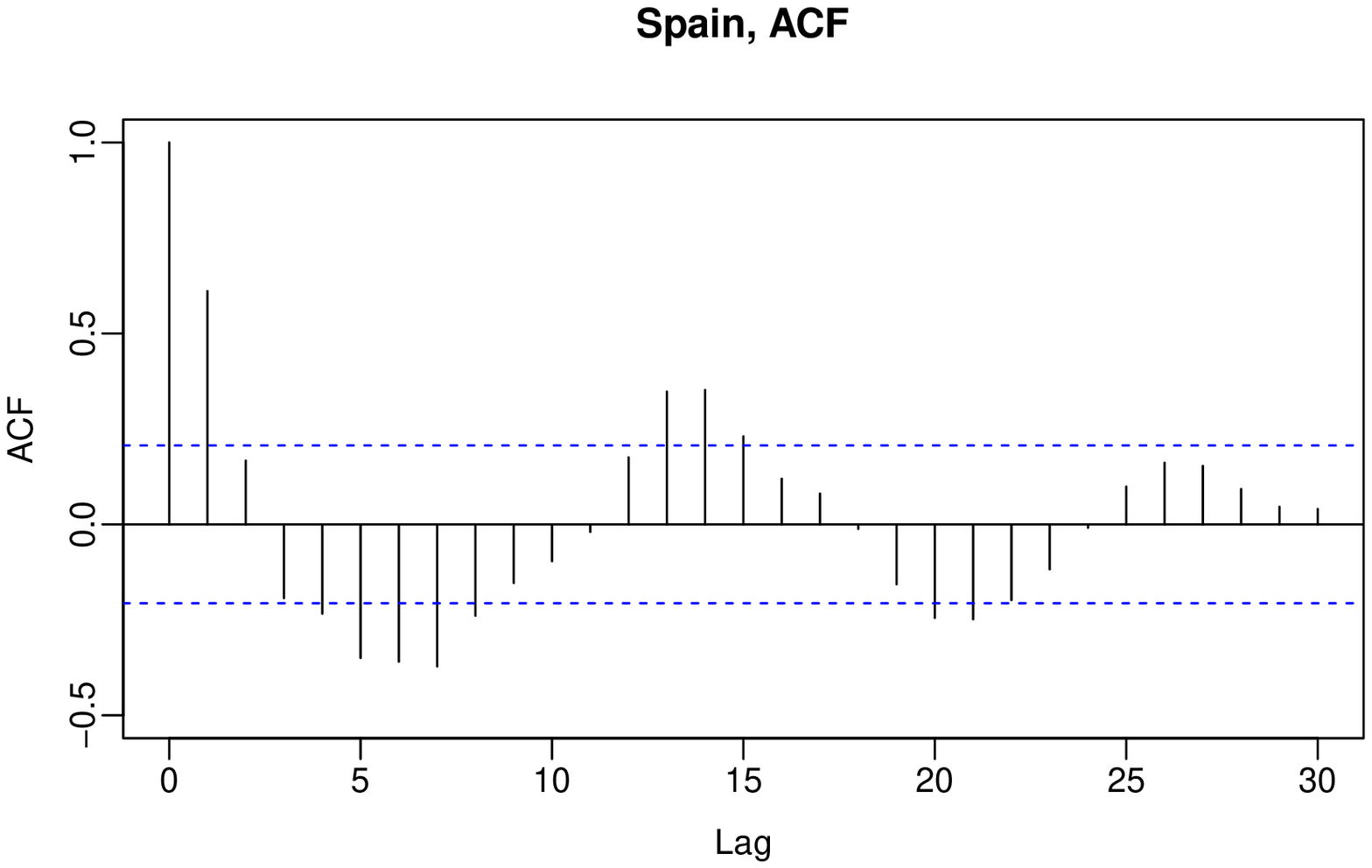}}
	\hfil
	\subfigure{\hspace{-6mm}
		\includegraphics[width=0.52\linewidth]{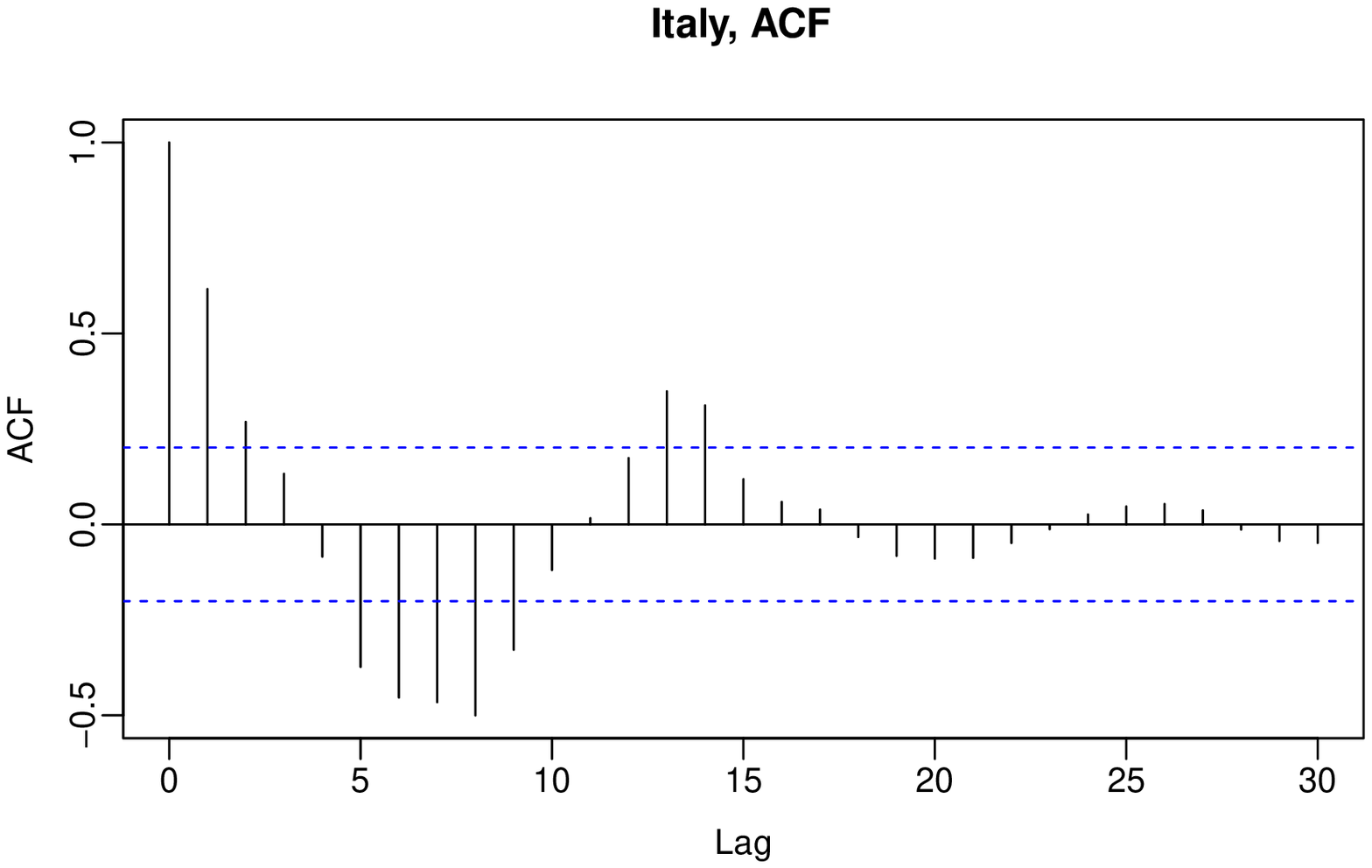}}
	\hfil\vspace{-8mm}
	
	\subfigure{\hspace{-5mm}
		\includegraphics[width=0.52\linewidth]{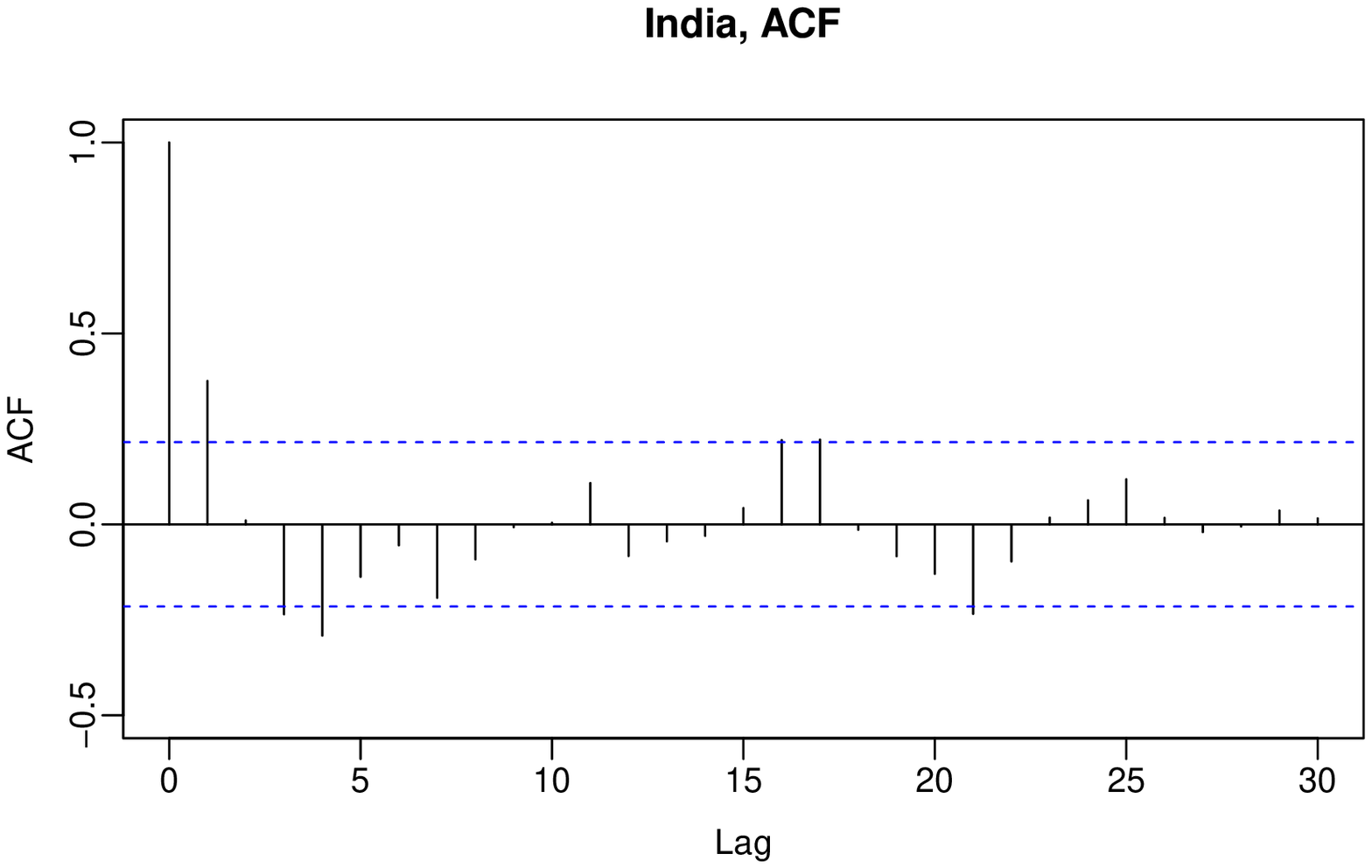}}
	\hfil
	\subfigure{\hspace{-6mm}
		\includegraphics[width=0.52\linewidth]{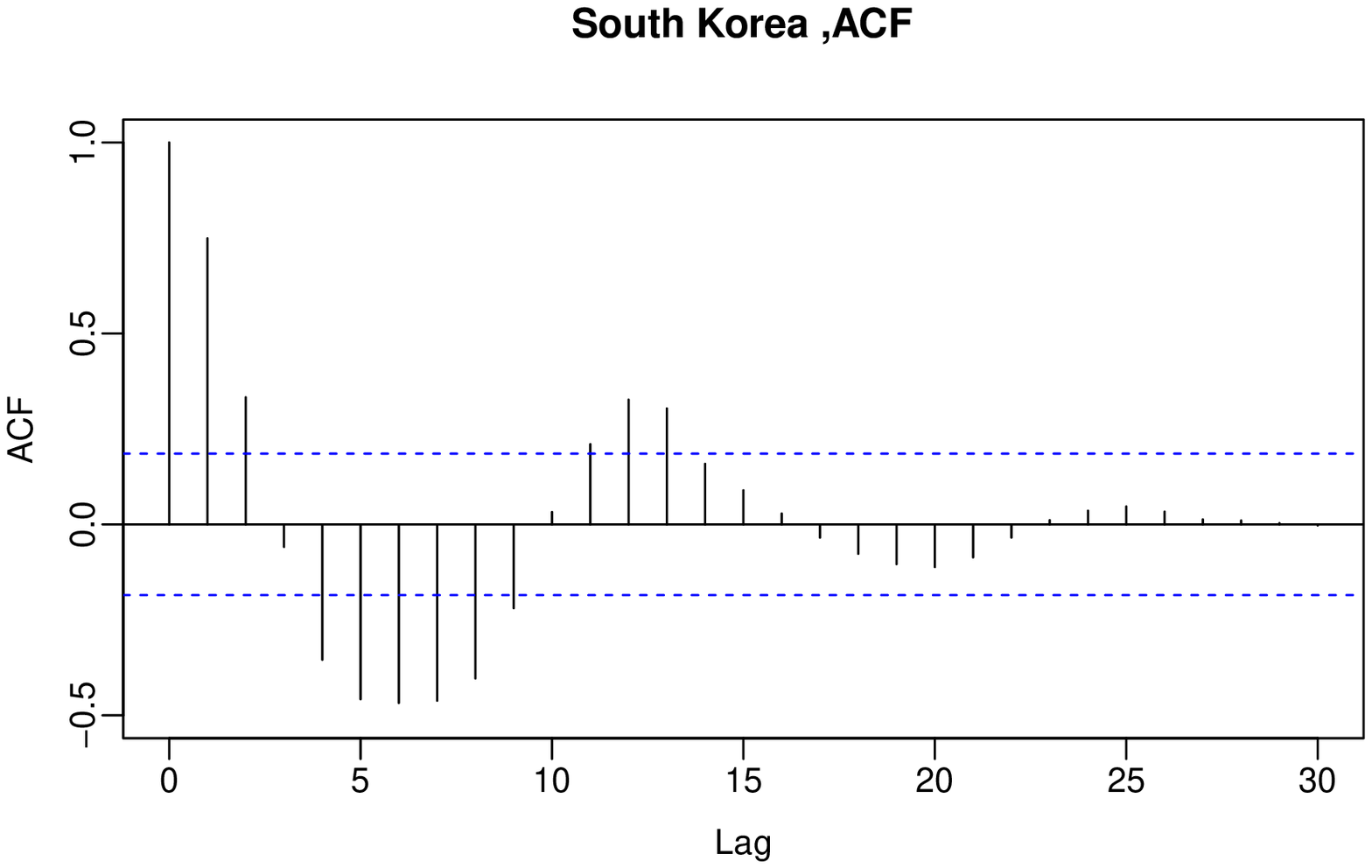}}
	\hfil\vspace{-8mm}
	\caption{ACF plot of residuals for 8 representative countries }
	\label{ACF}
\end{figure}

\begin{figure}[H]
	\centering
	\subfigure{	\hspace{-6mm}	
		\includegraphics[width=0.52\linewidth]{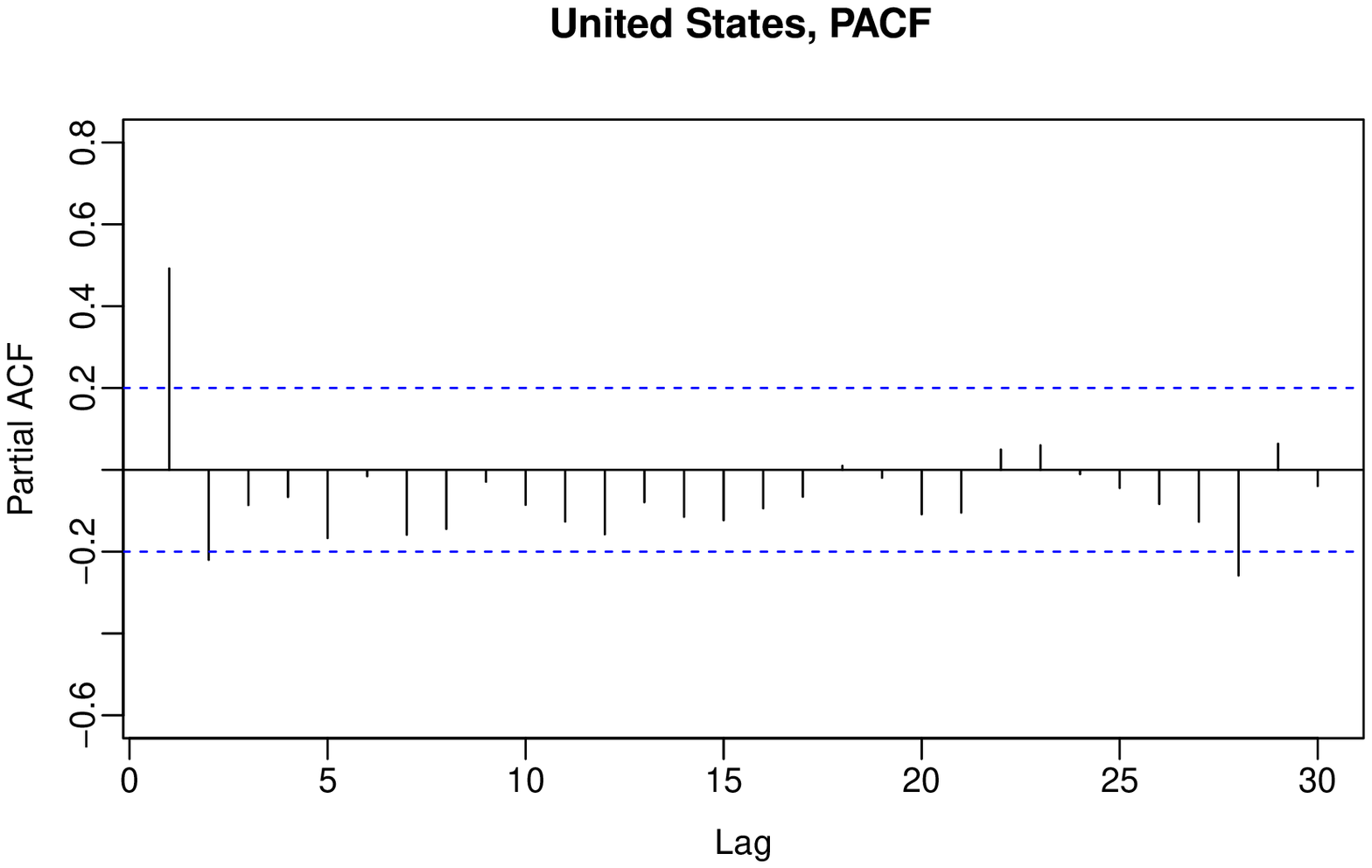}}
	\hfil
	\subfigure{	\hspace{-7mm}
		\includegraphics[width=0.52\linewidth]{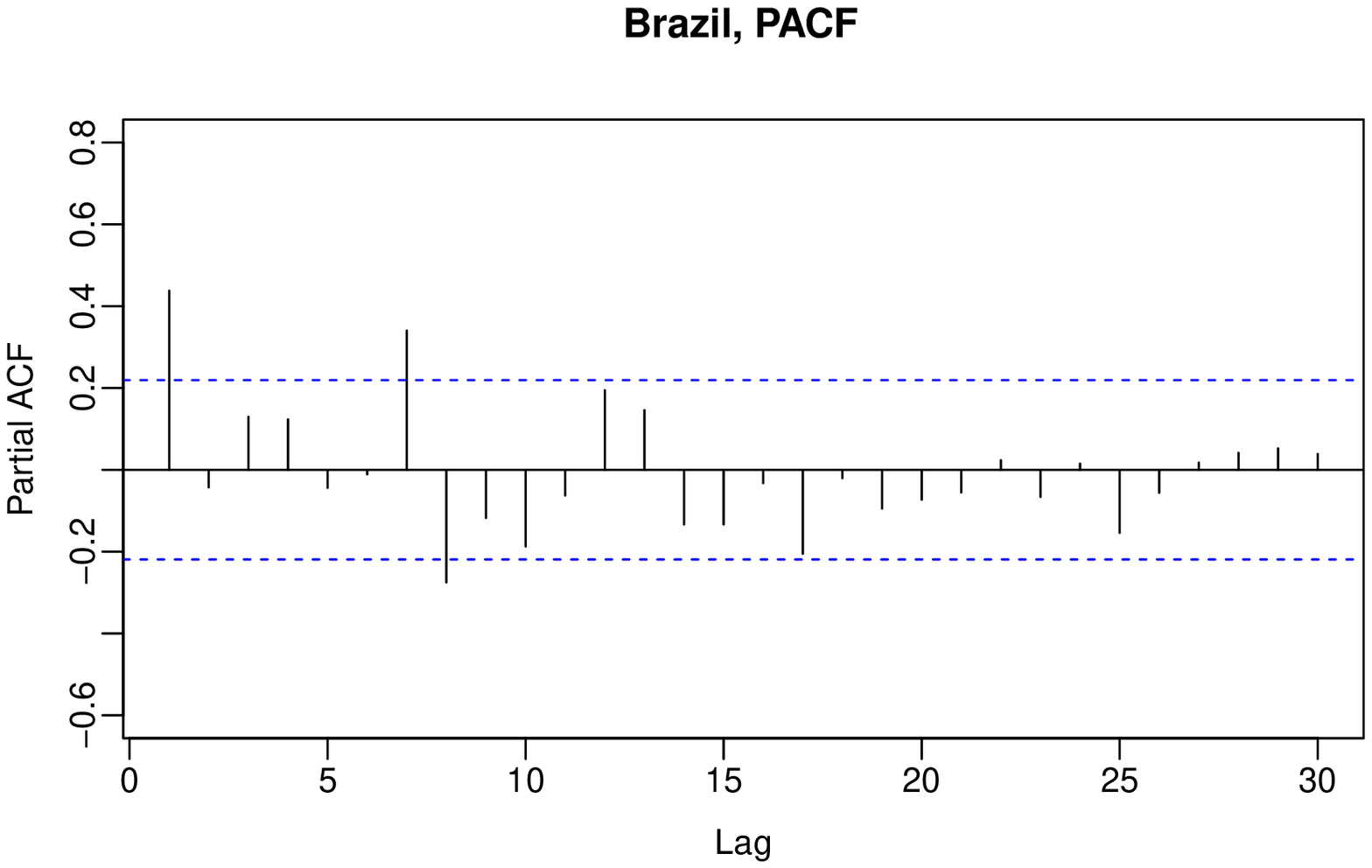}}
	\hfil\vspace{-8mm}
	\subfigure{\hspace{-5mm}
		\includegraphics[width=0.52\linewidth]{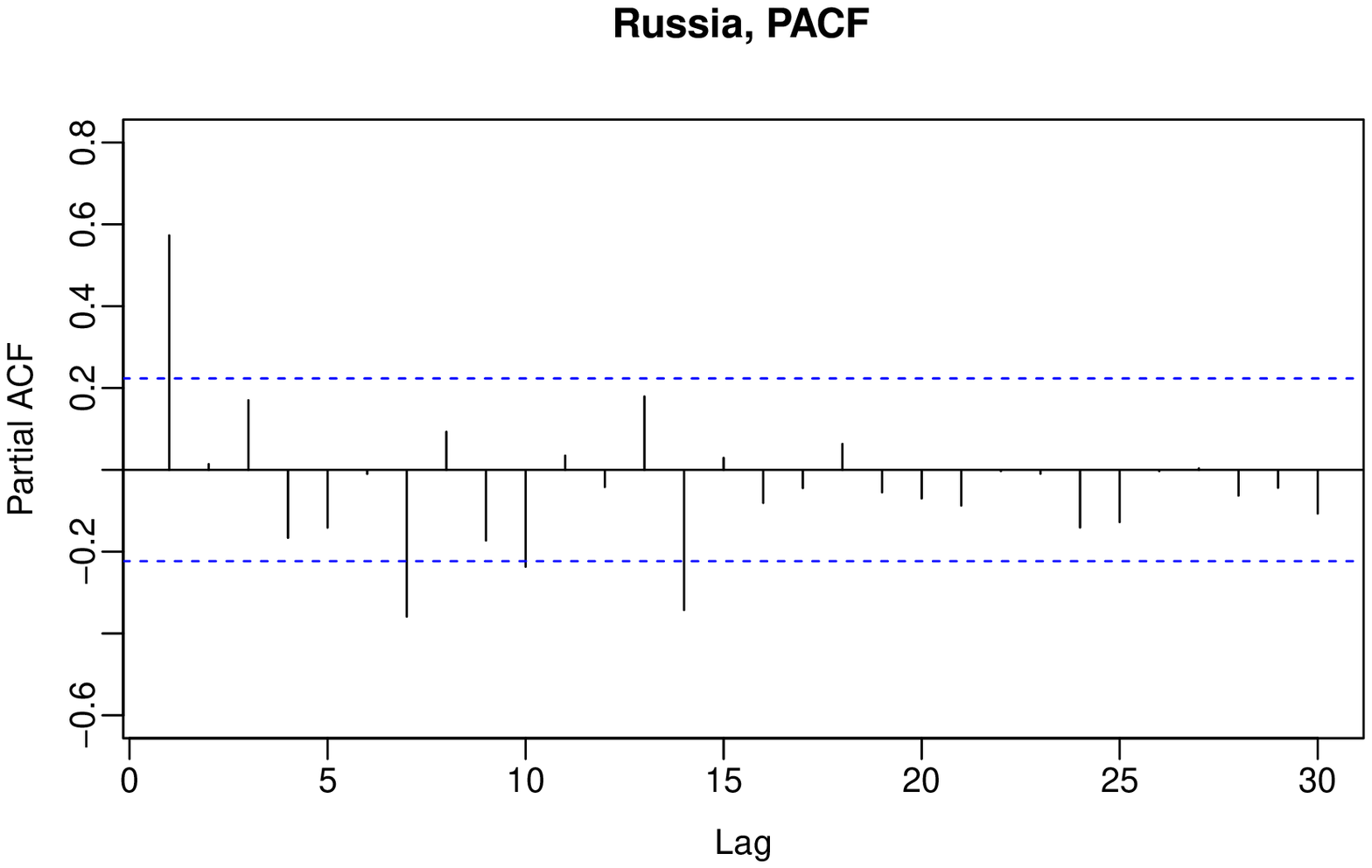}}
	\hfil
	\subfigure{\hspace{-6mm}
		\includegraphics[width=0.52\linewidth]{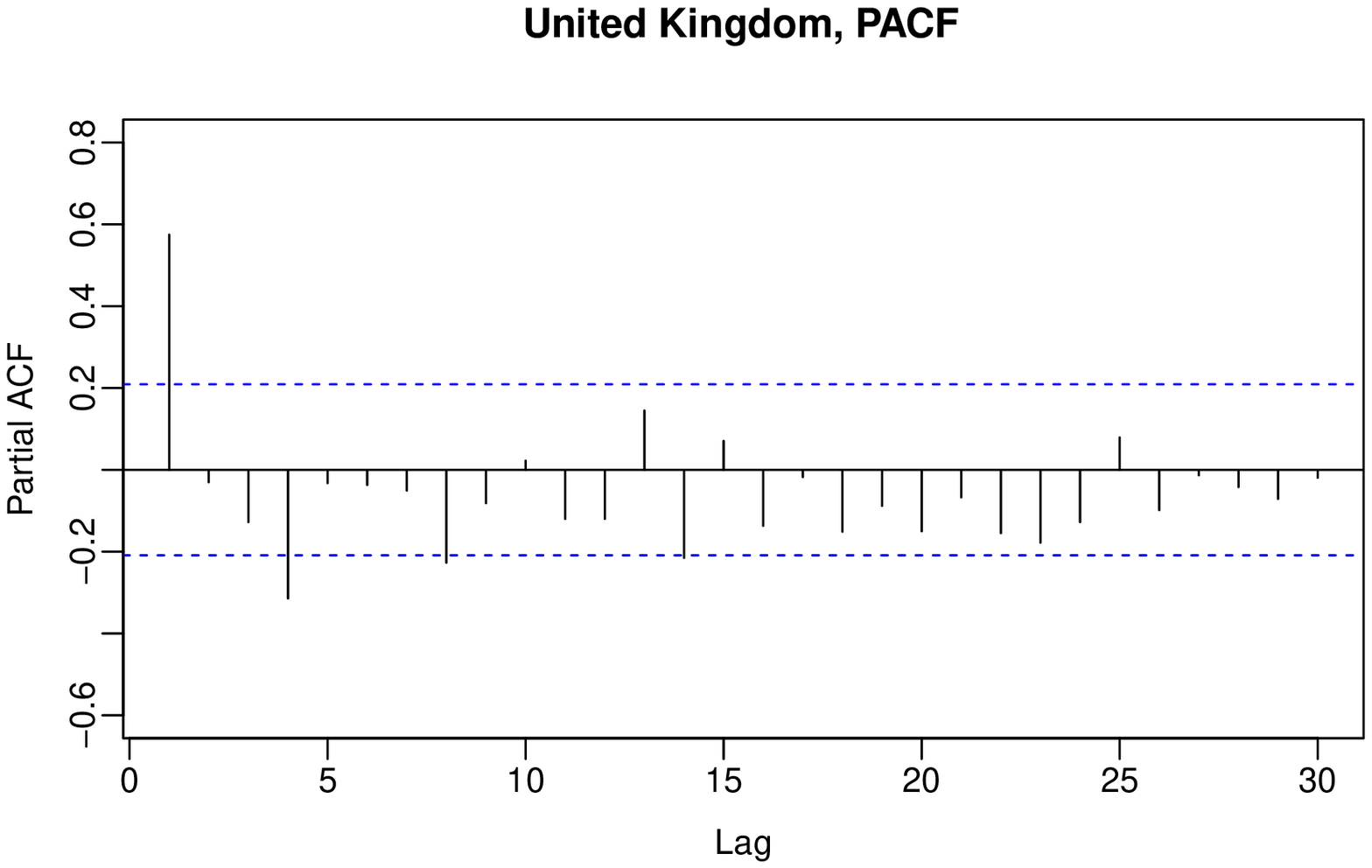}}	
	\hfil\vspace{-8mm}
	\subfigure{\hspace{-5mm}
		\includegraphics[width=0.52\linewidth]{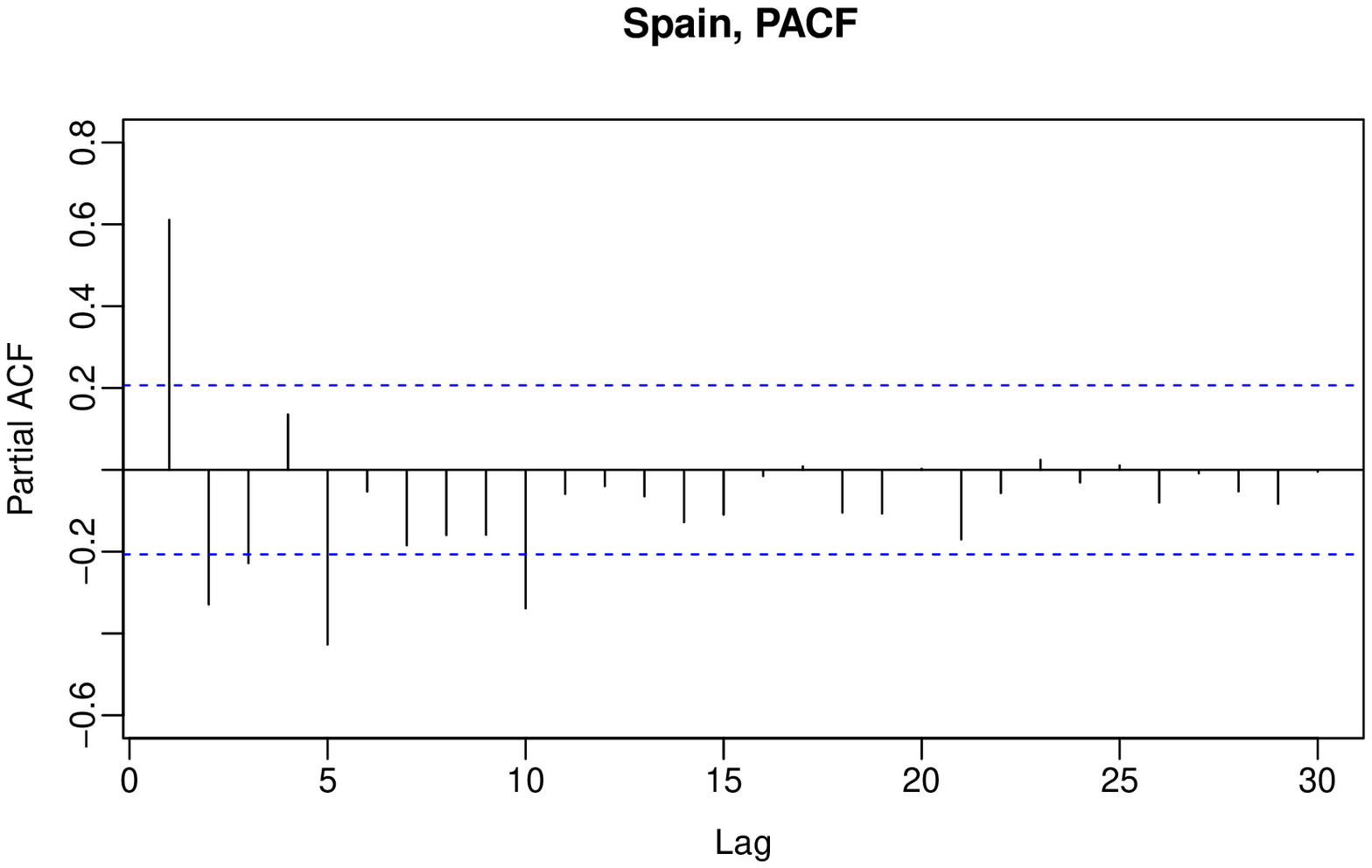}}
	\hfil
	\subfigure{\hspace{-6mm}
		\includegraphics[width=0.52\linewidth]{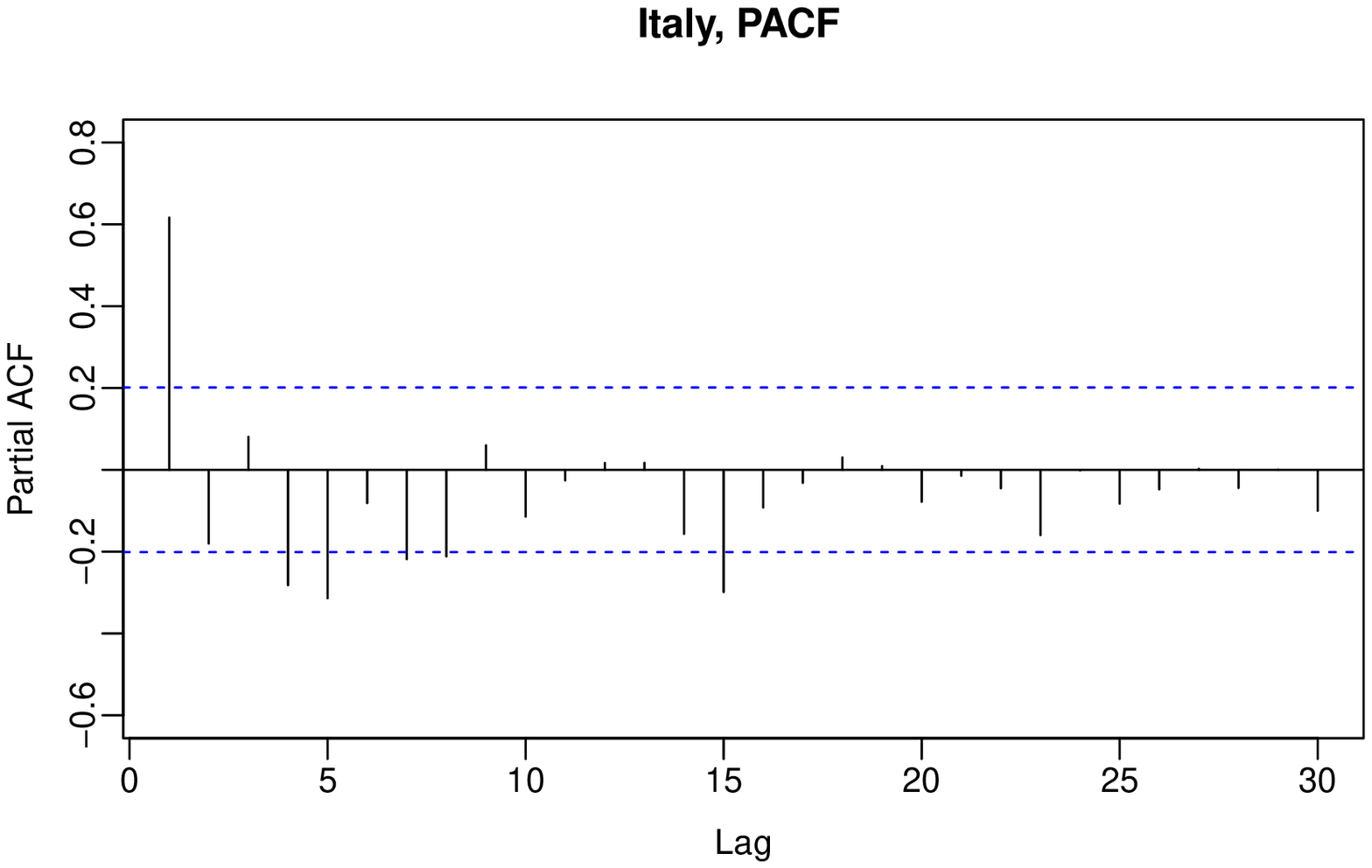}}
	\hfil\vspace{-8mm}
	
	\subfigure{\hspace{-5mm}
		\includegraphics[width=0.52\linewidth]{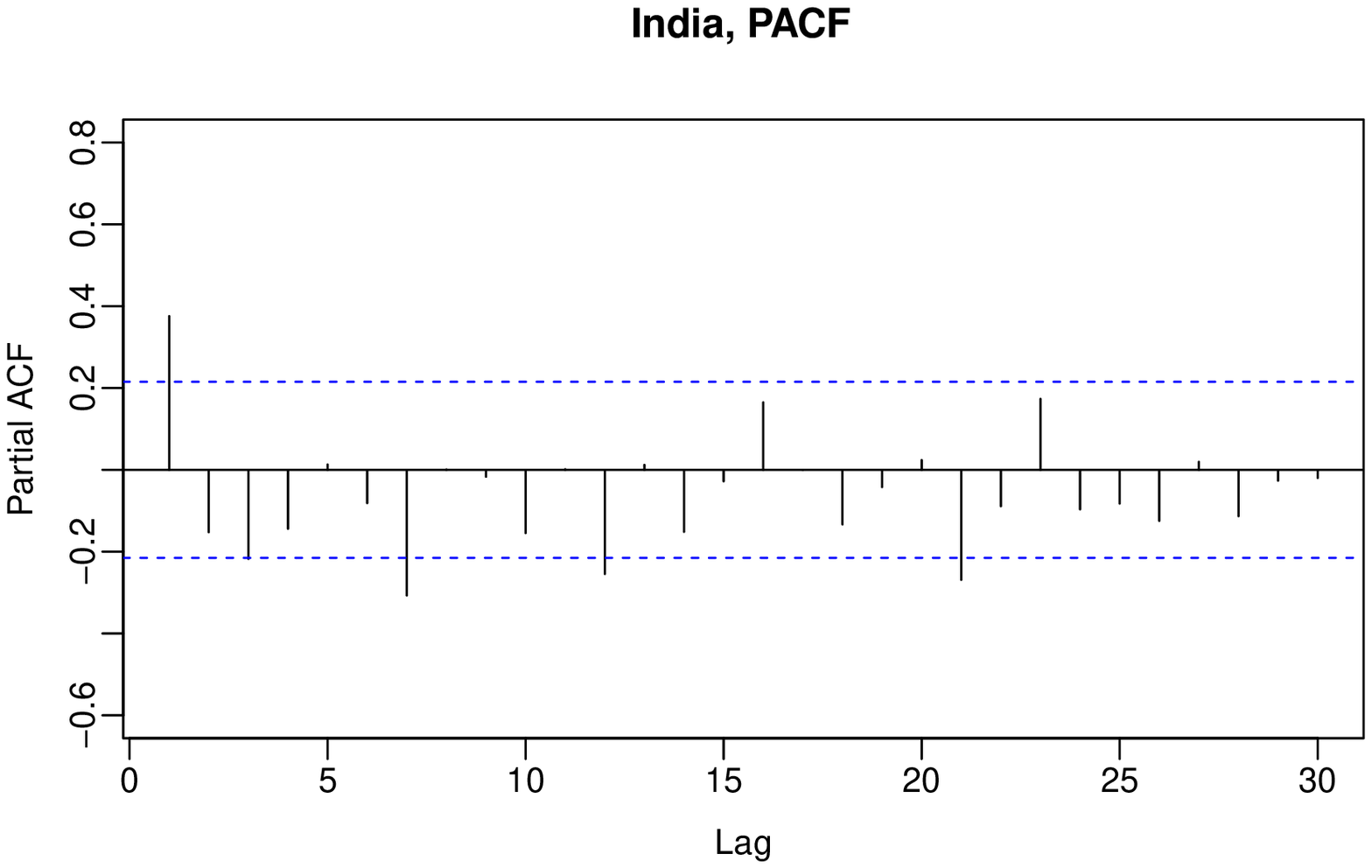}}
	\hfil
	\subfigure{\hspace{-6mm}
		\includegraphics[width=0.52\linewidth]{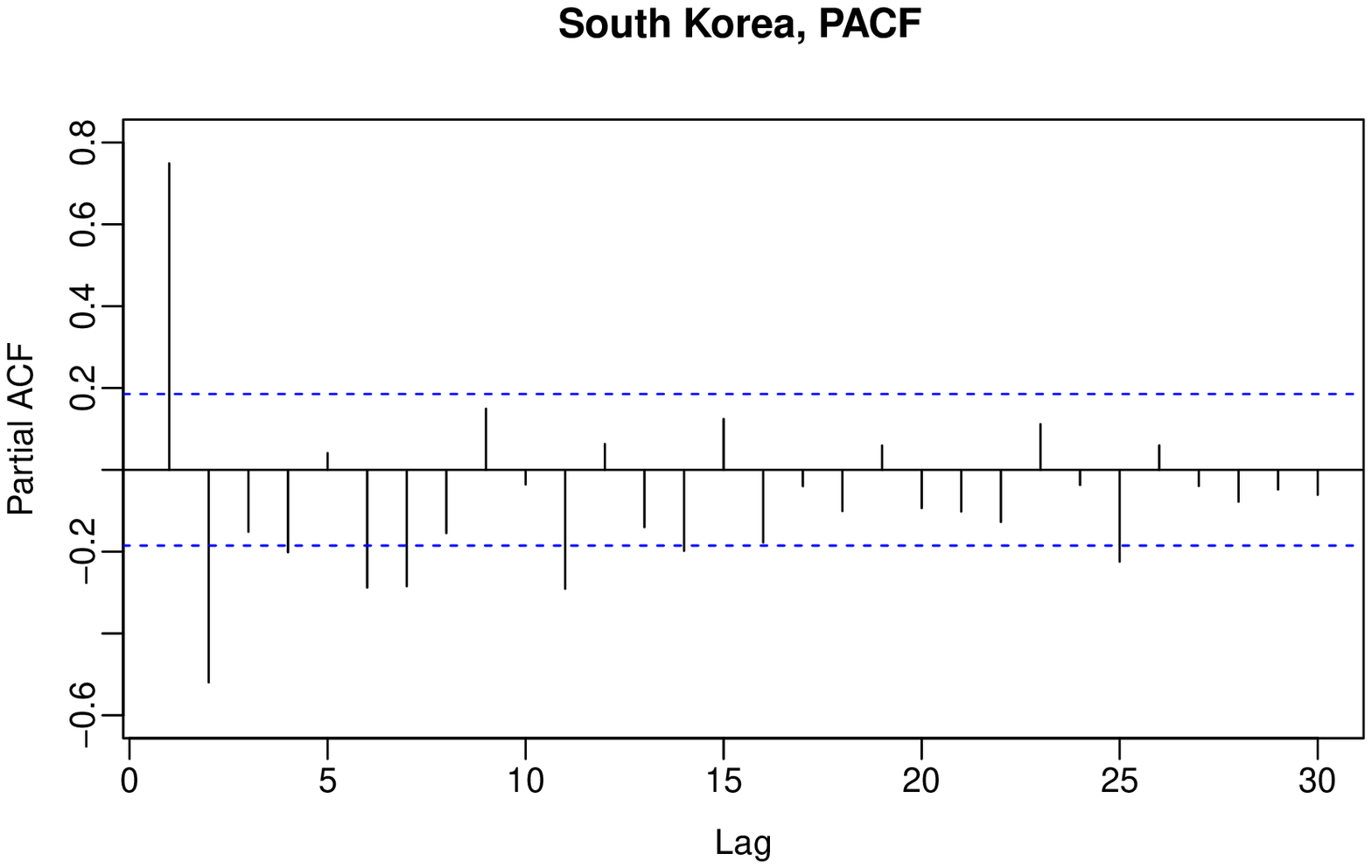}}
	\hfil\vspace{-8mm}
	\caption{PACF plot of residuals for 8 representative countries }
	\label{PACF}
\end{figure}

\end{document}